\documentclass[a4paper,11pt]{article}

\usepackage{jheppub} 
\usepackage{fancyhdr}
\usepackage{graphicx}
\usepackage{epstopdf}
\usepackage{psfrag}
\usepackage{xspace}

\usepackage{braket}

\DeclareSymbolFont{extraup}{U}{zavm}{m}{n}
\DeclareMathSymbol{\varheart}{\mathalpha}{extraup}{86}
\DeclareMathSymbol{\vardiamond}{\mathalpha}{extraup}{87}

\usepackage{amsfonts}
\newcommand{\Yes}{\checkmark}
\usepackage{pifont}
\newcommand{\No}{\hspace{1pt}\ding{55}}
\newcommand{\SARAH}{{\tt SARAH}\xspace}
\newcommand{\MATHEMATICA}{{\tt MATHEMATICA}\xspace}

\newcommand{\be}{\begin{equation}}
\newcommand{\ee}{\end{equation}}
\newcommand{\bea}{\begin{eqnarray}}

\newcommand{\eea}{\end{eqnarray}}

\newcommand{\refe}[1]{Eqn.~(\ref{#1})}

\newcommand{\Rmnum}[1]{\expandafter\@slowromancap\romannumeral #1@}

\begin{document} 


 \title{\boldmath  Natural Supersymmetry and Unification\\ in Five Dimensions }
 
 \begin{flushright}
WITS-MITP-005\\
 LYCEN-2015-03
 \end{flushright}

\author[\bigstar,\diamondsuit]{Ammar Abdalgabar,}
\author[\bigstar]{Alan S. Cornell,}
\author[\clubsuit,\dagger]{Aldo Deandrea}
\author[\spadesuit]{and Moritz McGarrie}


\affiliation[\bigstar]{National Institute for Theoretical Physics; School of Physics and Mandelstam Institute for Theoretical Physics, University of the Witwatersrand, Johannesburg, Wits 2050, South Africa}
\affiliation[\diamondsuit]{Department of Physics, Sudan University of Science and Technology,
Khartoum 407, Sudan}
\affiliation[\clubsuit]{Universit\'e de Lyon, F-69622 Lyon, France; Universit\'e Lyon 1, CNRS/IN2P3, UMR5822 IPNL, F-69622 Villeurbanne Cedex, France}
\affiliation[\dagger]{Institut Universitaire de France, 103 boulevard Saint-Michel, 75005 Paris, France}
\affiliation[\spadesuit]{Faculty of Physics, University of Warsaw, Ho\.za 69, 00-681 Warsaw, Poland}
\emailAdd{ammar.abdalgabar@wits.ac.za}
\emailAdd{alan.cornell@wits.ac.za}
\emailAdd{deandrea@ipnl.in2p3.fr}
\emailAdd{moritz.mcgarrie@fuw.edu.pl}


\abstract{We explore unification and natural supersymmetry in a five dimensional extension of the standard model in which the extra dimension may be large, of the order of 1-10 TeV. Power law running generates a TeV scale $A_t$ term allowing for the observed 125 GeV Higgs and allowing for stop masses below $2$ TeV, compatible with a natural SUSY spectrum. We supply the full one-loop RGEs for various models and use metastability to give a prediction that the gluino mass should be lighter than $3.5$ TeV for $A_{t}\leq-2.5$ TeV, for such a compactification scale, with brane localised 3rd generation matter.  We discuss why models in which the 1st and 2nd generation of matter fields are located in the bulk are likely to be ruled out.  We also look at electroweak symmetry breaking in these models.} 

\keywords{Large A-term, extra dimension, light third-generation squarks}

\maketitle
\flushbottom


\section{Introduction} \label{sec:intro}

\par In the context of supersymmetry and through the prism of the naturalness aesthetic, the discovery of a Standard Model-like scalar particle of mass $m_h \sim 125$ GeV \cite{Aad:2012tfa,Chatrchyan:2012ufa}, and no direct evidence so far of superparticles has motivated renewed interest in non-minimal extensions of the Supersymmetric Standard Model (SSM) that can help to compellingly explain such results. Within the Minimal-SSM (MSSM), for the lightest CP even charge neutral scalar to be the discovered scalar then requires either multi-TeV stops, which is disfavoured from naturalness, an enhancement to the tree-level Higgs mass such as for example \cite{Batra:2003nj,Maloney:2004rc,Bharucha:2013ela,McGarrie:2014xxa}, or a near maximal mixing scenario whereby $|A_t(M_z)|\gtrsim  1$ TeV. There are few models that compellingly achieve a large enough $A_t$ if one first assumes $A_t$ to vanish at some initial supersymmetry breaking scale. Even if one obtains such a large $A_t$, one must still explain why stops are lighter than their first and second generation counterpart squarks, consistent with colliders bounds \cite{ATLAS3rd,CMS3rd}. One such framework that can address both problems is a \emph{five dimensional}-SSM. 

\par In five dimensional (5D) supersymmetric standard models, power law running for a sufficiently low compactification radius $R$, generates at low energies a large enough $A_t$ to explain the observed Higgs mass \cite{Abdalgabar:2014bfa}. Furthermore, through spatially localising different generations along the extra dimension(s), one can explain geometrically why the third generation can be consistently lighter than its first and second generation counterparts \cite{Abdalgabar:2014bfa}.  

\par This framework is sufficiently compelling that it should understandably endure further scrutiny. In particular,  five dimensional theories are effective field theories with a cutoff and are (often over-dramatically) defined as non-renormalisable as many parameters such as gauge couplings, can be sensitive to this UV scale.  It is therefore important to confirm that results and conclusions made at one loop that are sensitive to this scale are still consistent and under control at two (and higher) loops. For instance one might be concerned that one loop linear sensitivity to the cutoff behaving as $\Lambda R$ do not result in terms of the form $(\Lambda R)^2$ at two-loop, which would then indicate a break-down of perturbation theory at renormalisation scales of the order of the compactification radius \cite{Oliver:2003cy}. Whilst this might be of concern to non-supersymmetric theories, the five dimensional SSM is reinterpreted in the language of  $\mathcal{N}=2$ four dimensional supersymmetry. This additional supersymmetry and the protection it affords, helps to reduce such terms \cite{Masip:2000yw,Hossenfelder:2004up}, at least for gauge couplings. The effect remains but has opposite sign for both Yukawa couplings and their soft breaking trilinear counterparts and so is still under complete control. For the case of bulk matter and in particular the top Yukawa in the bulk, a Landau pole appears and one must then seriously consider that either perturbation theory is problematic for these models (just as one would in any four dimensional theory with a Landau pole), or that a compelling explanation of how the top Yukawa may arise must be found such that this pathology may be avoided. 

\par There are further criteria for our model to be truly compelling: We require that it is supersymmetric and that supersymmetry is softly broken, that the superpotential is renormalisable and that the theory's gauge couplings \emph{unify} in the five dimensional description with a large enough extra dimensional scale as to make the extra dimensional features practically relevant to the phenomenology of the model. In other words we require a $1/R \sim 1$ to $10^3$ TeV scale extra dimension and not simply an (almost) GUT scale extra dimension. Such a criteria is useful to rule out certain models, for instance by this criteria one can straightforwardly rule out flat extra dimensional models in which the 1st and 2nd generation are in the bulk, with the 3rd generation either in the bulk or on a brane, as such a model can only unify with an extra dimensional scale of the order of the GUT scale, a topic we discuss in more detail later.

\par The outline of the paper is as follows: in section \ref{sec:model} we describe the models in detail and discuss unification. In section \ref{sec:explore} we describe our boundary conditions and how the four dimensional (4D) and 5D renormalisation group equations (RGEs) are matched and solved. We discuss the various energy scales of the model and then look at the running of various parameters including the gaugino mass spectrum and trilinear soft breaking terms. In section \ref{sec:Higgsmass} we explore how to obtain the correct 125 GeV Higgs mass, with stops lighter than 2 TeV. In section \ref{sec:conclude} we give our conclusions. We also include two detailed appendices, appendix \ref{RGES4D} including all the one-loop and two-loop RGEs of the four dimensional low energy model, of which we used the one-loop RGEs in the plots, and appendix \ref{RGES5D} includes the one-loop RGEs for the five dimensional models 1 and 2 of the main paper. The conventions and notation of this paper follow closely that of \cite{Abdalgabar:2014bfa}, which are based on conventions found in \cite{Hebecker:2001ke,Mirabelli:1997aj,McGarrie:2010kh,Cornell:2011fw,McGarrie:2013hca}. 


\section{5D-SSM with additional states: unification}\label{sec:model}
\begin{table}
\begin{center} 
\begin{tabular}{|c|c|c|c|c|} 
\hline 
Superfields & Brane& Bulk & $U(1)_Y\times SU(2)_L \times SU(3)_c$ \\
\hline 
\(\hat{q}^{f}\) &\Yes&-&  \((\frac{1}{6},{\bf 2},  {\bf 3}) \)  \\ 
\(\hat{d}^{f}\) &\Yes&-& \((\frac{1}{3},
{\bf 1},  {\bf \overline{3}}) \)  \\ 
\(\hat{u}^{f}\) &\Yes&-& \((-\frac{2}{3},
{\bf 1},  {\bf \overline{3}}) \)  \\
\(\hat{l}^{f}\) & \Yes&-& \((-\frac{1}{2}, {\bf 2}, {\bf 1}) \) \\
\(\hat{e}^{f}\) & \Yes&-& \((1, {\bf 1},
 {\bf 1}) \)  \\  \hline\hline
\(\hat{H}_d\)   & - &\Yes  & \( (-\frac{1}{2},{\bf 2},  {\bf 1}) \) \\ 
\(\hat{H}_u\) &  - &\Yes      & \( (\frac{1}{2},{\bf 2},  {\bf 1}) \)  \\ 
 \hline
\(\hat{F}_-\)   & - &\Yes     & \( (-1,{\bf 1},  {\bf 1}) \) \\ 
\(\hat{F}_+\) & - &\Yes       & \( (1,{\bf 1},  {\bf 1}) \)  \\ 
 \hline\hline
\(\hat{B}_V\)   & - &\Yes     & \( (0,{\bf 1},  {\bf 1}) \) \\ 
\(\hat{W}_V\) & - &\Yes       & \( (0,{\bf 3},  {\bf 1}) \)  \\ 
\(\hat{G}_V\)   & - &\Yes     & \( (0,{\bf 1},  {\bf 8}) \) \\ 
 \hline\hline
\end{tabular} \caption{The matter content of {\bf model 1}. All superfields of chiral fermions live on a brane and all Higgs-type superfields and gauge vector fields live in the bulk. The superscript $f=1,2,3$ denotes the generations. Neutrino superfields may be included straightforwardly. The gauge couplings of this model unify as in figure \ref{fig:alphas5D} (top left). 
\label{Table:Model1}}
\end{center} 
\end{table}
\par A TeV scale supersymmetric standard model in which the gauge coupling is precisely unified is proposed in \cite{Delgado:1998qr}. The key idea is to add two new hypermultiplets $F^{\pm}$ which are singlets under $SU(3)_c\times SU(2)_L$ and charged under $U(1)_Y$ with $Y_{F^{\pm}}=\pm 1$. The SSM chiral fermions are located on a boundary and in the 5D picture do not have Kaluza-Klein (KK) modes. The SSM Higgs chiral multiplets live in the bulk and we embed them as hypermultiplets in 5D.  The gauge fields and the additional states also live in the bulk as listed in table \ref{Table:Model1}: we call this {\bf model 1}. We will also explore our own model in which the third generation of superfields lives in the bulk, as in table \ref{Table:Model2}: we call this {\bf model 2} and this too may unify.  We compute and collate all supersymmetric and soft-term RGEs. These new states modify the beta function coefficient $b_1$ and lead to precision unification at one-loop.  The superpotential for both models is given by 
\begin{align} 
W = & \,  Y_u\,\hat{u}\,\epsilon_{ij} \hat{q}^i\,\hat{H}^j_u\,- Y_d
\,\hat{d}\,\epsilon_{ij} \hat{q}^i\,\hat{H}^j_d\,- Y_e \,\hat{e}\,\epsilon_{ij}
\hat{l}^i\,\hat{H}^j_d +\mu H_u H_d + \acute{\mu} F^{-} F^{+}\,\,.
\end{align} \label{eq:MSSM}
\begin{table}
\begin{center} 
\begin{tabular}{|c|c|c|c|c|} 
\hline 
Superfields & Brane& Bulk & $U(1)_Y\times SU(2)_L \times SU(3)_c$ \\
\hline 
\(\hat{q}^{1,2}\) &\Yes&-&  \((\frac{1}{6},{\bf 2},  {\bf 3}) \)  \\ 
\(\hat{d}^{1,2}\) &\Yes&-& \((\frac{1}{3},
{\bf 1},  {\bf \overline{3}}) \)  \\ 
\(\hat{u}^{1,2}\) &\Yes&-& \((-\frac{2}{3},
{\bf 1},  {\bf \overline{3}}) \)  \\
\(\hat{l}^{1,2}\) & \Yes&-& \((-\frac{1}{2}, {\bf 2}, {\bf 1}) \) \\
\(\hat{e}^{1,2}\) & \Yes&-& \((1, {\bf 1},
 {\bf 1}) \)  \\ 
\hline 
\(\hat{q}^{3}\) & - &\Yes&  \((\frac{1}{6},{\bf 2},  {\bf 3}) \)  \\ 
\(\hat{d}^{3}\) & - &\Yes& \((\frac{1}{3},
{\bf 1},  {\bf \overline{3}}) \)  \\ 
\(\hat{u}^{3}\) & - &\Yes& \((-\frac{2}{3},
{\bf 1},  {\bf \overline{3}}) \)  \\
\(\hat{l}^{3}\) &  - &\Yes& \((-\frac{1}{2}, {\bf 2}, {\bf 1}) \) \\
\(\hat{e}^{3}\) &  - &\Yes& \((1, {\bf 1},
 {\bf 1}) \)  \\  \hline\hline
\(\hat{H}_d\)   & - &\Yes  & \( (-\frac{1}{2},{\bf 2},  {\bf 1}) \) \\ 
\(\hat{H}_u\) &  - &\Yes      & \( (\frac{1}{2},{\bf 2},  {\bf 1}) \)  \\ 
 \hline
\(\hat{F}_-\)   & - &\Yes     & \( (-1,{\bf 1},  {\bf 1}) \) \\ 
\(\hat{F}_+\) & - &\Yes       & \( (1,{\bf 1},  {\bf 1}) \)  \\ 
 \hline\hline
\(\hat{B}_V\)   & - &\Yes     & \( (0,{\bf 1},  {\bf 1}) \) \\ 
\(\hat{W}_V\) & - &\Yes       & \( (0,{\bf 3},  {\bf 1}) \)  \\ 
\(\hat{G}_V\)   & - &\Yes     & \( (0,{\bf 1},  {\bf 8}) \) \\ 
 \hline\hline
\end{tabular} \caption{The matter content of {\bf model 2}. In this case the third generation also live in the bulk. The gauge couplings of this model unify as in figure \ref{fig:alphas5D} (top right).
\label{Table:Model2}}
\end{center} 
\end{table}
\begin{table}
\begin{center} 
\begin{tabular}{|c|c|c|c|c|} 
\hline 
Superfields & Brane& Bulk & $U(1)_Y\times SU(2)_L \times SU(3)_c$ \\
\hline 
\(\hat{q}^{1,2}\) & - &\Yes&  \((\frac{1}{6},{\bf 2},  {\bf 3}) \)  \\ 
\(\hat{d}^{1,2}\) & - &\Yes& \((\frac{1}{3},
{\bf 1},  {\bf \overline{3}}) \)  \\ 
\(\hat{u}^{1,2}\)& - &\Yes& \((-\frac{2}{3},
{\bf 1},  {\bf \overline{3}}) \)  \\
\(\hat{l}^{1,2}\) & - &\Yes& \((-\frac{1}{2}, {\bf 2}, {\bf 1}) \) \\
\(\hat{e}^{1,2}\)& - &\Yes& \((1, {\bf 1},
 {\bf 1}) \)  \\ 
\hline 
\(\hat{q}^{3}\) &\Yes &-&  \((\frac{1}{6},{\bf 2},  {\bf 3}) \)  \\ 
\(\hat{d}^{3}\) &\Yes &-& \((\frac{1}{3},
{\bf 1},  {\bf \overline{3}}) \)  \\ 
\(\hat{u}^{3}\) &\Yes &-& \((-\frac{2}{3},
{\bf 1},  {\bf \overline{3}}) \)  \\
\(\hat{l}^{3}\) &\Yes &-&\((-\frac{1}{2}, {\bf 2}, {\bf 1}) \) \\
\(\hat{e}^{3}\)&\Yes &-& \((1, {\bf 1},
 {\bf 1}) \)  \\  \hline\hline
\(\hat{H}_d\)   & - &\Yes  & \( (-\frac{1}{2},{\bf 2},  {\bf 1}) \) \\ 
\(\hat{H}_u\) &  - &\Yes      & \( (\frac{1}{2},{\bf 2},  {\bf 1}) \)  \\ 
 \hline
\(\hat{F}_-\)   & - &\Yes     & \( (-1,{\bf 1},  {\bf 1}) \) \\ 
\(\hat{F}_+\) & - &\Yes       & \( (1,{\bf 1},  {\bf 1}) \)  \\ 
 \hline\hline
\(\hat{B}_V\)   & - &\Yes     & \( (0,{\bf 1},  {\bf 1}) \) \\ 
\(\hat{W}_V\) & - &\Yes       & \( (0,{\bf 3},  {\bf 1}) \)  \\ 
\(\hat{G}_V\)   & - &\Yes     & \( (0,{\bf 1},  {\bf 8}) \) \\ 
 \hline\hline
\end{tabular} \caption{The matter content of {\bf model 3}. In this case the 1st and 2nd generation live in the bulk. The gauge couplings of this scenario do not unify, as in figure \ref{fig:alphas5D} (bottom).
\label{Table:Model3}}
\end{center} 
\end{table}

\subsection{Gauge coupling unification}

\par A sufficient condition for unification in a five dimensional model is \cite{Dienes:1998vh,Dienes:1998vg} that
\be
R_{ij}=\frac{b^{(5D)}_i-b^{(5D)}_j}{b^{SSM}_i-b^{SSM}_j}
\ee
does not depend on $(i,j)$, where  $b^{5D}_i$ are the five dimensional beta function coefficients, at one-loop. The $\beta$-function of an $SU(N)$ gauge theory at one-loop is
\be
\beta_g = \mu \frac{dg}{d\mu}=- \frac{g^3}{ 16 \pi^2} \left(\frac{11}{3} T(Adj) - \frac{2}{3} T_{fer}(R) - \frac{1}{3} T_{sc}(R)  \right) =  \frac{b_g^{(1-loop)} \, g^3}{ (2 \pi)^4}
\ee
for gauge fields, Weyl fermions and complex scalars respectively. $R$ is the representation and in particular $T(\textrm{Ad})=N  \   \textrm{and}  \ T(\Box) = \frac{1}{2} \,.$

\par For a $U(1)$ theory \cite{Aitchison:2007fn} the gauge field is uncharged, there is also an overall normalisation constant which can be fixed to embed the particular $U(1)$ in a larger group. Such that focusing on the $U(1)$ of the SSM one finds
\be
b_1 = \frac{3}{5}\left( \frac{2}{3} \sum_f Y_f^2 + \frac{1}{3}\sum_s Y_s^2  \right)\ \ \ \ \text{or} \ \ \ b_1 = \frac{3}{5}\left(  \sum_{\Phi} Y_{\Phi}^2   \right) \; ,
\ee
the latter is for chiral superfields, and the $Y$'s are hypercharges, where the hypercharge is rescaled by $g_1\equiv \sqrt{5/3}g'$ as usual in unified models \cite{Mohapatra:1997sp}. The results for various models may be found in table \ref{tableRGES}.  In a number of these scenarios additional matter is required to obtain unification, or indeed the extra dimensional scale $1/R>10^{10}$ GeV, which for phenomenological purposes is essentially four dimensional and so not of  interest.
\begin{table}[t]
\footnotesize
\begin{center} 
\begin{tabular}{|c|c|c|c|}
\hline
Scenario & $(b_1,b_2,b_3)$ &  Refs:& $1/R$-GUT \\
\hline
$4D$ SM & $(\frac{41}{10},-19/6,-7)$ &   & - \\
$4D$ MSSM & $(\frac{33}{5},1,-3)$ &  \cite{Martin:1997ns,Drees:2004jm} & -\\
$5D$ MSSM: Chiral Higgses in the bulk & $(\frac{3}{5},-3,-6)$ & \cite{Dienes:1998vg,Masip:2000yw} & $\star$(does not exist)  \\
$5D$ MSSM: Hyper Higgses in the bulk& $(\frac{6}{5},-2,-6)$ & \cite{Delgado:1998qr,Abdalgabar:2014bfa} &  $\sim 10^{10}$ GeV \\
$5D$ MSSM-UED & $(\frac{66}{5},10,6)$ & \cite{Bhattacharyya:2006ym,Abdalgabar:2013laa} &$\geq5\times10^{10}$ GeV\\
$5D$ 3rd Gen \& Hyper Higgses in the bulk & $(\frac{26}{5},2,-2)$ &  \cite{Bhattacharyya:2010rm}&$\sim 10^{10}$ GeV\\
$5D$ 1st,2nd  Gen \& Hyper Higgses in the bulk & $(\frac{40}{5},4,2)$ &      & \No \\
$5D$ Gauge only in the bulk & $(0,-4,-6)$ &  &$\sim 10^{10}$ GeV\\
\hline
$4D$ $SSMF^{\pm}$ & $(\frac{39}{5},1,-3)$ &  &- \\
$5DSSMF^{\pm}$:Hyper Higgses in bulk& $(\frac{18}{5},-2,-6)$ &   {\bf model 1} &  $\geq1$ TeV\\
$5DSSMF^{\pm}$:3rd Gen \& Hyper Higgses in bulk & $(\frac{38}{5},2,-2)$ &    {\bf model 2}  & $\geq1$ TeV\\
$5DSSMF^{\pm}$:1st,2nd Gen \& Hyper Higgses in bulk & $(\frac{52}{5},4,2)$ &   {\bf model 3}    & \No \\
\hline
\hline
$4D$ MSSM+Dirac& $(\frac{33}{5},-1,0)$ &   &-\\
$4D$ M-Dirac-SSM & $(\frac{48}{5},4,0)$ &  \cite{Benakli:2014cia}&- \\
$5D$ MSYM only in the bulk & $(0,0,0)$ & \cite{McGarrie:2013hca} & any \\
$5D$ MSYM Hyper-Higgs in the bulk & $(\frac{6}{5},2,0)$ & &\No \\
\hline
\end{tabular}
\end{center}
\caption{The one-loop beta function coefficients of the gauge couplings for various scenarios. Requiring gauge coupling unification puts a bound on the inverse 
radius of the extra dimension in five dimensional models, which is estimated in the right-most column.}
\label{tableRGES}
\end{table}
\par
\begin{figure}[tb]
\begin{center}
\includegraphics[width=7.5cm,angle=0]{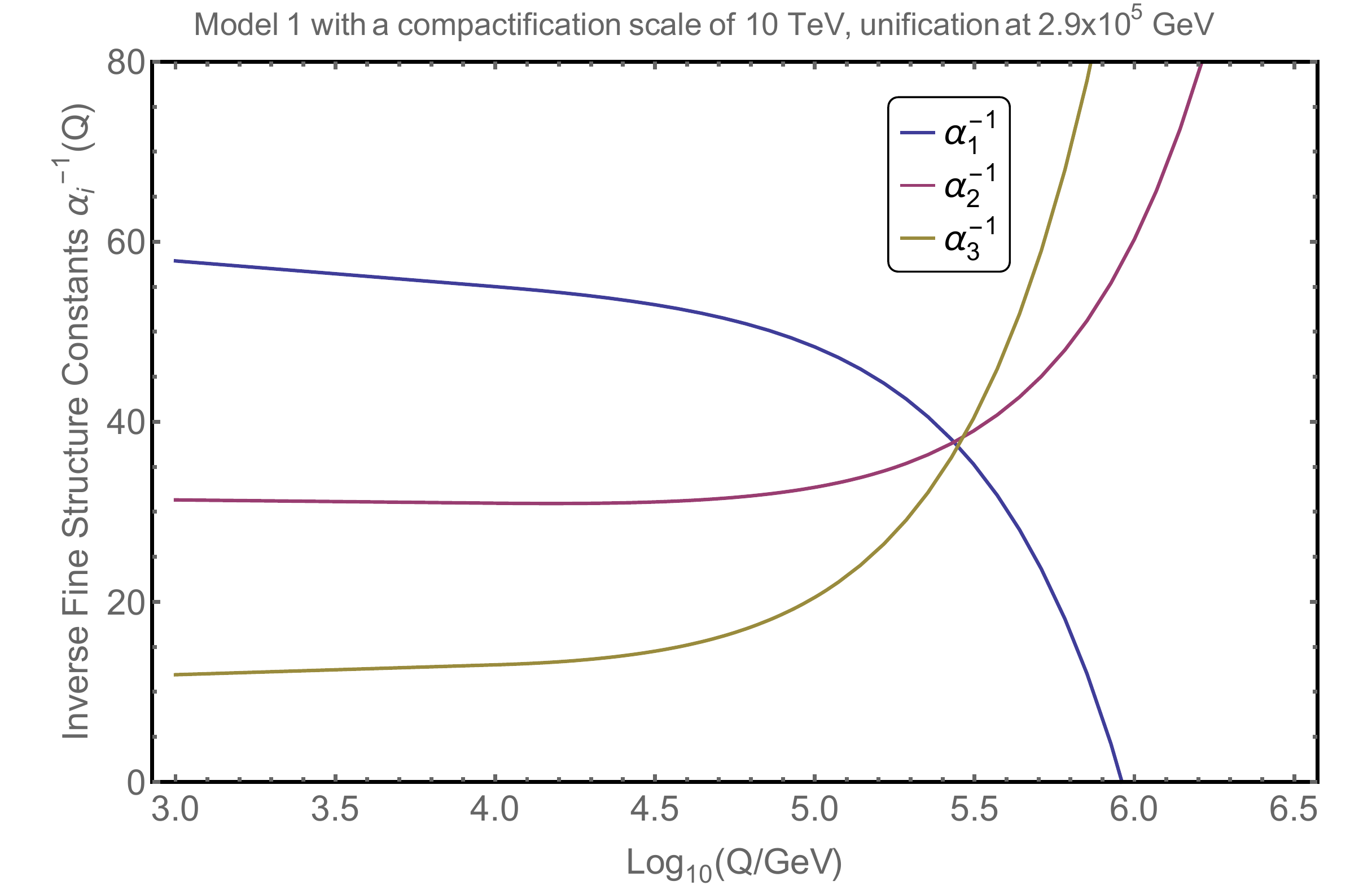}
\includegraphics[width=7.5cm,angle=0]{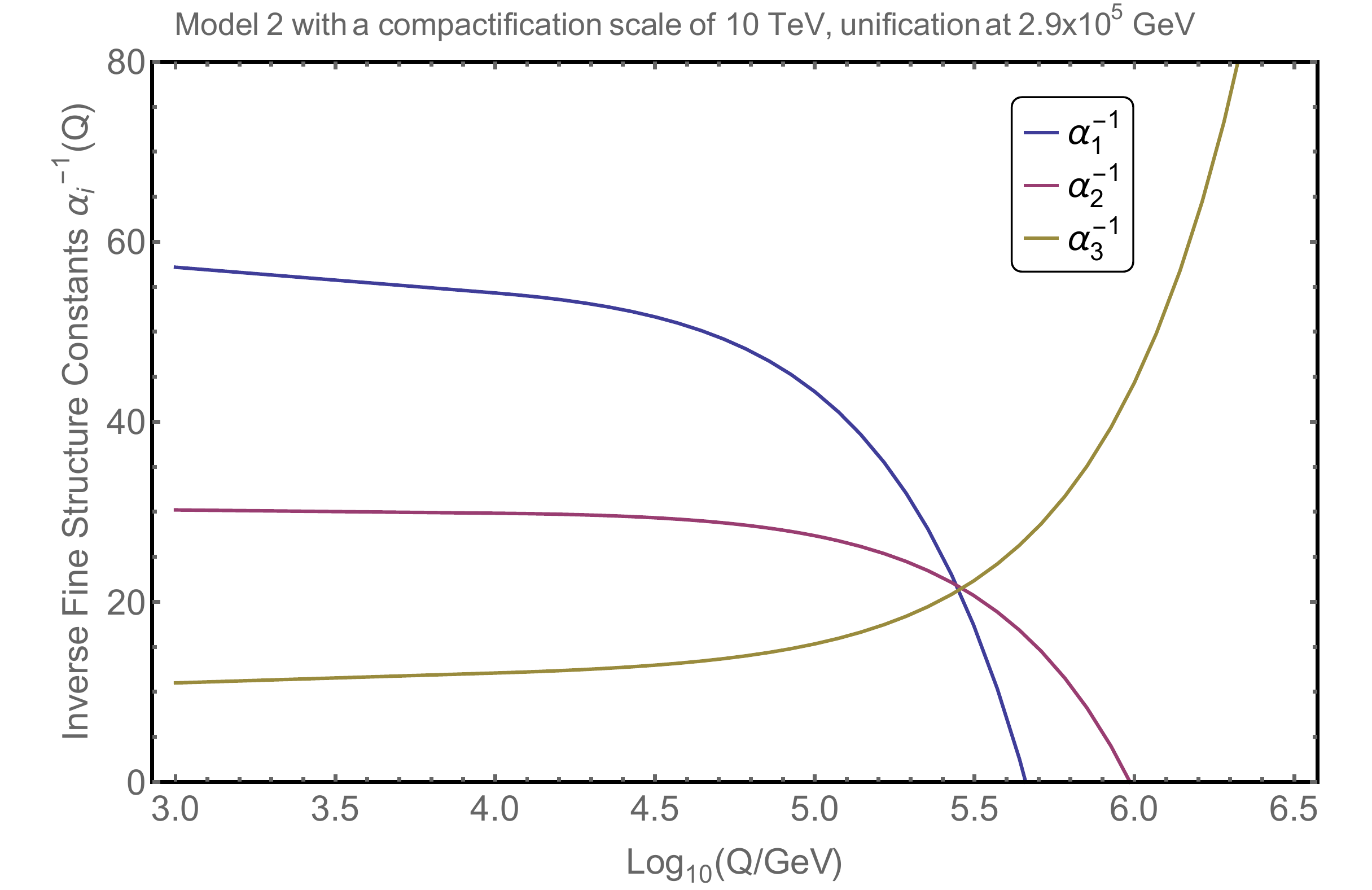}\qquad
\includegraphics[width=7cm,angle=0]{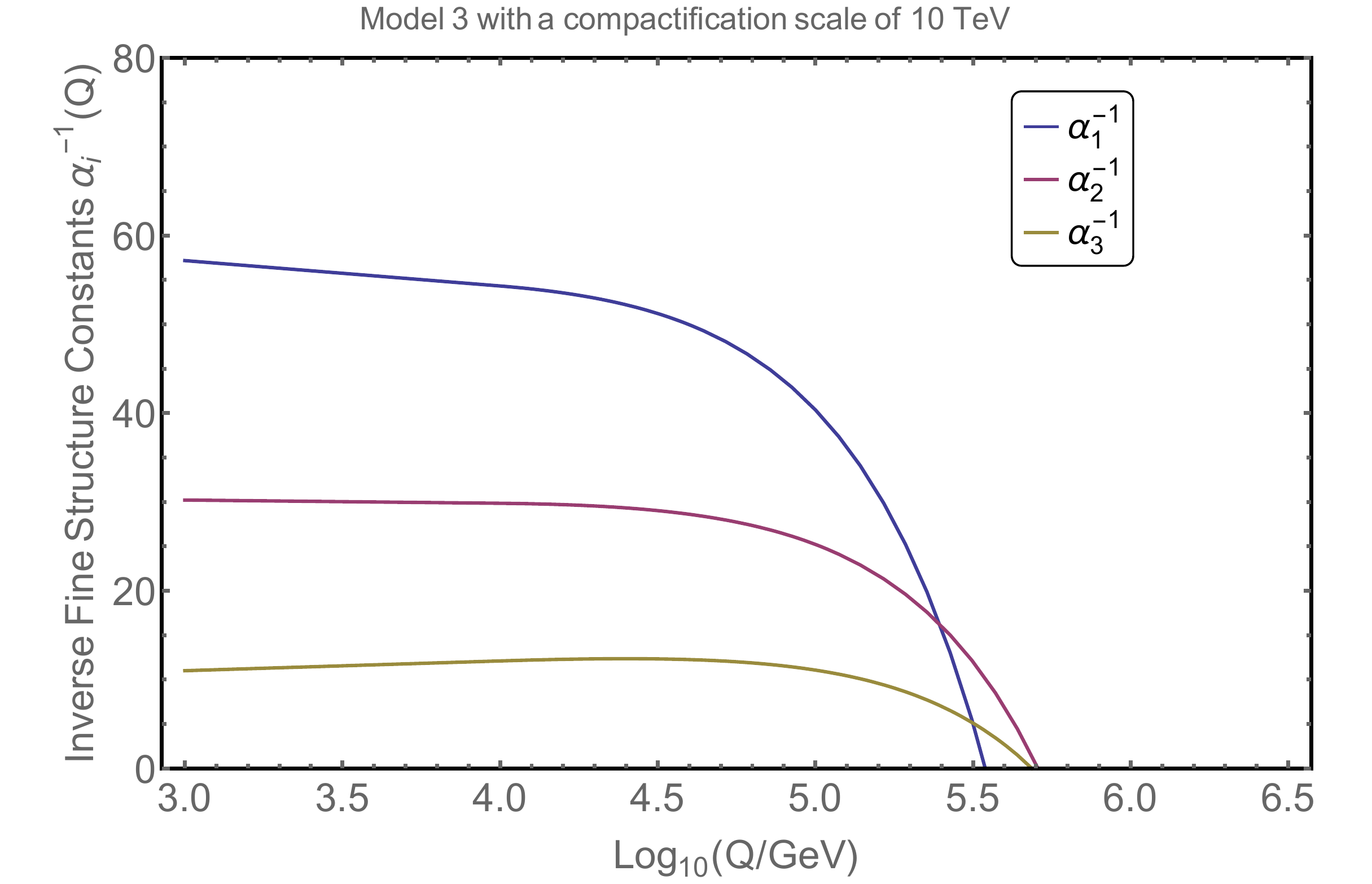}
\caption{{\it Running of the inverse fine structure constants $\alpha^{-1}_i(Q)$, for three different models with compactification scales 10 TeV  as a function of  $Log_{10}$(Q/GeV).}} 
\label{fig:alphas5D}
\end{center}
\end{figure}

 A useful comment is appropriate here that the additional matter of the $5D$ MSSM-UED  scenario means that all beta function coefficients are positive. This forces $1/R \gtrsim  10^{10}$ GeV for unification to still be possible \cite{Bhattacharyya:2006ym}. Low scale (supersymmetric) extra dimensions therefore require that most of the MSSM matter does not live in the bulk. Our preferred scenarios are therefore ones in which the matter multiplets all live on a brane ({\bf model 1}) or where the 1st and 2nd generation live on an opposite brane to the 3rd generation, or where only the third generation lives in the bulk ({\bf model 2}). In either case the Higgses can live in the bulk or on a brane. Additional fields may be added to accomplish precision unification at low scales \cite{Paccetti:2002xs}. This leads to three options: the models 1 and 2 that we consider in this paper (plotted in figure \ref{fig:alphas5D}) and one might also be able to combine a 4D M-Dirac-SSM  \cite{Benakli:2014cia} with a maximal super Yang-Mills theory only in the bulk \cite{McGarrie:2013hca} to, rather remarkably, achieve unification for any and all sizes of inverse radius. In this theory the gauge couplings only run in the four dimensional theory as the beta functions for the gauge couplings vanish exactly to all orders in perturbation theory in the maximal super Yang-Mills theory. As a result there are no power law contributions for gauge couplings (but there may be for the Yukawas and soft terms) and an inverse radius of a few TeV is possible with gauge coupling unification at $10^{17}$ GeV, which is very counter-intuitive. The effective cutoff of a five dimensional theory is essentially defined as the scale at which some parameter, such as the gauge couplings, hit a Landau pole: as no Landau pole arises this allows for the range of validity of this theory to extend further.

\par A case that fails to unify (which is denoted as {\bf model 3} in figure \ref{fig:alphas5D} (bottom)) is a scenario in which the 1st and 2nd generation of matter fields (as well as the Higgs doublets and gauge theories) live in the bulk, listed in table \ref{Table:Model3}. Such a theory has a hard time unifying its gauge couplings as in the near four dimensional limit it does not even reduce to the MSSM, due to the additional $F^{\pm}$.


\section{Exploring the models}\label{sec:explore}

\par In this section we explore the typical scales of the models, we describe how we solve the various RGEs and the boundary conditions that we use and then look at many of the running parameters  of the model, such as trilinear soft breaking parameters and the gaugino mass spectrum.

\subsection{Typical scales of the models}

\par It is useful to set the mass and energy scales in which we wish to consider these models. We wish for a large extra dimension, which then leads us to fix the gauge coupling unification scale and the scale of the cut off, where the gauge couplings hit a Landau pole (see figure \ref{fig:alphas5D}):
\be
\frac{1}{R} \sim 10~\text{TeV} \ , \  M_{GUT}\sim 300~\text{TeV}   \ , \ \Lambda \sim 1,000~\text{TeV}.
\ee
Although they differ in magnitude, this is natural in that fixing any one of these determines the other two. Next we wish for a gluino mass above collider exclusions and to determine the Higgs mass correctly to be $m_h=125$ GeV from a sizeable $A_t$. We find (see for instance figure \ref{fig:gauginotrilinear5D})
\be
M_3=900  \ \text{GeV} \  \text{leads to}  \ A_t\sim -700 \  \text{GeV}   \ \ , \ \  M_3=1700  \ \text{GeV} \  \text{leads to}  \ A_t\sim -1250 \  \text{GeV} .
\ee
Strong exclusion limits on the gluino arise from ATLAS and CMS null searches for jets plus missing energy, for example ~$m_{\tilde{g}} > 1600$ GeV for $m_{\tilde{q}_{1,2}} > 2000$ GeV~\cite{ATLAS-CONF-2013-047,Chatrchyan:2012lia}, although this can be lowered if one wishes to also include R-parity violation with our models, hence the  $M_3=900$ GeV case. Conversely, allowing for an upper bound on the top trilinear coupling, from considering metastability of the electroweak vacuum,
\be
  A_t=-2  \ \text{TeV} \  \text{leads to}  \ M_3\sim 2.77 \  \text{TeV} \ \ \text{and} \ \ A_t=-2.5  \ \text{TeV} \  \text{leads to}  \ M_3\sim 3.5 \  \text{TeV}.
\ee
To allow for the correct Higgs mass $m_h=125$ GeV, the electroweak parameters should be in the range 
\be
\tan \beta \subset (5,30), \ \  \mu \leq 1 \text{TeV},
\ee
represented in figure \ref{fig:Higgs5D}. We do not expect $\tan \beta$ to be much larger, due to $B_s \rightarrow X_s \gamma$ flavour constraints and $\mu$ is bounded by naturalness considerations of the renormalisation group effects on the Higgs tadpole equations (minimisation of the scalar potential).
\begin{figure}[tb]
\begin{center}
\includegraphics[width=7.5cm,angle=0]{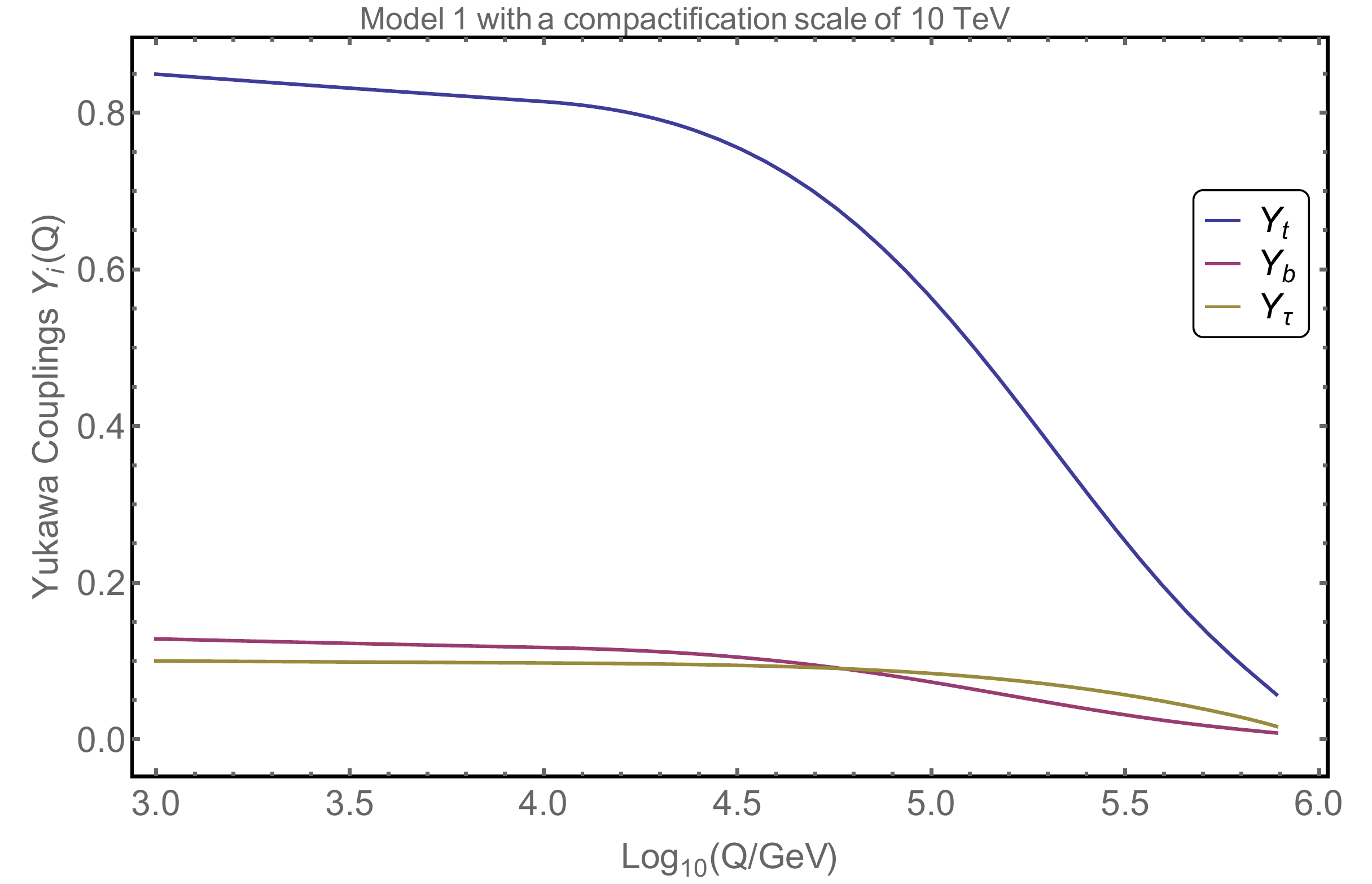}
\includegraphics[width=7.5cm,angle=0]{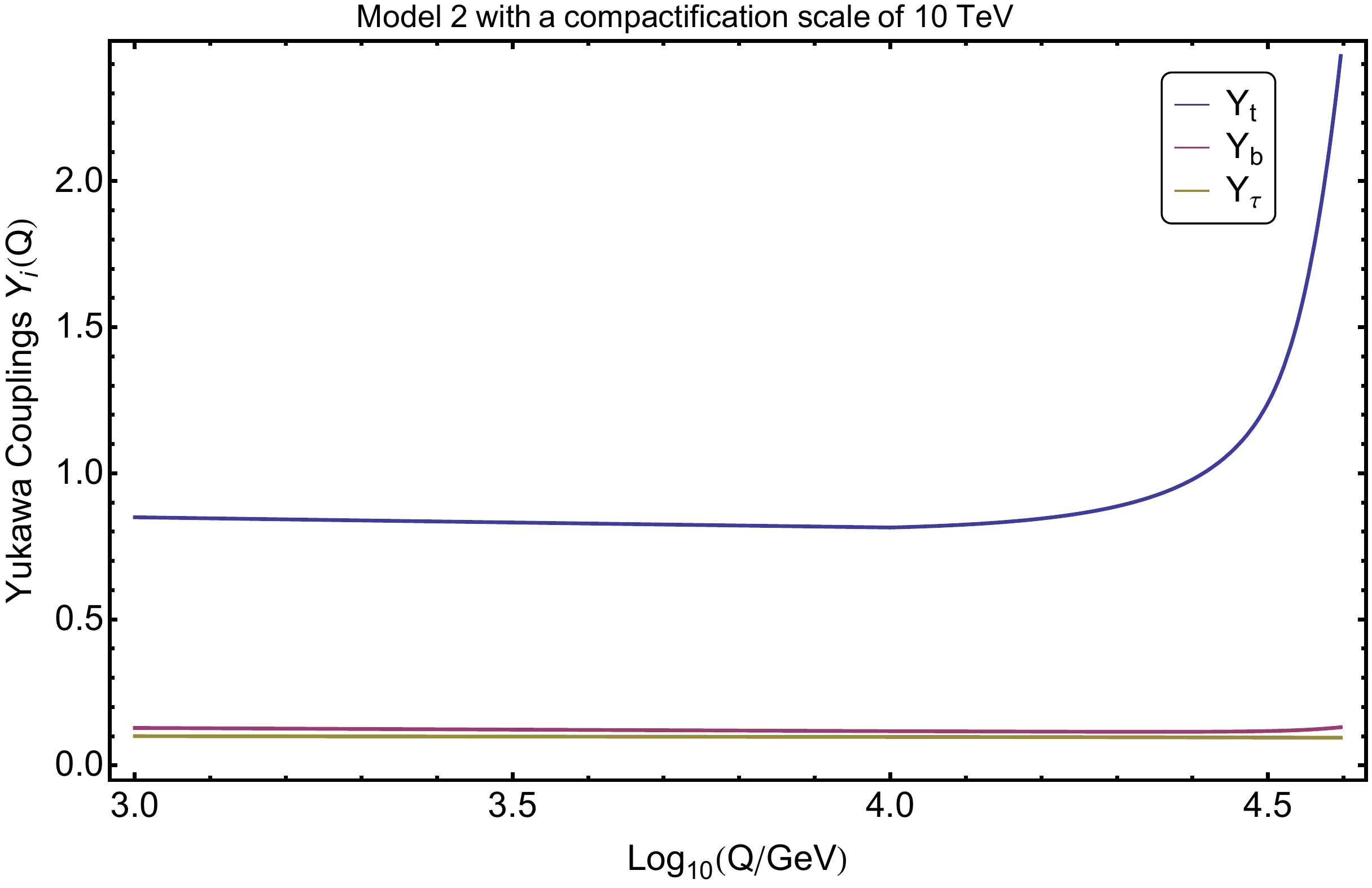}\qquad
\caption{{\it Running of the Yukawa couplings $Y_{i}(Q)$, for two different models with compactification scales 10 TeV  as a function of  $Log_{10}$(Q/GeV). The top Yukawa coupling typically hits a Landau pole before the GUT scale when the 3rd generation matter is located in the bulk (right).}} 
\label{fig:yukawas5D}
\end{center}
\end{figure}

\subsection{Implementation and results}

\par To obtain our results we computed by hand the various RGEs of the four dimensional (zero mode) description that both {\bf model 1} and {\bf 2} (tables \ref{Table:Model1} and \ref{Table:Model2}) reduce to at low energies. We then confirmed these with the output of an implementation of the four dimensional regime in \SARAH \cite{Staub:2009bi,Staub:2010jh,Staub:2012pb,Staub:2013tta}.  We then computed, by hand only, the one-loop RGEs for each of model, {\bf model 1} and {\bf 2}, including all the additional fields of the KK sector. Using \MATHEMATICA we solve the combined set of RGEs and match the four and five dimensional RGEs at the matching compactification scale such that at low energies the theory is described by the four dimensional RGEs only.  
\begin{table}
\begin{center} 
\begin{tabular}{|c|c|c|} 
\hline 
Parameter & Value  & Name \\
\hline 
$Q_{0}$&1000& (SUSY Scale) \\
$g_1(Q_{0})$& 0.360945804 & g1 \\
$g_2(Q_{0})$& 0.633371083 & g2 \\
$g_3(Q_{0})$& 1.02739852& g3 \\
\hline
$\tan \beta$ &10& (Tan beta) \\
$Y_t(Q_{0})$& 0.849348847  & (Top Yukawa)\\
$Y_b(Q_{0})$& 0.128188819 & (Bottom Yukawa)\\
$Y_{\tau}(Q_{0})$& 0.0999653768  & (Tau Yukawa)\\
 \hline\hline
\end{tabular} \caption{A table of the boundary conditions used in our study.
\label{Table:boundaryconditions}}
\end{center} 
\end{table}

\par
\begin{figure}[!thb]
\begin{center}
\includegraphics[width=7.5cm,angle=0]{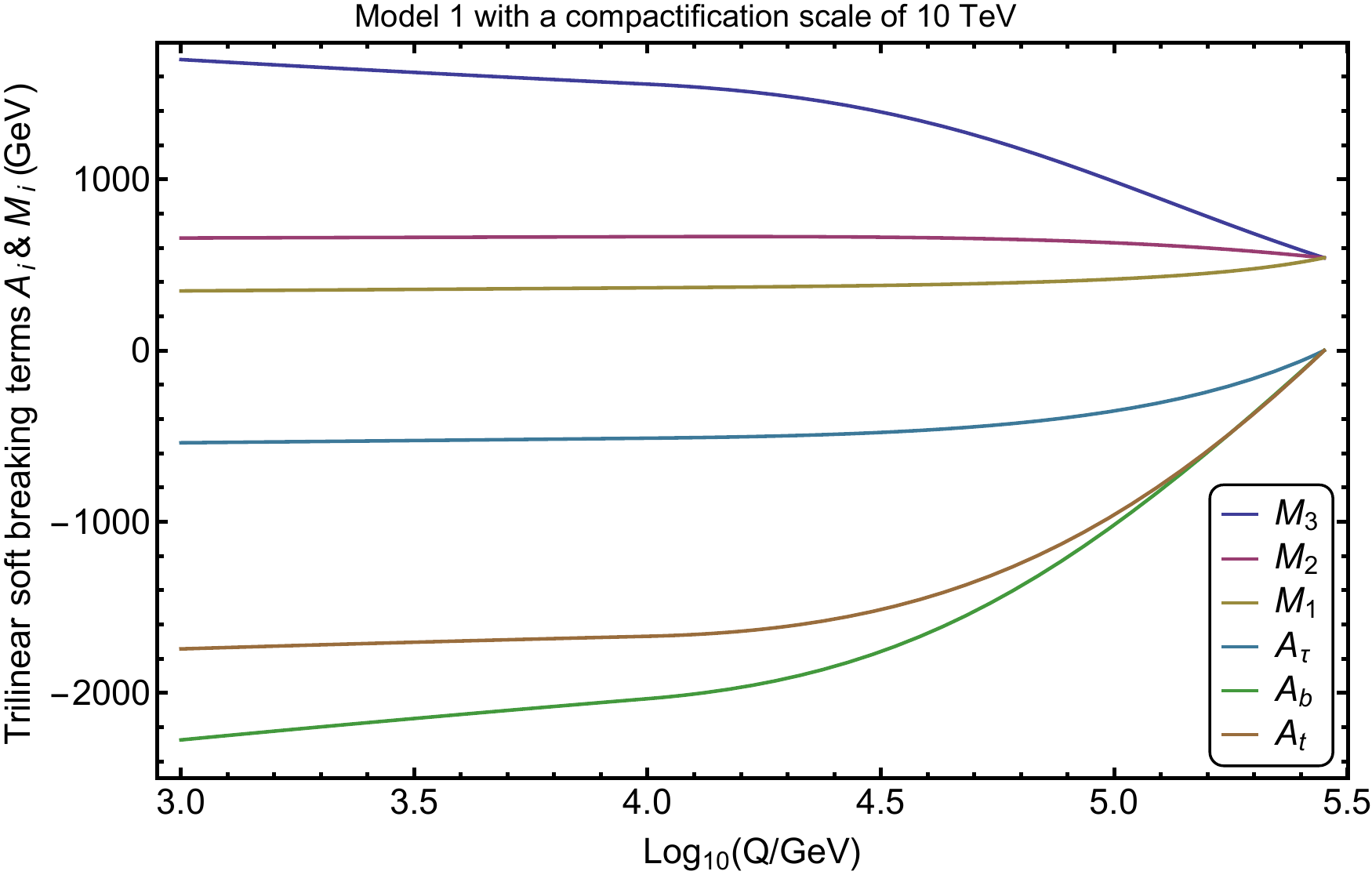}
\includegraphics[width=7.5cm,angle=0]{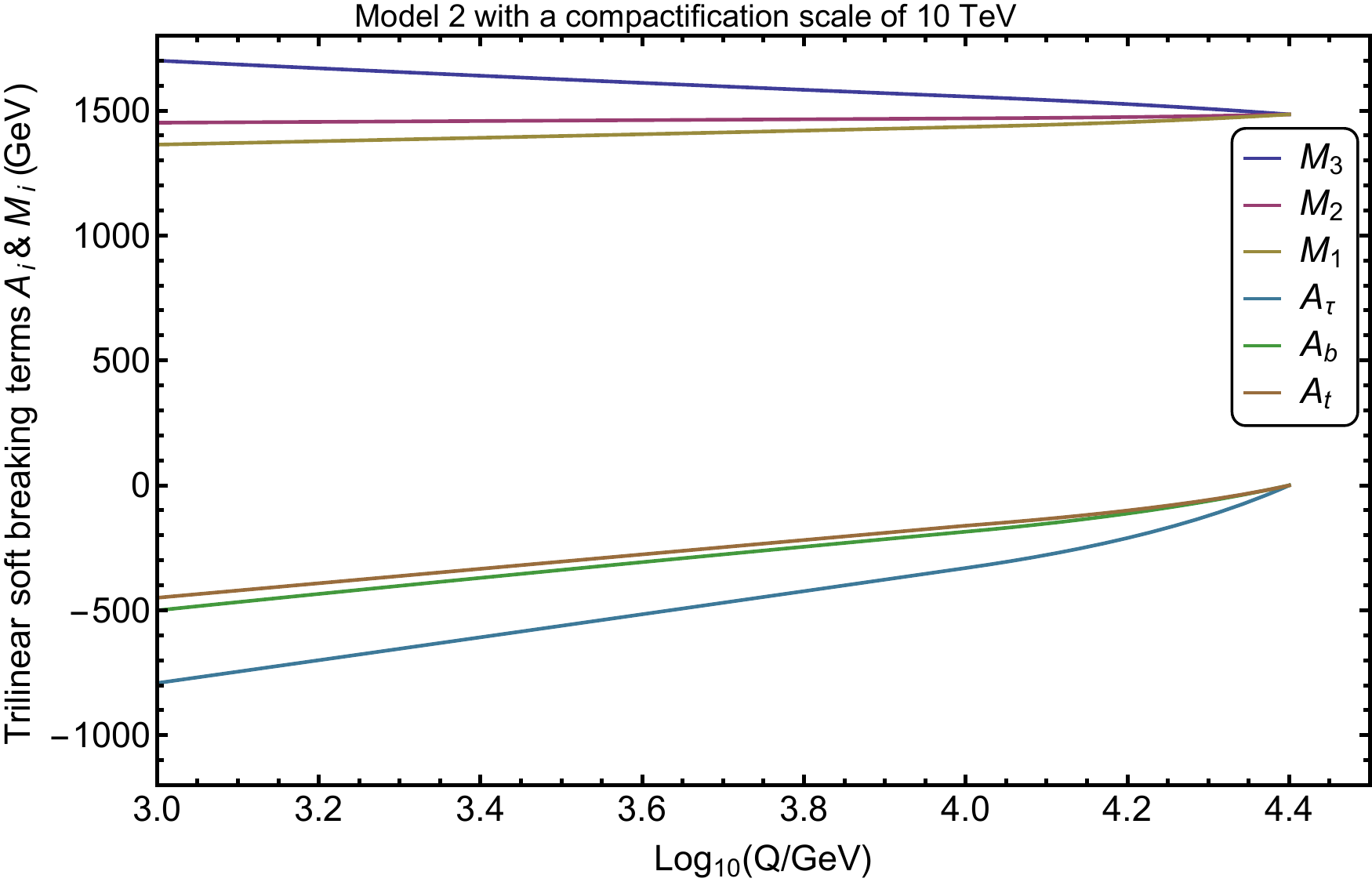}
\caption{{\it Running of the gaugino masses and trilinear couplings $M_{i}(Q)$ and  $A_i(Q)$, for the  two different models with compactification scales 10 TeV, as a function of $Log_{10}$(Q/GeV).}} 
\label{fig:gauginotrilinear5D}
\end{center}
\end{figure}

\begin{figure}[!thb]
\begin{center}
\includegraphics[width=7.5cm,angle=0]{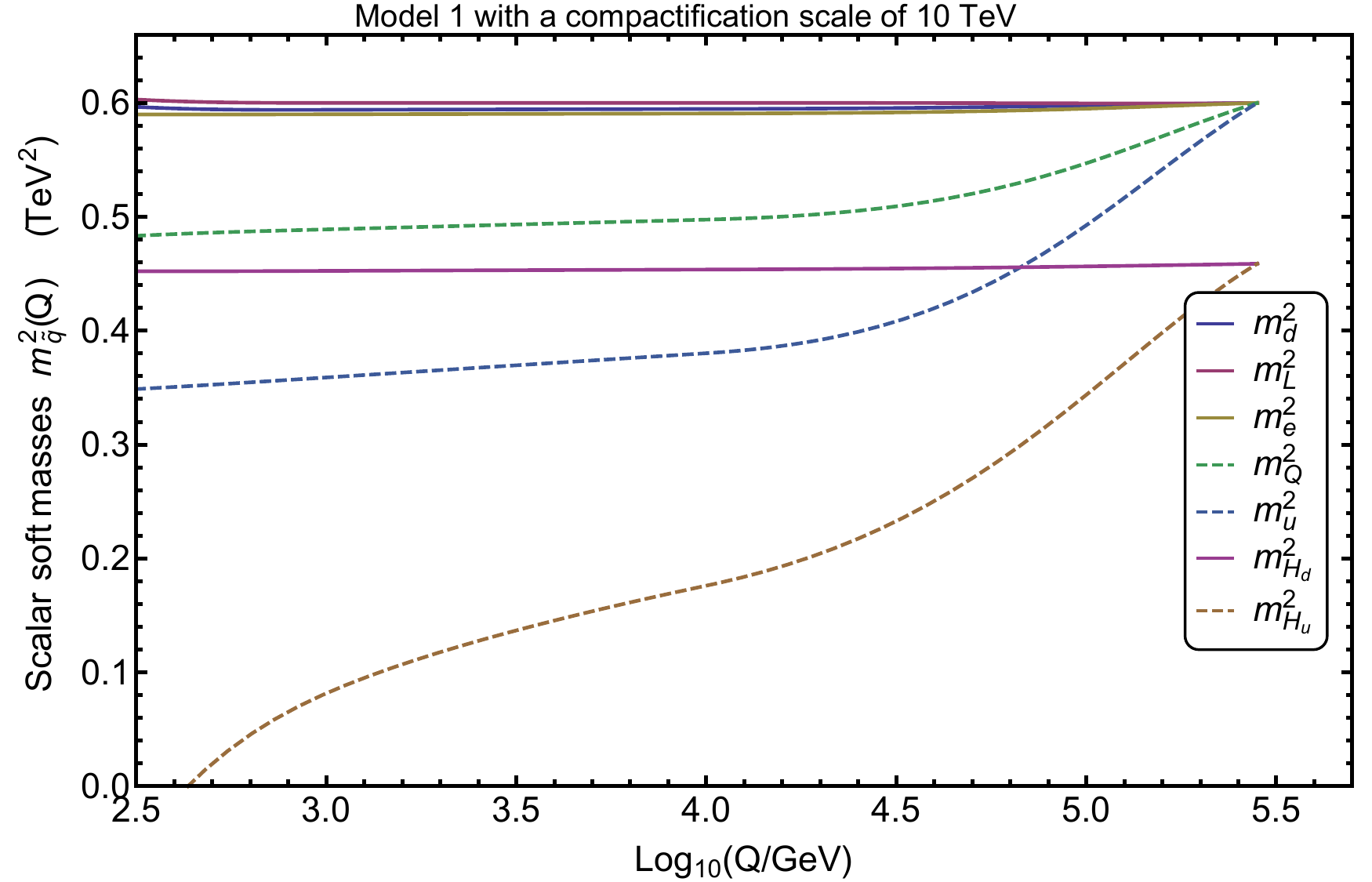}
\includegraphics[width=7.5cm,angle=0]{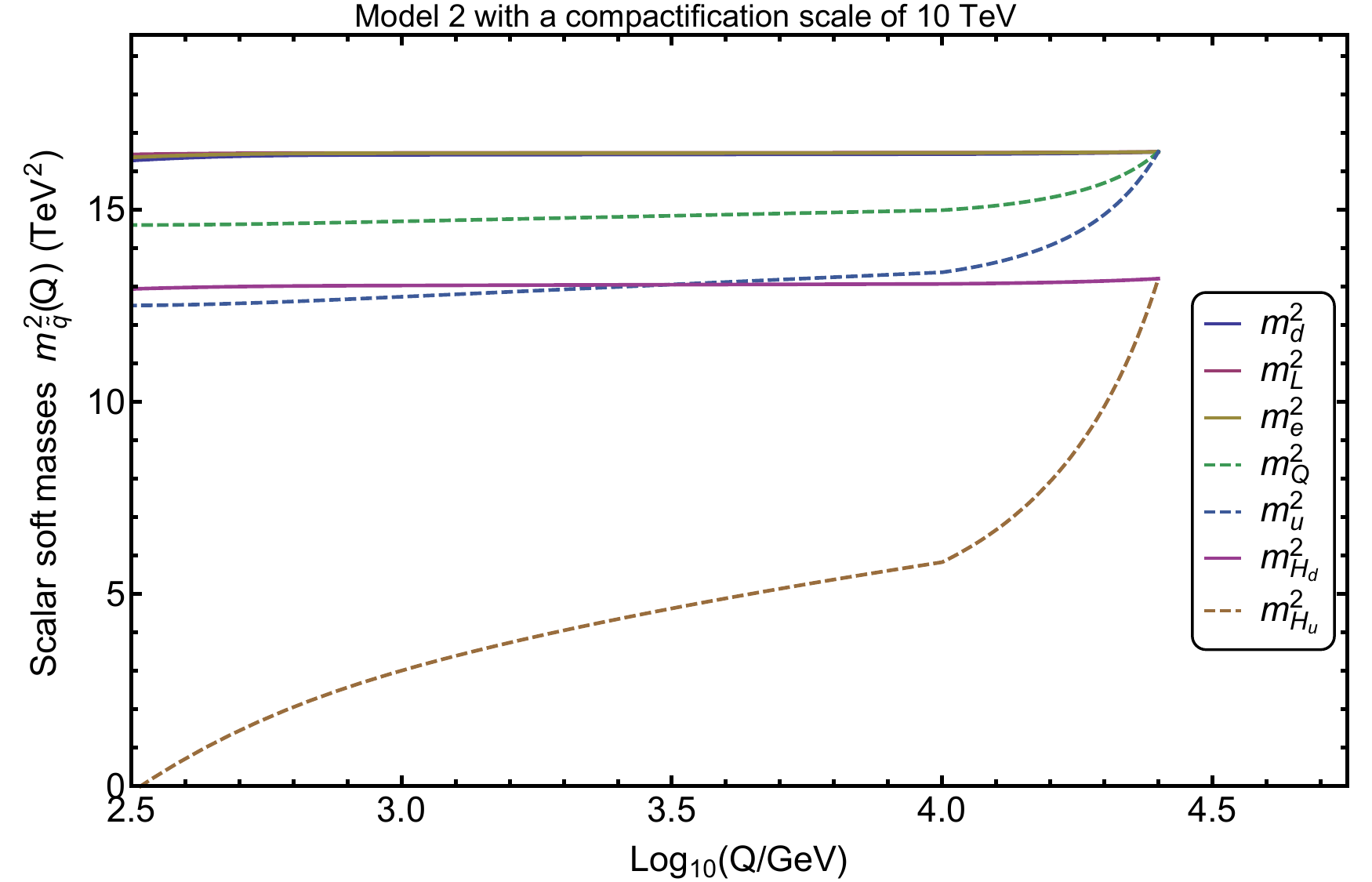}
\caption{{\it Running of the various soft masses for the two models.}} 
\label{fig:soft5D}
\end{center}
\end{figure}

\begin{figure}[!thb]
\begin{center}
\includegraphics[width=7.5cm,angle=0]{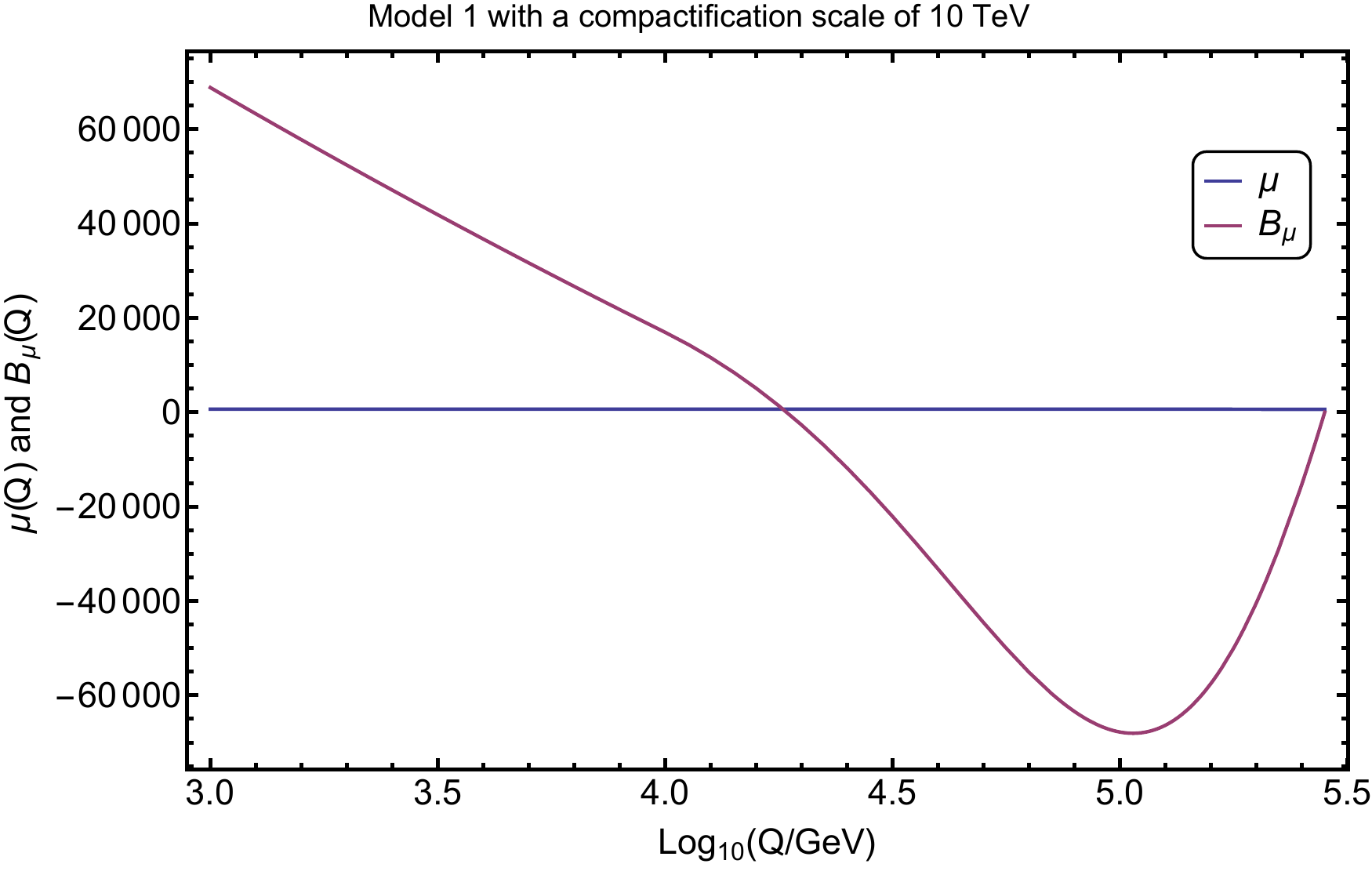}
\includegraphics[width=7.5cm,angle=0]{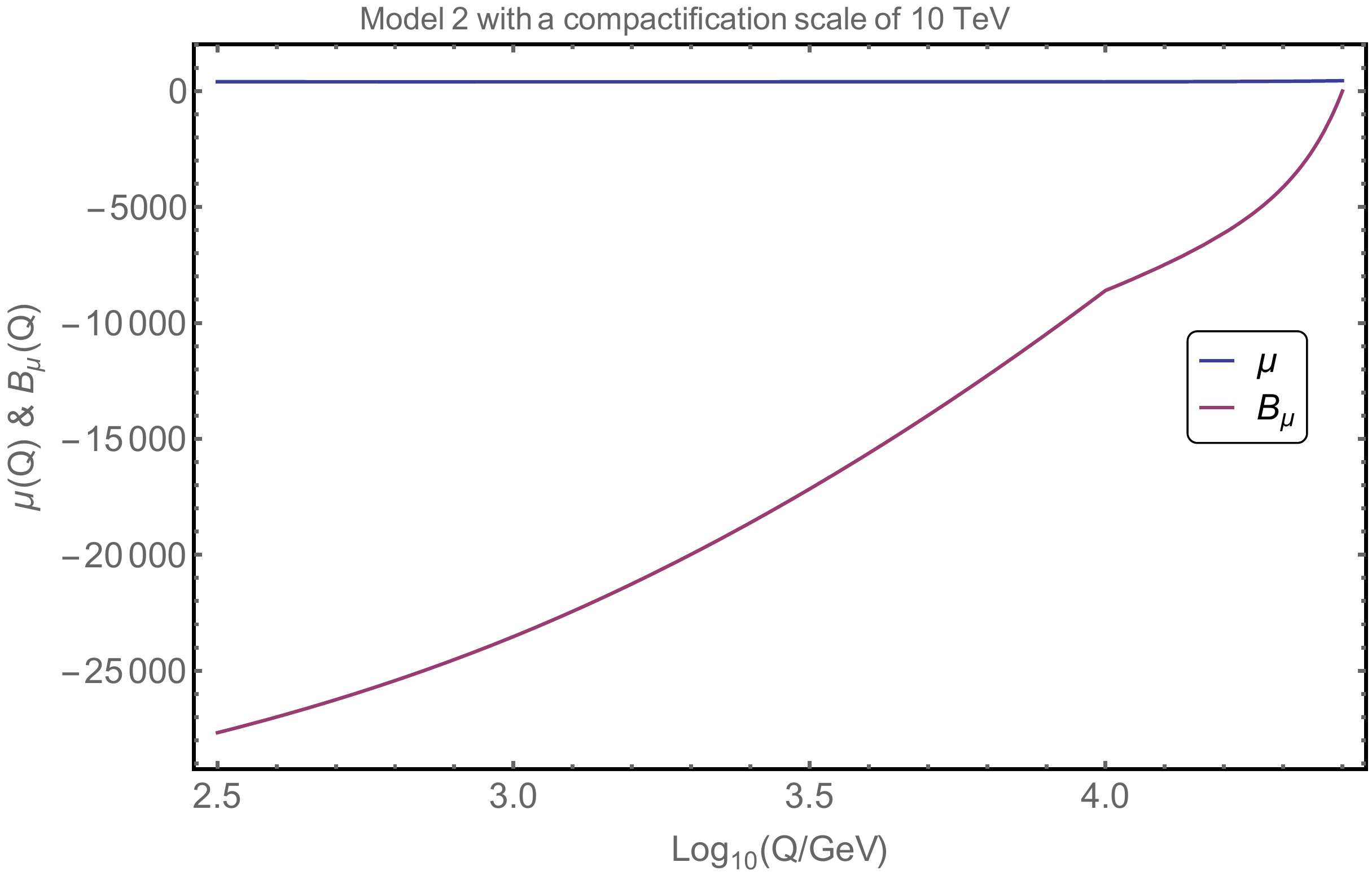}
\caption{{\it The running of $\mu$ and $B_{\mu}$ for the two models.}} 
\label{fig:mubmu}
\end{center}
\end{figure}

 Once we have a combined set of RGEs, we must specify a set of boundary conditions.  In this case we must simply specify all boundary conditions at the same scale (rather than, for example having a set of boundary conditions at both the GUT/SUSY-breaking scale and at the electroweak scale), which we took to be $t$=3, or $Q=10^3$ GeV (where we define $t=Log_{10}Q$).  The gauge couplings and Yukawa couplings are easily obtained by running up from $m_Z$ and are listed in table \ref{Table:boundaryconditions}, for example in figure \ref{fig:yukawas5D} for $\tan \beta\sim 10$.  Regarding the soft breaking terms we made some specific choices which we enforce by choosing a low-scale boundary value such that it holds true once run up to the high scale.  We also make the assumption that the SUSY breaking scale is equal to the GUT scale, but of course other scenarios should be considered. For {\bf model 1}:
\begin{itemize}
\item  We assume supersymmetry breaking occurs at the unification scale, which is found by finding the scale at which $g_1=g_2=g_3$, which is lowered compared to the 4D MSSM, by the effects of the compactification. This is picture in figure \ref{fig:alphas5D} (top left).
\item We specify the value of the gluino mass, $M_3(Q)$, at $Q=10^3$ GeV. We then find the bino and wino soft masses $M_1$ and $M_2$ such that all gaugino masses $M_1=M_2=M_3$ at the GUT scale. This is pictured in figure \ref{fig:gauginotrilinear5D}.
\item We take the trilinear soft breaking terms, $A_{u/d/e}$, to vanish at the unification scale, also pictured in figure \ref{fig:gauginotrilinear5D}.
\item We take $\mu(t=3) \sim 500$ GeV and $B_{\mu}(M_{GUT})=0$, as pictured in figure \ref{fig:mubmu} (left).
\end{itemize}
\noindent The results are rather different for {\bf model 2}:
\begin{itemize}
\item  We found the scale at which $g_1=g_2=g_3$, which is lowered compared to the 4D MSSM, pictured in figure \ref{fig:alphas5D} (top right).
\item The top Yukawa coupling hits a Landau pole just after $t=4.595$, as pictured in figure \ref{fig:yukawas5D} (right).
\end{itemize}
The result was that we could not set the supersymmetry breaking scale at $M_{GUT}$ and instead chose the supersymmetry breaking scale to occur below the top Yukawa Landau pole, at $t=4.4$.  We then chose for the plots in {\bf model 2}:
\begin{itemize}
\item  We choose the gaugino masses to unify $M_3(t=4.4)=M_2(t=4.4)=M_1(t=4.4)$ and let $M_3(t=3)=1700$ GeV.
\item $A_{u/d/e}(t=4.4)$ are set to vanish and this model does not develop a TeV scale $A_t(t=3)$, as pictured in figure \ref{fig:gauginotrilinear5D} (right).
\item Whilst electroweak symmetry breaking is possible starting from the condition $m_{H_d}^2=m_{H_u}^2$,  it does not automatically arise from using $(m_0^2+\mu^2)^{1/2}$, where $m_0^2$ would set the scalar soft mass boundary condition. This is pictured in figure \ref{fig:soft5D} (right), where a representative case is given that achieves the correct Higgs mass.
\item We take $\mu(t=3) \sim 500$ GeV and $B_{\mu}(M_{GUT})=0$, as pictured in figure \ref{fig:mubmu} (left).
\end{itemize}

\subsection{Two ways to accommodate natural supersymmetry}

\par The two models we explore in this paper can accommodate a natural spectrum of sparticles in two very different ways, whilst still obtaining the correctly observed Higgs mass:

In \textbf{model 2} the third generation are located in the bulk and feel the effects of supersymmetry more indirectly than the first and second generation. This will allow for a spectrum of light stops with a heavier first and second generation, above present collider exclusions.    One may use the NMSSM or D-terms to lift the Higgs mass to its correct value.

In \textbf{model 1} the Higgs mass is obtained through a TeV scale $A_t$ term that is generated entirely through RGE evolution, allowing for the correct Higgs mass with stops much below 2 TeV even within an MSSM-like Higgs sector, but does not yet explain any heirarchy between the generation of squarks. In this subsection we explain these details of each model further.

\subsubsection{The third generation in the bulk}

\par Exclusions on first and second generation squarks are presently nearing 2 TeV \cite{ATLASsqgl,CMSsqgl}, while the aesthetic of naturalness for the Higgs sector (and much weaker bounds on 3rd generation squarks of around 300-400 GeV \cite{ATLAS3rd,CMS3rd} from direct searches) favour a 3rd generation below a TeV. In order for this hierarchy to emerge at low scales it is likely to be imprinted in the soft SUSY breaking terms and not simply a renormalisation group effect. At the supersymmetry breaking scale this might imply that the soft terms, in the flavour basis, take the form,
\be
m_{\tilde{f}}^2 \sim \Lambda^2 \left(\begin{array}{ccc}
1& 0 & 0 \\
 0 & 1& 0 \\
 0 & 0 & 0\\
\end{array}\right)+...\label{eq:softheirarchy}
\ee
or indeed
\be
m_{\tilde{f}}^2 \sim \Lambda^2 \left(\begin{array}{ccc}
1& 0 & 0 \\
 0 & 1& 0 \\
 0 & 0 & -\varepsilon \\
\end{array}\right)+...
\ee
the $\varepsilon$ denoting any subleading effects, as one does not require exactly zero entries.  Some ideas have been put forward to explain such a heirarchy, see for instance \cite{Bharucha:2013ela,Brummer:2013upa,Abel:2014fka,Abdalgabar:2014bfa}, and we wish to advance the argument that a five dimensional model with the \emph{3rd generation in the bulk}, i.e our \textbf{model 2}, explains such a hierarchy.  

\par We put forward the idea that the first and second generation of squarks live on the \emph{same brane} as the source of supersymmetry breaking.  They will feel directly the effect of supersymmetry breaking and generate large soft breaking terms.  The 3rd generation is, however, located in the bulk and will feel the supersymmetry breaking indirectly through either gravity or gauge mediation.  This will lead to the boundary conditions in \refe{eq:softheirarchy}. For a calculation of gauge mediated soft terms from a \emph{brane to a bulk field} see \cite{McGarrie:2010kh,McGarrie:2011av}, for \emph{brane to other brane} see \cite{Mirabelli:1997aj,McGarrie:2010kh,McGarrie:2011av}. Such an effect is still felt directly by the gauginos (and the gravitino) and they will also have a large SUSY breaking soft mass, which have important RGE effects as discussed in this paper.

\subsubsection{A large $A_t$ term}

\par Our \textbf{model 1} does not geometrically explain why the first and second generation might be much heavier than the 3rd, but it does allow for a large $A_t$ term generated entirely through RGE evolution, and this can still allow for stops much below $2$ TeV and still obtain the correct Higgs mass from the usual MSSM Higgs sector. Therefore for \textbf{model 1}, we do not yet offer an explanation of the source of supersymmtry breaking. We discuss obtaining the correct Higgs mas in  \textbf{model 1} in the next section.


\section{The one-loop Higgs mass}\label{sec:Higgsmass}
\begin{figure}[!thb]
\begin{center}
\includegraphics[width=12cm,angle=0]{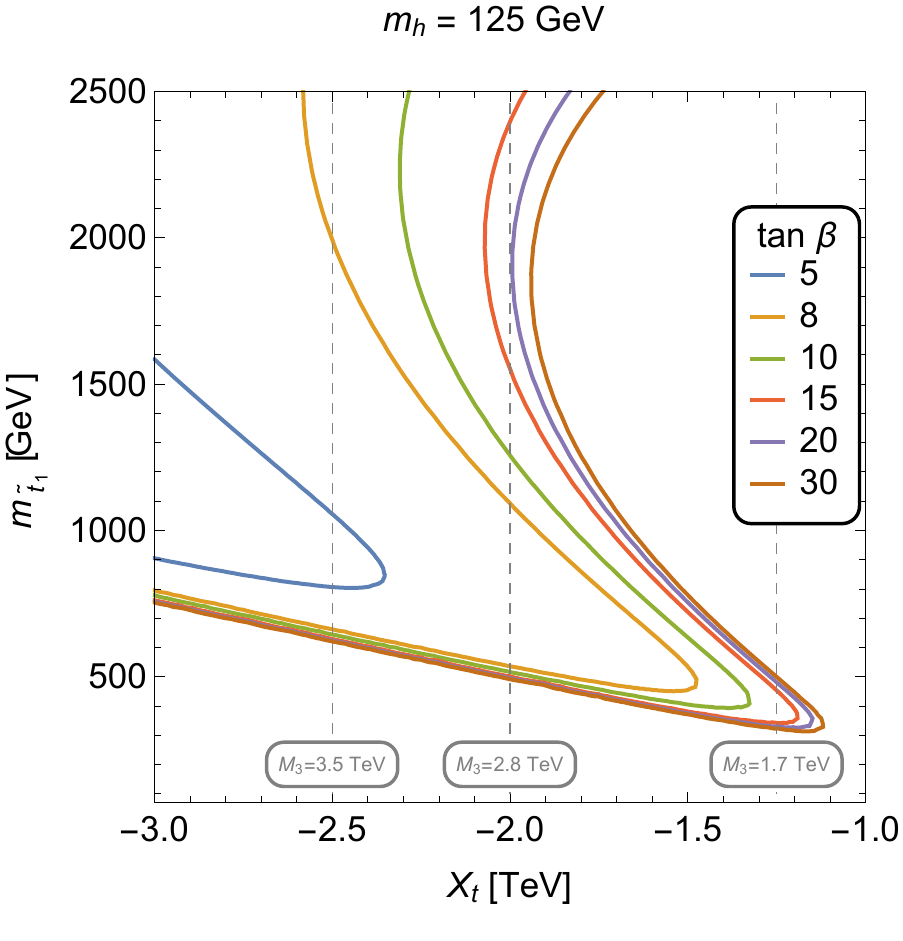}
\caption{{\it Contours of the lightest Higgs mass $m_h=125$~GeV in the plane ($m_{\tilde{t}_1},X_t$) for various values of $\tan \beta$. The dashed gray lines represent the gluino mass for values of $X_t$ found in {\bf model 1}. Stop masses below $2$~TeV are obtainable in our model due to the TeV-scale $A_t$ term. }} 
\label{fig:Higgs5D}
\end{center}
\end{figure}
\par The leading one-loop self energy contributions to the lightest CP even Higgs mass is given by ~\cite{Ellis:1991zd,Lopez:1991aw,Carena:1995bx,Haber:1996fp,Degrassi:2002fi} 
\be
m_{h,1}^2\simeq m_{z}^2\cos^2 2\beta +
\frac{3}{4\pi^2}\frac{m_t^4}{v^2_{ew}}\left[\ln
\frac{M^2_{S}}{m_t^2}+\frac{X^2_t}{M_S^2}\left(1-\frac{X^2_t}{12M_S^2}
\right)\right],\label{eq:higgsmass}
\ee
where $v_{ew}$ is the electroweak Higgs vev,  $X_t=A_t -\mu \cot \beta$ and $M^2_S=m_{\tilde{t}_1}m_{\tilde{t}_2}$.  If $\mu$ is a few $100$ GeV and $A_t\gg \mu$ then $X_{t}\sim A_t$. We plot, in figure \ref{fig:Higgs5D}, the Higgs mass formula for representative values of $A_t$ in our setup. This allows for a prediction of $\tan \beta$ and the stop squark masses which can be below $2$~TeV. One could also lift the tree-level Higgs mass with the $NMSSM +F^{\pm}$ or else through non decoupling D-terms (see for example \cite{Bharucha:2013ela,McGarrie:2014xxa}) which would require introducing an additional $U(1)$ or $SU(2)$ or both. Such an additional feature would be necessary for {\bf model 2} as a large $A_t$ does not arise in this case.


\section{Conclusions}\label{sec:conclude}

\par In this paper we explore various five dimensional extensions of the supersymmetric standard model that unify, with an inverse radius of the extra dimension of roughly a $10$ TeV scale.  Such models are compelling extensions of the MSSM in that they may achieve interesting phenomenological features such as additional $Z',W'$ and $G'$ bosons in the $1-10$ TeV range and achieve the correct $125$~GeV Higgs mass and a relatively natural sparticle spectrum for model 1, while for model 2 this spectrum is heavier, without sacrificing unification of gauge couplings. Such models achieves a natural spectrum by generating a TeV scale $A_t$ term from ``power-law'' running and unification of gauge couplings through the addition of two charged superfields $F^{\pm}$ in the bulk.  

\par In particular we look at two models that can achieve unification, either all chiral matter superfields on the boundaries, or just the third generation in the bulk and the first two on a boundary. In either case the Higgs doublet superfields $H_u,H_d$ and $F^{\pm}$ are located in the bulk along with all three gauge groups $SU(3)_c\times SU(2)_L\times U(1)_Y$. We also point out that five dimensional models in which the 1st and 2nd generation are located in the bulk cannot possibly achieve unification unless the inverse radius of the extra dimension is essentially at the GUT scale (and in any case not with this matter content), and so are entirely four dimensional from a phenomenological perspective. 

\par This paper can be extended in a number of ways and we discuss just a few. In many models of supersymmetry breaking, electroweak symmetry breaking is not optimal both in terms of fine tuning and in obtaining electroweak breaking from a given parameterisation of soft breaking terms at the high scale. These remain an interesting open question, and may benefit from further discoveries or exclusions in the Higgs sector,  at the LHC13/14. Our results are representative only, and clearly a more dedicated spectrum generator built using the RGEs and including threshold corrections will give more precise results, and we provide in this paper a concrete set of RGEs from which this spectrum generator can be constructed. Different supersymmetry breaking parameterisations and how flavour arises is also an interesting further direction to consider.


\section{Acknowledgements}
This work was partially supported by the Foundation for Polish Science International PhD Projects Programme co-financed by the EU European Regional Development Fund. This work has been partially supported by National Science Centre under research grant DEC-2012/04/A/ST2/00099. AD is partially supported by Institut Universitaire de France. We also acknowledge partial support from the Labex-LIO (Lyon Institute of Origins) under grant ANR-10-LABX-66 and FRAMA (FR3127, F\'ed\'eration de Recherche ``Andr\'e Marie Amp\`ere").


\appendix 

\section{Renormalisation group equations for 4D-SSM+$F^{\pm}$}\label{RGES4D}

\par In this appendix we document the one- and two-loop RGEs for the four dimensional low energy model for which the 5D models 1 and 2 are completions. Recall that the output of our implementation in the four dimensional regime was done using \SARAH \cite{Staub:2009bi,Staub:2010jh,Staub:2012pb,Staub:2013tta}, as such as we have used the same conventions and notations, where $T_i = A_i y_i$ ($i = t, b, \tau$ etc.) in these appendices.

\subsection{Anomalous Dimensions}
{\allowdisplaybreaks \begin{align} 
\gamma_{\hat{q}}^{(1)} & =  
-\frac{1}{30} \Big(45 g_{2}^{2}  + 80 g_{3}^{2}  + g_{1}^{2}\Big){\bf 1}  + {Y_{d}^{\dagger}  Y_d} + {Y_{u}^{\dagger}  Y_u}\\ 
\gamma_{\hat{q}}^{(2)} & =  
+\frac{1}{180} \Big(2 g_{1}^{2} \Big(16 g_{3}^{2}  + 9 g_{2}^{2} \Big) + 47 g_{1}^{4}  + 5 \Big(135 g_{2}^{4}  + 288 g_{2}^{2} g_{3}^{2}  -32 g_{3}^{4} \Big)\Big){\bf 1} \nonumber \\ 
 & \ \ \ \  +\frac{4}{5} g_{1}^{2} {Y_{u}^{\dagger}  Y_u} -2 {Y_{d}^{\dagger}  Y_d  Y_{d}^{\dagger}  Y_d} -2 {Y_{u}^{\dagger}  Y_u  Y_{u}^{\dagger}  Y_u} \nonumber \\  
 & \ \ \ \     + {Y_{d}^{\dagger}  Y_d} \Big(-3 \mbox{Tr}\Big({Y_d  Y_{d}^{\dagger}}\Big)  + \frac{2}{5} g_{1}^{2}  - \mbox{Tr}\Big({Y_e  Y_{e}^{\dagger}}\Big) \Big)-3 {Y_{u}^{\dagger}  Y_u} \mbox{Tr}\Big({Y_u  Y_{u}^{\dagger}}\Big) \\ 
\gamma_{\hat{l}}^{(1)} & =  
-\frac{3}{10} \Big(5 g_{2}^{2}  + g_{1}^{2}\Big){\bf 1}  + {Y_{e}^{\dagger}  Y_e}\\ 
\gamma_{\hat{l}}^{(2)} & =  
-2 {Y_{e}^{\dagger}  Y_e  Y_{e}^{\dagger}  Y_e}  + \frac{3}{100} \Big(125 g_{2}^{4}  + 30 g_{1}^{2} g_{2}^{2}  + 81 g_{1}^{4} \Big){\bf 1}  \nonumber \\   & \ \ \ \  + {Y_{e}^{\dagger}  Y_e} \Big(-3 \mbox{Tr}\Big({Y_d  Y_{d}^{\dagger}}\Big)  + \frac{6}{5} g_{1}^{2}  - \mbox{Tr}\Big({Y_e  Y_{e}^{\dagger}}\Big) \Big)\\ 
\gamma_{\hat{H}_d}^{(1)} & =  
3 \mbox{Tr}\Big({Y_d  Y_{d}^{\dagger}}\Big)  -\frac{3}{10} \Big(5 g_{2}^{2}  + g_{1}^{2}\Big) + \mbox{Tr}\Big({Y_e  Y_{e}^{\dagger}}\Big)\\ 
\gamma_{\hat{H}_d}^{(2)} & =  
+\frac{243}{100} g_{1}^{4} +\frac{9}{10} g_{1}^{2} g_{2}^{2} +\frac{15}{4} g_{2}^{4} -\frac{2}{5} \Big(-40 g_{3}^{2}  + g_{1}^{2}\Big)\mbox{Tr}\Big({Y_d  Y_{d}^{\dagger}}\Big) 
+\frac{6}{5} g_{1}^{2} \mbox{Tr}\Big({Y_e  Y_{e}^{\dagger}}\Big) \nonumber \\ 
& \ \ \ \  -9 \mbox{Tr}\Big({Y_d  Y_{d}^{\dagger}  Y_d  Y_{d}^{\dagger}}\Big)  -3 \mbox{Tr}\Big({Y_d  Y_{u}^{\dagger}  Y_u  Y_{d}^{\dagger}}\Big) -3 \mbox{Tr}\Big({Y_e  Y_{e}^{\dagger}  Y_e  Y_{e}^{\dagger}}\Big) \\ 
\gamma_{\hat{H}_u}^{(1)} & =  
-\frac{3}{10} \Big(-10 \mbox{Tr}\Big({Y_u  Y_{u}^{\dagger}}\Big)  + 5 g_{2}^{2}  + g_{1}^{2}\Big)\\ 
\gamma_{\hat{H}_u}^{(2)} & =  
-3 \mbox{Tr}\Big({Y_d  Y_{u}^{\dagger}  Y_u  Y_{d}^{\dagger}}\Big)  -9 \mbox{Tr}\Big({Y_u  Y_{u}^{\dagger}  Y_u  Y_{u}^{\dagger}}\Big)  + \frac{15}{4} g_{2}^{4}  + \frac{243}{100} g_{1}^{4}  \\  & \ \ \ \ + \frac{4}{5} \Big(20 g_{3}^{2}  + g_{1}^{2}\Big)\mbox{Tr}\Big({Y_u  Y_{u}^{\dagger}}\Big) \nonumber  + \frac{9}{10} g_{1}^{2} g_{2}^{2} \\ 
\gamma_{\hat{d}}^{(1)} & =  
2 {Y_d^*  Y_{d}^{T}}  -\frac{2}{15} \Big(20 g_{3}^{2}  + g_{1}^{2}\Big){\bf 1} \\ 
\gamma_{\hat{d}}^{(2)} & =  
+\frac{2}{225} \Big(-100 g_{3}^{4}  + 119 g_{1}^{4}  + 80 g_{1}^{2} g_{3}^{2} \Big){\bf 1} -2 \Big({Y_d^*  Y_{d}^{T}  Y_d^*  Y_{d}^{T}} + {Y_d^*  Y_{u}^{T}  Y_u^*  Y_{d}^{T}}\Big)\nonumber \\ 
 & \ \ \ \ +{Y_d^*  Y_{d}^{T}} \Big(-2 \mbox{Tr}\Big({Y_e  Y_{e}^{\dagger}}\Big)  + 6 g_{2}^{2}  -6 \mbox{Tr}\Big({Y_d  Y_{d}^{\dagger}}\Big)  + \frac{2}{5} g_{1}^{2} \Big)\\ 
\gamma_{\hat{u}}^{(1)} & =  
2 {Y_u^*  Y_{u}^{T}}  -\frac{8}{15} \Big(5 g_{3}^{2}  + g_{1}^{2}\Big){\bf 1} \\ 
\gamma_{\hat{u}}^{(2)} & =  
+\frac{8}{45} \Big(16 g_{1}^{2} g_{3}^{2}  + 25 g_{1}^{4}  -5 g_{3}^{4} \Big){\bf 1} -2 \Big({Y_u^*  Y_{d}^{T}  Y_d^*  Y_{u}^{T}} + {Y_u^*  Y_{u}^{T}  Y_u^*  Y_{u}^{T}}\Big)\nonumber \\ 
 & \ \ \ \ +{Y_u^*  Y_{u}^{T}} \Big(6 g_{2}^{2}  -6 \mbox{Tr}\Big({Y_u  Y_{u}^{\dagger}}\Big)  -\frac{2}{5} g_{1}^{2} \Big)\\ 
\gamma_{\hat{e}}^{(1)} & =  
2 {Y_e^*  Y_{e}^{T}}  -\frac{6}{5} g_{1}^{2} {\bf 1} \\ 
\gamma_{\hat{e}}^{(2)} & =  
-2 {Y_e^*  Y_{e}^{T}  Y_e^*  Y_{e}^{T}}  + \frac{54}{5} g_{1}^{4} {\bf 1}  + {Y_e^*  Y_{e}^{T}} \Big(-2 \mbox{Tr}\Big({Y_e  Y_{e}^{\dagger}}\Big)  + 6 g_{2}^{2}  -6 \mbox{Tr}\Big({Y_d  Y_{d}^{\dagger}}\Big)  -\frac{6}{5} g_{1}^{2} \Big)\\ 
\gamma_{\phi_F^+}^{(1)} & =  
-\frac{6}{5} g_{1}^{2}  \ \  , \ \ 
\gamma_{ \phi_F^+}^{(2)}  =  
\frac{54}{5} g_{1}^{4}  \ \  , \ \ 
\gamma_{\phi_F^-}^{(1)}  =  
-\frac{6}{5} g_{1}^{2}  \ \  , \ \ 
\gamma_{\phi_F^-}^{(2)}  =  
\frac{54}{5} g_{1}^{4} .
\end{align} } 

\subsection{Gauge Couplings}
{\allowdisplaybreaks  \begin{align} 
\beta_{g_1}^{(1)} & =  
\frac{39}{5} g_{1}^{3} \ \ \  , \ \  \beta_{g_2}^{(1)}  =   
g_{2}^{3}\  \ \ \  ,  \ \  \beta_{g_3}^{(1)}  =  
-3 g_{3}^{3} \\ 
\beta_{g_1}^{(2)} & =  
\frac{1}{25} g_{1}^{3} \Big(-130 \mbox{Tr}\Big({Y_u  Y_{u}^{\dagger}}\Big)  + 135 g_{2}^{2}  + 271 g_{1}^{2}  + 440 g_{3}^{2}  -70 \mbox{Tr}\Big({Y_d  Y_{d}^{\dagger}}\Big)  -90 \mbox{Tr}\Big({Y_e  Y_{e}^{\dagger}}\Big) \Big)\\ 
\beta_{g_2}^{(2)} & =  
\frac{1}{5} g_{2}^{3} \Big(-10 \mbox{Tr}\Big({Y_e  Y_{e}^{\dagger}}\Big)  + 120 g_{3}^{2}  + 125 g_{2}^{2}  -30 \mbox{Tr}\Big({Y_d  Y_{d}^{\dagger}}\Big)  -30 \mbox{Tr}\Big({Y_u  Y_{u}^{\dagger}}\Big)  + 9 g_{1}^{2} \Big)\\ 
\beta_{g_3}^{(2)} & =  
\frac{1}{5} g_{3}^{3} \Big(11 g_{1}^{2}  -20 \mbox{Tr}\Big({Y_d  Y_{d}^{\dagger}}\Big)  -20 \mbox{Tr}\Big({Y_u  Y_{u}^{\dagger}}\Big)  + 45 g_{2}^{2}  + 70 g_{3}^{2} \Big)
\end{align}} 

\subsection{Gaugino Mass Parameters}
{\allowdisplaybreaks  \begin{align} 
\beta_{M_1}^{(1)} & =  
\frac{78}{5} g_{1}^{2} M_1  \ \ \ , \ \ \ \beta_{M_2}^{(1)} =  
2 g_{2}^{2} M_2 \ \ \ \ , \ \ \ \  \beta_{M_3}^{(1)}  =  
-6 g_{3}^{2} M_3 \\ 
\beta_{M_1}^{(2)} & =  
\frac{2}{25} g_{1}^{2} \Big(542 g_{1}^{2} M_1 +135 g_{2}^{2} M_1 +440 g_{3}^{2} M_1 +440 g_{3}^{2} M_3 +135 g_{2}^{2} M_2 -70 M_1 \mbox{Tr}\Big({Y_d  Y_{d}^{\dagger}}\Big) \nonumber \\ 
&  \ \ \ \   -90 M_1 \mbox{Tr}\Big({Y_e  Y_{e}^{\dagger}}\Big)-130 M_1 \mbox{Tr}\Big({Y_u  Y_{u}^{\dagger}}\Big) +70 \mbox{Tr}\Big({Y_{d}^{\dagger}  T_d}\Big) +90 \mbox{Tr}\Big({Y_{e}^{\dagger}  T_e}\Big) +130 \mbox{Tr}\Big({Y_{u}^{\dagger}  T_u}\Big) \Big)\\ 
\beta_{M_2}^{(2)} & =  
\frac{2}{5} g_{2}^{2} \Big(9 g_{1}^{2} M_1 +120 g_{3}^{2} M_3 +9 g_{1}^{2} M_2 +250 g_{2}^{2} M_2 +120 g_{3}^{2} M_2 -30 M_2 \mbox{Tr}\Big({Y_d  Y_{d}^{\dagger}}\Big) \nonumber \\ 
&  \ \ \ \ -10 M_2 \mbox{Tr}\Big({Y_e  Y_{e}^{\dagger}}\Big) -30 M_2 \mbox{Tr}\Big({Y_u  Y_{u}^{\dagger}}\Big) +30 \mbox{Tr}\Big({Y_{d}^{\dagger}  T_d}\Big) +10 \mbox{Tr}\Big({Y_{e}^{\dagger}  T_e}\Big) +30 \mbox{Tr}\Big({Y_{u}^{\dagger}  T_u}\Big) \Big)\\ 
\beta_{M_3}^{(2)} & =  
\frac{2}{5} g_{3}^{2} \Big(11 g_{1}^{2} M_1 +11 g_{1}^{2} M_3 +45 g_{2}^{2} M_3 +140 g_{3}^{2} M_3 +45 g_{2}^{2} M_2 -20 M_3 \mbox{Tr}\Big({Y_d  Y_{d}^{\dagger}}\Big)  \nonumber \\ 
&  \ \ \ \ -20 M_3 \mbox{Tr}\Big({Y_u  Y_{u}^{\dagger}}\Big)+20 \mbox{Tr}\Big({Y_{d}^{\dagger}  T_d}\Big) +20 \mbox{Tr}\Big({Y_{u}^{\dagger}  T_u}\Big) \Big)
\end{align}} 

\subsection{Trilinear Superpotential Parameters}
{\allowdisplaybreaks  \begin{align} 
\beta_{Y_d}^{(1)} & =  
3 {Y_d  Y_{d}^{\dagger}  Y_d}  + Y_d \Big(-3 g_{2}^{2}  + 3 \mbox{Tr}\Big({Y_d  Y_{d}^{\dagger}}\Big)  -\frac{16}{3} g_{3}^{2}  -\frac{7}{15} g_{1}^{2}  + \mbox{Tr}\Big({Y_e  Y_{e}^{\dagger}}\Big)\Big) + {Y_d  Y_{u}^{\dagger}  Y_u}\\ 
\beta_{Y_d}^{(2)} & =  
+\frac{4}{5} g_{1}^{2} {Y_d  Y_{u}^{\dagger}  Y_u} -4 {Y_d  Y_{d}^{\dagger}  Y_d  Y_{d}^{\dagger}  Y_d} -2 {Y_d  Y_{u}^{\dagger}  Y_u  Y_{d}^{\dagger}  Y_d} -2 {Y_d  Y_{u}^{\dagger}  Y_u  Y_{u}^{\dagger}  Y_u} \nonumber \\ 
& \ \ \ \ +{Y_d  Y_{d}^{\dagger}  Y_d} \Big(-3 \mbox{Tr}\Big({Y_e  Y_{e}^{\dagger}}\Big)  + 6 g_{2}^{2}  -9 \mbox{Tr}\Big({Y_d  Y_{d}^{\dagger}}\Big)  + \frac{4}{5} g_{1}^{2} \Big)-3 {Y_d  Y_{u}^{\dagger}  Y_u} \mbox{Tr}\Big({Y_u  Y_{u}^{\dagger}}\Big) \nonumber \\ 
& \ \ \ \ +Y_d \Big(\frac{1687}{450} g_{1}^{4} +g_{1}^{2} g_{2}^{2} +\frac{15}{2} g_{2}^{4} +\frac{8}{9} g_{1}^{2} g_{3}^{2} +8 g_{2}^{2} g_{3}^{2} -\frac{16}{9} g_{3}^{4} -\frac{2}{5} \Big(-40 g_{3}^{2}  + g_{1}^{2}\Big)\mbox{Tr}\Big({Y_d  Y_{d}^{\dagger}}\Big) \nonumber \\ 
& \ \ \ \ +\frac{6}{5} g_{1}^{2} \mbox{Tr}\Big({Y_e  Y_{e}^{\dagger}}\Big) -9 \mbox{Tr}\Big({Y_d  Y_{d}^{\dagger}  Y_d  Y_{d}^{\dagger}}\Big) -3 \mbox{Tr}\Big({Y_d  Y_{u}^{\dagger}  Y_u  Y_{d}^{\dagger}}\Big) -3 \mbox{Tr}\Big({Y_e  Y_{e}^{\dagger}  Y_e  Y_{e}^{\dagger}}\Big) \Big)\\ 
\beta_{Y_e}^{(1)} & =  
3 {Y_e  Y_{e}^{\dagger}  Y_e}  + Y_e \Big(-3 g_{2}^{2}  + 3 \mbox{Tr}\Big({Y_d  Y_{d}^{\dagger}}\Big)  -\frac{9}{5} g_{1}^{2}  + \mbox{Tr}\Big({Y_e  Y_{e}^{\dagger}}\Big)\Big)\\ 
\beta_{Y_e}^{(2)} & =  
-4 {Y_e  Y_{e}^{\dagger}  Y_e  Y_{e}^{\dagger}  Y_e} +{Y_e  Y_{e}^{\dagger}  Y_e} \Big(-3 \mbox{Tr}\Big({Y_e  Y_{e}^{\dagger}}\Big)  + 6 g_{2}^{2}     -9 \mbox{Tr}\Big({Y_d  Y_{d}^{\dagger}}\Big) \Big)\nonumber \\ 
& \ \ \ \ +Y_e \Big(\frac{783}{50} g_{1}^{4} +\frac{9}{5} g_{1}^{2} g_{2}^{2} +\frac{15}{2} g_{2}^{4} -\frac{2}{5} \Big(-40 g_{3}^{2}  + g_{1}^{2}\Big)\mbox{Tr}\Big({Y_d  Y_{d}^{\dagger}}\Big) +\frac{6}{5} g_{1}^{2} \mbox{Tr}\Big({Y_e  Y_{e}^{\dagger}}\Big) \nonumber \\ 
& \ \ \ \ -9 \mbox{Tr}\Big({Y_d  Y_{d}^{\dagger}  Y_d  Y_{d}^{\dagger}}\Big) -3 \mbox{Tr}\Big({Y_d  Y_{u}^{\dagger}  Y_u  Y_{d}^{\dagger}}\Big) -3 \mbox{Tr}\Big({Y_e  Y_{e}^{\dagger}  Y_e  Y_{e}^{\dagger}}\Big) \Big)\\ 
\beta_{Y_u}^{(1)} & =  
3 {Y_u  Y_{u}^{\dagger}  Y_u}  -\frac{1}{15} Y_u \Big(13 g_{1}^{2}  + 45 g_{2}^{2}  -45 \mbox{Tr}\Big({Y_u  Y_{u}^{\dagger}}\Big)  + 80 g_{3}^{2} \Big) + {Y_u  Y_{d}^{\dagger}  Y_d}\\ 
\beta_{Y_u}^{(2)} & =  
+\frac{2}{5} g_{1}^{2} {Y_u  Y_{u}^{\dagger}  Y_u} +6 g_{2}^{2} {Y_u  Y_{u}^{\dagger}  Y_u} -2 {Y_u  Y_{d}^{\dagger}  Y_d  Y_{d}^{\dagger}  Y_d} -2 {Y_u  Y_{d}^{\dagger}  Y_d  Y_{u}^{\dagger}  Y_u} \nonumber \\ 
 & \ \ \ \ -4 {Y_u  Y_{u}^{\dagger}  Y_u  Y_{u}^{\dagger}  Y_u} +{Y_u  Y_{d}^{\dagger}  Y_d} \Big(-3 \mbox{Tr}\Big({Y_d  Y_{d}^{\dagger}}\Big)  + \frac{2}{5} g_{1}^{2}  - \mbox{Tr}\Big({Y_e  Y_{e}^{\dagger}}\Big) \Big)-9 {Y_u  Y_{u}^{\dagger}  Y_u} \mbox{Tr}\Big({Y_u  Y_{u}^{\dagger}}\Big) \nonumber \\ 
 & \ \ \ \ +Y_u \Big(\frac{3211}{450} g_{1}^{4} +g_{1}^{2} g_{2}^{2} +\frac{15}{2} g_{2}^{4} +\frac{136}{45} g_{1}^{2} g_{3}^{2} +8 g_{2}^{2} g_{3}^{2} -\frac{16}{9} g_{3}^{4} +\frac{4}{5} \Big(20 g_{3}^{2}  + g_{1}^{2}\Big)\mbox{Tr}\Big({Y_u  Y_{u}^{\dagger}}\Big) \nonumber \\ 
 & \ \ \ \ -3 \mbox{Tr}\Big({Y_d  Y_{u}^{\dagger}  Y_u  Y_{d}^{\dagger}}\Big) -9 \mbox{Tr}\Big({Y_u  Y_{u}^{\dagger}  Y_u  Y_{u}^{\dagger}}\Big) \Big)
\end{align}} 

\subsection{Bilinear Superpotential Parameters}
 \begin{align} 
\beta_{\mu}^{(1)} & =  
3 \mu \mbox{Tr}\Big({Y_d  Y_{d}^{\dagger}}\Big)  -\frac{3}{5} \mu \Big(5 g_{2}^{2}  -5 \mbox{Tr}\Big({Y_u  Y_{u}^{\dagger}}\Big)  + g_{1}^{2}\Big) + \mu \mbox{Tr}\Big({Y_e  Y_{e}^{\dagger}}\Big) \\ 
\beta_{\mu}^{(2)} & =  
\frac{1}{50} \mu \Big(243 g_{1}^{4} +90 g_{1}^{2} g_{2}^{2} +375 g_{2}^{4} -20 \Big(-40 g_{3}^{2}  + g_{1}^{2}\Big)\mbox{Tr}\Big({Y_d  Y_{d}^{\dagger}}\Big) +60 g_{1}^{2} \mbox{Tr}\Big({Y_e  Y_{e}^{\dagger}}\Big) \nonumber \\ 
 & \ \ \ \ +40 g_{1}^{2} \mbox{Tr}\Big({Y_u  Y_{u}^{\dagger}}\Big) +800 g_{3}^{2} \mbox{Tr}\Big({Y_u  Y_{u}^{\dagger}}\Big) -450 \mbox{Tr}\Big({Y_d  Y_{d}^{\dagger}  Y_d  Y_{d}^{\dagger}}\Big) -300 \mbox{Tr}\Big({Y_d  Y_{u}^{\dagger}  Y_u  Y_{d}^{\dagger}}\Big)\nonumber \\ 
 &  \ \ \ \ -150 \mbox{Tr}\Big({Y_e  Y_{e}^{\dagger}  Y_e  Y_{e}^{\dagger}}\Big) -450 \mbox{Tr}\Big({Y_u  Y_{u}^{\dagger}  Y_u  Y_{u}^{\dagger}}\Big) \Big)\\ 
\beta_{\nu}^{(1)} & =  
-\frac{12}{5} g_{1}^{2} \nu \ \ \ \  ,  \ \ \ \ 
\beta_{\nu}^{(2)}  =  
\frac{108}{5} g_{1}^{4} \nu 
\end{align}

\subsection{Trilinear Soft-Breaking Parameters}
{\allowdisplaybreaks  \begin{align} 
\beta_{T_d}^{(1)} & =  
+4 {Y_d  Y_{d}^{\dagger}  T_d} +2 {Y_d  Y_{u}^{\dagger}  T_u} +5 {T_d  Y_{d}^{\dagger}  Y_d} +{T_d  Y_{u}^{\dagger}  Y_u}-\frac{7}{15} g_{1}^{2} T_d -3 g_{2}^{2} T_d -\frac{16}{3} g_{3}^{2} T_d +3 T_d \mbox{Tr}\Big({Y_d  Y_{d}^{\dagger}}\Big) \nonumber \\ 
 & \ \ \ \ +T_d \mbox{Tr}\Big({Y_e  Y_{e}^{\dagger}}\Big) +Y_d \Big(2 \mbox{Tr}\Big({Y_{e}^{\dagger}  T_e}\Big)  + 6 g_{2}^{2} M_2  + 6 \mbox{Tr}\Big({Y_{d}^{\dagger}  T_d}\Big)  + \frac{14}{15} g_{1}^{2} M_1  + \frac{32}{3} g_{3}^{2} M_3 \Big)\\ 
\beta_{T_d}^{(2)} & =  
+\frac{6}{5} g_{1}^{2} {Y_d  Y_{d}^{\dagger}  T_d} +6 g_{2}^{2} {Y_d  Y_{d}^{\dagger}  T_d} -\frac{8}{5} g_{1}^{2} M_1 {Y_d  Y_{u}^{\dagger}  Y_u} +\frac{8}{5} g_{1}^{2} {Y_d  Y_{u}^{\dagger}  T_u} \nonumber \\ 
 & \ \ \ \ +\frac{6}{5} g_{1}^{2} {T_d  Y_{d}^{\dagger}  Y_d} +12 g_{2}^{2} {T_d  Y_{d}^{\dagger}  Y_d} +\frac{4}{5} g_{1}^{2} {T_d  Y_{u}^{\dagger}  Y_u} -6 {Y_d  Y_{d}^{\dagger}  Y_d  Y_{d}^{\dagger}  T_d} \nonumber \\ 
 & \ \ \ \ -8 {Y_d  Y_{d}^{\dagger}  T_d  Y_{d}^{\dagger}  Y_d} -2 {Y_d  Y_{u}^{\dagger}  Y_u  Y_{d}^{\dagger}  T_d} -4 {Y_d  Y_{u}^{\dagger}  Y_u  Y_{u}^{\dagger}  T_u} -4 {Y_d  Y_{u}^{\dagger}  T_u  Y_{d}^{\dagger}  Y_d} \nonumber \\ 
 & \ \ \ \ -4 {Y_d  Y_{u}^{\dagger}  T_u  Y_{u}^{\dagger}  Y_u} -6 {T_d  Y_{d}^{\dagger}  Y_d  Y_{d}^{\dagger}  Y_d} -4 {T_d  Y_{u}^{\dagger}  Y_u  Y_{d}^{\dagger}  Y_d} -2 {T_d  Y_{u}^{\dagger}  Y_u  Y_{u}^{\dagger}  Y_u} \nonumber \\ 
 & \ \ \ \ +\frac{1687}{450} g_{1}^{4} T_d +g_{1}^{2} g_{2}^{2} T_d +\frac{15}{2} g_{2}^{4} T_d +\frac{8}{9} g_{1}^{2} g_{3}^{2} T_d +8 g_{2}^{2} g_{3}^{2} T_d -\frac{16}{9} g_{3}^{4} T_d \nonumber \\ 
 & \ \ \ \ -12 {Y_d  Y_{d}^{\dagger}  T_d} \mbox{Tr}\Big({Y_d  Y_{d}^{\dagger}}\Big) -15 {T_d  Y_{d}^{\dagger}  Y_d} \mbox{Tr}\Big({Y_d  Y_{d}^{\dagger}}\Big) -\frac{2}{5} g_{1}^{2} T_d \mbox{Tr}\Big({Y_d  Y_{d}^{\dagger}}\Big) \nonumber \\ 
 & \ \ \ \ +16 g_{3}^{2} T_d \mbox{Tr}\Big({Y_d  Y_{d}^{\dagger}}\Big) -4 {Y_d  Y_{d}^{\dagger}  T_d} \mbox{Tr}\Big({Y_e  Y_{e}^{\dagger}}\Big) -5 {T_d  Y_{d}^{\dagger}  Y_d} \mbox{Tr}\Big({Y_e  Y_{e}^{\dagger}}\Big) \nonumber \\ 
 & \ \ \ \ +\frac{6}{5} g_{1}^{2} T_d \mbox{Tr}\Big({Y_e  Y_{e}^{\dagger}}\Big) -6 {Y_d  Y_{u}^{\dagger}  T_u} \mbox{Tr}\Big({Y_u  Y_{u}^{\dagger}}\Big) -3 {T_d  Y_{u}^{\dagger}  Y_u} \mbox{Tr}\Big({Y_u  Y_{u}^{\dagger}}\Big) \nonumber \\ 
 & \ \ \ \ -\frac{2}{5} {Y_d  Y_{d}^{\dagger}  Y_d} \Big(15 \mbox{Tr}\Big({Y_{e}^{\dagger}  T_e}\Big)  + 30 g_{2}^{2} M_2  + 45 \mbox{Tr}\Big({Y_{d}^{\dagger}  T_d}\Big)  + 4 g_{1}^{2} M_1 \Big)-6 {Y_d  Y_{u}^{\dagger}  Y_u} \mbox{Tr}\Big({Y_{u}^{\dagger}  T_u}\Big) \nonumber \\ 
 & \ \ \ \ -9 T_d \mbox{Tr}\Big({Y_d  Y_{d}^{\dagger}  Y_d  Y_{d}^{\dagger}}\Big) -3 T_d \mbox{Tr}\Big({Y_d  Y_{u}^{\dagger}  Y_u  Y_{d}^{\dagger}}\Big) -3 T_d \mbox{Tr}\Big({Y_e  Y_{e}^{\dagger}  Y_e  Y_{e}^{\dagger}}\Big) \nonumber \\ 
 & \ \ \ \ +Y_d \Big(-\frac{3374}{225} g_{1}^{4} M_1 -2 g_{1}^{2} g_{2}^{2} M_1 -\frac{16}{9} g_{1}^{2} g_{3}^{2} M_1 -\frac{16}{9} g_{1}^{2} g_{3}^{2} M_3 -16 g_{2}^{2} g_{3}^{2} M_3 +\frac{64}{9} g_{3}^{4} M_3 \nonumber \\ 
 & \ \ \ \ -2 g_{1}^{2} g_{2}^{2} M_2 -30 g_{2}^{4} M_2 -16 g_{2}^{2} g_{3}^{2} M_2 +\frac{4}{5} \Big(-40 g_{3}^{2} M_3  + g_{1}^{2} M_1 \Big)\mbox{Tr}\Big({Y_d  Y_{d}^{\dagger}}\Big) \nonumber \\ 
 & \ \ \ \ -\frac{12}{5} g_{1}^{2} M_1 \mbox{Tr}\Big({Y_e  Y_{e}^{\dagger}}\Big) -\frac{4}{5} g_{1}^{2} \mbox{Tr}\Big({Y_{d}^{\dagger}  T_d}\Big) +32 g_{3}^{2} \mbox{Tr}\Big({Y_{d}^{\dagger}  T_d}\Big) +\frac{12}{5} g_{1}^{2} \mbox{Tr}\Big({Y_{e}^{\dagger}  T_e}\Big) \nonumber \\ 
 & \ \ \ \ -36 \mbox{Tr}\Big({Y_d  Y_{d}^{\dagger}  T_d  Y_{d}^{\dagger}}\Big) -6 \mbox{Tr}\Big({Y_d  Y_{u}^{\dagger}  T_u  Y_{d}^{\dagger}}\Big) -12 \mbox{Tr}\Big({Y_e  Y_{e}^{\dagger}  T_e  Y_{e}^{\dagger}}\Big) -6 \mbox{Tr}\Big({Y_u  Y_{d}^{\dagger}  T_d  Y_{u}^{\dagger}}\Big) \Big)\\ 
\beta_{T_e}^{(1)} & =  
+4 {Y_e  Y_{e}^{\dagger}  T_e} +5 {T_e  Y_{e}^{\dagger}  Y_e} -\frac{9}{5} g_{1}^{2} T_e -3 g_{2}^{2} T_e +3 T_e \mbox{Tr}\Big({Y_d  Y_{d}^{\dagger}}\Big) +T_e \mbox{Tr}\Big({Y_e  Y_{e}^{\dagger}}\Big) \nonumber \\ 
 & \ \ \ \ +Y_e \Big(2 \mbox{Tr}\Big({Y_{e}^{\dagger}  T_e}\Big)  + 6 g_{2}^{2} M_2  + 6 \mbox{Tr}\Big({Y_{d}^{\dagger}  T_d}\Big)  + \frac{18}{5} g_{1}^{2} M_1 \Big)\\ 
\beta_{T_e}^{(2)} & =  
+\frac{6}{5} g_{1}^{2} {Y_e  Y_{e}^{\dagger}  T_e} +6 g_{2}^{2} {Y_e  Y_{e}^{\dagger}  T_e} -\frac{6}{5} g_{1}^{2} {T_e  Y_{e}^{\dagger}  Y_e} +12 g_{2}^{2} {T_e  Y_{e}^{\dagger}  Y_e} \nonumber \\ 
 & \ \ \ \ -6 {Y_e  Y_{e}^{\dagger}  Y_e  Y_{e}^{\dagger}  T_e} -8 {Y_e  Y_{e}^{\dagger}  T_e  Y_{e}^{\dagger}  Y_e} -6 {T_e  Y_{e}^{\dagger}  Y_e  Y_{e}^{\dagger}  Y_e} +\frac{783}{50} g_{1}^{4} T_e +\frac{9}{5} g_{1}^{2} g_{2}^{2} T_e +\frac{15}{2} g_{2}^{4} T_e \nonumber \\ 
 & \ \ \ \ -12 {Y_e  Y_{e}^{\dagger}  T_e} \mbox{Tr}\Big({Y_d  Y_{d}^{\dagger}}\Big) -15 {T_e  Y_{e}^{\dagger}  Y_e} \mbox{Tr}\Big({Y_d  Y_{d}^{\dagger}}\Big) -\frac{2}{5} g_{1}^{2} T_e \mbox{Tr}\Big({Y_d  Y_{d}^{\dagger}}\Big) \nonumber \\ 
 & \ \ \ \ +16 g_{3}^{2} T_e \mbox{Tr}\Big({Y_d  Y_{d}^{\dagger}}\Big) -4 {Y_e  Y_{e}^{\dagger}  T_e} \mbox{Tr}\Big({Y_e  Y_{e}^{\dagger}}\Big) -5 {T_e  Y_{e}^{\dagger}  Y_e} \mbox{Tr}\Big({Y_e  Y_{e}^{\dagger}}\Big) \nonumber \\ 
 & \ \ \ \ +\frac{6}{5} g_{1}^{2} T_e \mbox{Tr}\Big({Y_e  Y_{e}^{\dagger}}\Big) -6 {Y_e  Y_{e}^{\dagger}  Y_e} \Big(2 g_{2}^{2} M_2  + 3 \mbox{Tr}\Big({Y_{d}^{\dagger}  T_d}\Big)  + \mbox{Tr}\Big({Y_{e}^{\dagger}  T_e}\Big)\Big)-9 T_e \mbox{Tr}\Big({Y_d  Y_{d}^{\dagger}  Y_d  Y_{d}^{\dagger}}\Big) \nonumber \\ 
 & \ \ \ \ -3 T_e \mbox{Tr}\Big({Y_d  Y_{u}^{\dagger}  Y_u  Y_{d}^{\dagger}}\Big) -3 T_e \mbox{Tr}\Big({Y_e  Y_{e}^{\dagger}  Y_e  Y_{e}^{\dagger}}\Big) \nonumber \\ 
 & \ \ \ \ -\frac{2}{25} Y_e \Big(783 g_{1}^{4} M_1 +45 g_{1}^{2} g_{2}^{2} M_1 +45 g_{1}^{2} g_{2}^{2} M_2 +375 g_{2}^{4} M_2 -10 \Big(-40 g_{3}^{2} M_3  + g_{1}^{2} M_1 \Big)\mbox{Tr}\Big({Y_d  Y_{d}^{\dagger}}\Big) \nonumber \\ 
 & \ \ \ \ +30 g_{1}^{2} M_1 \mbox{Tr}\Big({Y_e  Y_{e}^{\dagger}}\Big) +10 g_{1}^{2} \mbox{Tr}\Big({Y_{d}^{\dagger}  T_d}\Big) -400 g_{3}^{2} \mbox{Tr}\Big({Y_{d}^{\dagger}  T_d}\Big) -30 g_{1}^{2} \mbox{Tr}\Big({Y_{e}^{\dagger}  T_e}\Big) \nonumber \\ 
 & \ \ \ \ +450 \mbox{Tr}\Big({Y_d  Y_{d}^{\dagger}  T_d  Y_{d}^{\dagger}}\Big) +75 \mbox{Tr}\Big({Y_d  Y_{u}^{\dagger}  T_u  Y_{d}^{\dagger}}\Big) +150 \mbox{Tr}\Big({Y_e  Y_{e}^{\dagger}  T_e  Y_{e}^{\dagger}}\Big) +75 \mbox{Tr}\Big({Y_u  Y_{d}^{\dagger}  T_d  Y_{u}^{\dagger}}\Big) \Big)\\ 
\beta_{T_u}^{(1)} & =  
+2 {Y_u  Y_{d}^{\dagger}  T_d} +4 {Y_u  Y_{u}^{\dagger}  T_u} +{T_u  Y_{d}^{\dagger}  Y_d}+5 {T_u  Y_{u}^{\dagger}  Y_u} -\frac{13}{15} g_{1}^{2} T_u -3 g_{2}^{2} T_u -\frac{16}{3} g_{3}^{2} T_u \nonumber \\ 
 & \ \ \ \ +3 T_u \mbox{Tr}\Big({Y_u  Y_{u}^{\dagger}}\Big) +Y_u \Big(6 g_{2}^{2} M_2  + 6 \mbox{Tr}\Big({Y_{u}^{\dagger}  T_u}\Big)  + \frac{26}{15} g_{1}^{2} M_1  + \frac{32}{3} g_{3}^{2} M_3 \Big)\\ 
\beta_{T_u}^{(2)} & =  
+\frac{4}{5} g_{1}^{2} {Y_u  Y_{d}^{\dagger}  T_d} -\frac{4}{5} g_{1}^{2} M_1 {Y_u  Y_{u}^{\dagger}  Y_u} -12 g_{2}^{2} M_2 {Y_u  Y_{u}^{\dagger}  Y_u} +\frac{6}{5} g_{1}^{2} {Y_u  Y_{u}^{\dagger}  T_u} \nonumber \\ 
 & \ \ \ \ +6 g_{2}^{2} {Y_u  Y_{u}^{\dagger}  T_u} +\frac{2}{5} g_{1}^{2} {T_u  Y_{d}^{\dagger}  Y_d} +12 g_{2}^{2} {T_u  Y_{u}^{\dagger}  Y_u} -4 {Y_u  Y_{d}^{\dagger}  Y_d  Y_{d}^{\dagger}  T_d} \nonumber \\ 
 & \ \ \ \ -2 {Y_u  Y_{d}^{\dagger}  Y_d  Y_{u}^{\dagger}  T_u} -4 {Y_u  Y_{d}^{\dagger}  T_d  Y_{d}^{\dagger}  Y_d} -4 {Y_u  Y_{d}^{\dagger}  T_d  Y_{u}^{\dagger}  Y_u} -6 {Y_u  Y_{u}^{\dagger}  Y_u  Y_{u}^{\dagger}  T_u} \nonumber \\ 
 & \ \ \ \ -8 {Y_u  Y_{u}^{\dagger}  T_u  Y_{u}^{\dagger}  Y_u} -2 {T_u  Y_{d}^{\dagger}  Y_d  Y_{d}^{\dagger}  Y_d} -4 {T_u  Y_{d}^{\dagger}  Y_d  Y_{u}^{\dagger}  Y_u} -6 {T_u  Y_{u}^{\dagger}  Y_u  Y_{u}^{\dagger}  Y_u} +\frac{3211}{450} g_{1}^{4} T_u \nonumber \\ 
 & \ \ \ \ +g_{1}^{2} g_{2}^{2} T_u +\frac{15}{2} g_{2}^{4} T_u +\frac{136}{45} g_{1}^{2} g_{3}^{2} T_u +8 g_{2}^{2} g_{3}^{2} T_u -\frac{16}{9} g_{3}^{4} T_u -6 {Y_u  Y_{d}^{\dagger}  T_d} \mbox{Tr}\Big({Y_d  Y_{d}^{\dagger}}\Big) \nonumber \\ 
 & \ \ \ \ -3 {T_u  Y_{d}^{\dagger}  Y_d} \mbox{Tr}\Big({Y_d  Y_{d}^{\dagger}}\Big) -2 {Y_u  Y_{d}^{\dagger}  T_d} \mbox{Tr}\Big({Y_e  Y_{e}^{\dagger}}\Big) - {T_u  Y_{d}^{\dagger}  Y_d} \mbox{Tr}\Big({Y_e  Y_{e}^{\dagger}}\Big) \nonumber \\ 
 & \ \ \ \ -12 {Y_u  Y_{u}^{\dagger}  T_u} \mbox{Tr}\Big({Y_u  Y_{u}^{\dagger}}\Big) -15 {T_u  Y_{u}^{\dagger}  Y_u} \mbox{Tr}\Big({Y_u  Y_{u}^{\dagger}}\Big) +\frac{4}{5} g_{1}^{2} T_u \mbox{Tr}\Big({Y_u  Y_{u}^{\dagger}}\Big) \nonumber \\ 
 & \ \ \ \ +16 g_{3}^{2} T_u \mbox{Tr}\Big({Y_u  Y_{u}^{\dagger}}\Big) -\frac{2}{5} {Y_u  Y_{d}^{\dagger}  Y_d} \Big(15 \mbox{Tr}\Big({Y_{d}^{\dagger}  T_d}\Big)  + 2 g_{1}^{2} M_1  + 5 \mbox{Tr}\Big({Y_{e}^{\dagger}  T_e}\Big) \Big)\nonumber \\ 
 & \ \ \ \ -18 {Y_u  Y_{u}^{\dagger}  Y_u} \mbox{Tr}\Big({Y_{u}^{\dagger}  T_u}\Big) -3 T_u \mbox{Tr}\Big({Y_d  Y_{u}^{\dagger}  Y_u  Y_{d}^{\dagger}}\Big) -9 T_u \mbox{Tr}\Big({Y_u  Y_{u}^{\dagger}  Y_u  Y_{u}^{\dagger}}\Big) \nonumber \\ 
 & \ \ \ \ -\frac{2}{225} Y_u \Big(3211 g_{1}^{4} M_1 +225 g_{1}^{2} g_{2}^{2} M_1 +680 g_{1}^{2} g_{3}^{2} M_1 +680 g_{1}^{2} g_{3}^{2} M_3 +1800 g_{2}^{2} g_{3}^{2} M_3 -800 g_{3}^{4} M_3 \nonumber \\ 
 & \ \ \ \ +225 g_{1}^{2} g_{2}^{2} M_2 +3375 g_{2}^{4} M_2 +1800 g_{2}^{2} g_{3}^{2} M_2 +180 \Big(20 g_{3}^{2} M_3  + g_{1}^{2} M_1 \Big)\mbox{Tr}\Big({Y_u  Y_{u}^{\dagger}}\Big) \nonumber \\ 
 & \ \ \ \ -180 \Big(20 g_{3}^{2}  + g_{1}^{2}\Big)\mbox{Tr}\Big({Y_{u}^{\dagger}  T_u}\Big) +675 \mbox{Tr}\Big({Y_d  Y_{u}^{\dagger}  T_u  Y_{d}^{\dagger}}\Big) +675 \mbox{Tr}\Big({Y_u  Y_{d}^{\dagger}  T_d  Y_{u}^{\dagger}}\Big) \nonumber \\ 
 & \ \ \ \ +4050 \mbox{Tr}\Big({Y_u  Y_{u}^{\dagger}  T_u  Y_{u}^{\dagger}}\Big) \Big)
\end{align}} 

\subsection{Vacuum expectation values}

{\allowdisplaybreaks  \begin{align} 
\beta_{v_d}^{(1)} & =  
\frac{1}{20} v_d \Big(-20 \mbox{Tr}\Big({Y_e  Y_{e}^{\dagger}}\Big)  + 3 \Big(5 g_{2}^{2}  + g_{1}^{2}\Big)\Big(1 + \xi \Big) -60 \mbox{Tr}\Big({Y_d  Y_{d}^{\dagger}}\Big) \Big)\\ 
\beta_{v_d}^{(2)} & =  
\frac{1}{400} v_d \Big(-486 g_{1}^{4} -180 g_{1}^{2} g_{2}^{2} -1200 g_{2}^{4} -9 g_{1}^{4} \xi -90 g_{1}^{2} g_{2}^{2} \xi +875 g_{2}^{4} \xi +9 g_{1}^{4} \xi^{2} +90 g_{1}^{2} g_{2}^{2} \xi^{2} \nonumber \\ 
 & \ \ \ \ -225 g_{2}^{4} \xi^{2} -40 \Big(5 \Big(32 g_{3}^{2}  + 9 g_{2}^{2} \xi \Big) + g_{1}^{2} \Big(9 \xi  -4\Big)\Big)\mbox{Tr}\Big({Y_d  Y_{d}^{\dagger}}\Big) \nonumber \\ 
 &  \ \ \ \ -120 \Big(5 g_{2}^{2} \xi  + g_{1}^{2} \Big(4 + \xi\Big)\Big)\mbox{Tr}\Big({Y_e  Y_{e}^{\dagger}}\Big) \nonumber \\ 
 & \ \ \ \ +3600 \mbox{Tr}\Big({Y_d  Y_{d}^{\dagger}  Y_d  Y_{d}^{\dagger}}\Big) +1200 \mbox{Tr}\Big({Y_d  Y_{u}^{\dagger}  Y_u  Y_{d}^{\dagger}}\Big) +1200 \mbox{Tr}\Big({Y_e  Y_{e}^{\dagger}  Y_e  Y_{e}^{\dagger}}\Big) \Big)\\ 
\beta_{v_u}^{(1)} & =  
\frac{3}{20} v_u \Big(-20 \mbox{Tr}\Big({Y_u  Y_{u}^{\dagger}}\Big)  + \Big(5 g_{2}^{2}  + g_{1}^{2}\Big)\Big(1 + \xi\Big)\Big)\\ 
\beta_{v_u}^{(2)} & =  
\frac{1}{400} v_u \Big(-486 g_{1}^{4} -180 g_{1}^{2} g_{2}^{2} -1200 g_{2}^{4} -9 g_{1}^{4} \xi -90 g_{1}^{2} g_{2}^{2} \xi +875 g_{2}^{4} \xi +9 g_{1}^{4} \xi^{2} +90 g_{1}^{2} g_{2}^{2} \xi^{2} \nonumber \\ 
 & \ \ \ \ -225 g_{2}^{4} \xi^{2} -40 \Big(5 \Big(32 g_{3}^{2}  + 9 g_{2}^{2} \xi \Big) + g_{1}^{2} \Big(9 \xi  + 8\Big)\Big)\mbox{Tr}\Big({Y_u  Y_{u}^{\dagger}}\Big)\nonumber \\ 
 &  \ \ \ \ +1200 \mbox{Tr}\Big({Y_d  Y_{u}^{\dagger}  Y_u  Y_{d}^{\dagger}}\Big) +3600 \mbox{Tr}\Big({Y_u  Y_{u}^{\dagger}  Y_u  Y_{u}^{\dagger}}\Big) \Big)
\end{align}} 
Note that $\xi$ is the gauge-fixing parameter, where we are using the $R_\xi$ gauge.

\subsection{Bilinear Soft-Breaking Parameters}

{\allowdisplaybreaks  \begin{align} 
\beta_{B_{\mu}}^{(1)} & =  
+\frac{6}{5} g_{1}^{2} M_1 \mu +6 g_{2}^{2} M_2 \mu +B_{\mu} \Big(-3 g_{2}^{2}  + 3 \mbox{Tr}\Big({Y_d  Y_{d}^{\dagger}}\Big)  + 3 \mbox{Tr}\Big({Y_u  Y_{u}^{\dagger}}\Big)  -\frac{3}{5} g_{1}^{2}  + \mbox{Tr}\Big({Y_e  Y_{e}^{\dagger}}\Big)\Big)\nonumber \\ 
 & \ \ \ \ +6 \mu \mbox{Tr}\Big({Y_{d}^{\dagger}  T_d}\Big) +2 \mu \mbox{Tr}\Big({Y_{e}^{\dagger}  T_e}\Big) +6 \mu \mbox{Tr}\Big({Y_{u}^{\dagger}  T_u}\Big) \\ 
\beta_{B_{\mu}}^{(2)} & =  
+B_{\mu} \Big(\frac{243}{50} g_{1}^{4} +\frac{9}{5} g_{1}^{2} g_{2}^{2} +\frac{15}{2} g_{2}^{4} -\frac{2}{5} \Big(-40 g_{3}^{2}  + g_{1}^{2}\Big)\mbox{Tr}\Big({Y_d  Y_{d}^{\dagger}}\Big) +\frac{6}{5} g_{1}^{2} \mbox{Tr}\Big({Y_e  Y_{e}^{\dagger}}\Big) +\frac{4}{5} g_{1}^{2} \mbox{Tr}\Big({Y_u  Y_{u}^{\dagger}}\Big) \nonumber \\ 
 & \ \ \ \ +16 g_{3}^{2} \mbox{Tr}\Big({Y_u  Y_{u}^{\dagger}}\Big) -9 \mbox{Tr}\Big({Y_d  Y_{d}^{\dagger}  Y_d  Y_{d}^{\dagger}}\Big) -6 \mbox{Tr}\Big({Y_d  Y_{u}^{\dagger}  Y_u  Y_{d}^{\dagger}}\Big) -3 \mbox{Tr}\Big({Y_e  Y_{e}^{\dagger}  Y_e  Y_{e}^{\dagger}}\Big) -9 \mbox{Tr}\Big({Y_u  Y_{u}^{\dagger}  Y_u  Y_{u}^{\dagger}}\Big) \Big)\nonumber \\ 
 & \ \ \ \ -\frac{2}{25} \mu \Big(243 g_{1}^{4} M_1 +45 g_{1}^{2} g_{2}^{2} M_1 +45 g_{1}^{2} g_{2}^{2} M_2 +375 g_{2}^{4} M_2 -10 \Big(-40 g_{3}^{2} M_3  + g_{1}^{2} M_1 \Big)\mbox{Tr}\Big({Y_d  Y_{d}^{\dagger}}\Big) \nonumber \\ 
 & \ \ \ \ +30 g_{1}^{2} M_1 \mbox{Tr}\Big({Y_e  Y_{e}^{\dagger}}\Big) +20 g_{1}^{2} M_1 \mbox{Tr}\Big({Y_u  Y_{u}^{\dagger}}\Big) +400 g_{3}^{2} M_3 \mbox{Tr}\Big({Y_u  Y_{u}^{\dagger}}\Big) +10 g_{1}^{2} \mbox{Tr}\Big({Y_{d}^{\dagger}  T_d}\Big) \nonumber \\ 
 & \ \ \ \ -400 g_{3}^{2} \mbox{Tr}\Big({Y_{d}^{\dagger}  T_d}\Big) -30 g_{1}^{2} \mbox{Tr}\Big({Y_{e}^{\dagger}  T_e}\Big) -20 g_{1}^{2} \mbox{Tr}\Big({Y_{u}^{\dagger}  T_u}\Big) -400 g_{3}^{2} \mbox{Tr}\Big({Y_{u}^{\dagger}  T_u}\Big) \nonumber \\ 
 & \ \ \ \ +450 \mbox{Tr}\Big({Y_d  Y_{d}^{\dagger}  T_d  Y_{d}^{\dagger}}\Big) +150 \mbox{Tr}\Big({Y_d  Y_{u}^{\dagger}  T_u  Y_{d}^{\dagger}}\Big) +150 \mbox{Tr}\Big({Y_e  Y_{e}^{\dagger}  T_e  Y_{e}^{\dagger}}\Big) +150 \mbox{Tr}\Big({Y_u  Y_{d}^{\dagger}  T_d  Y_{u}^{\dagger}}\Big) \nonumber \\ 
 & \ \ \ \ +450 \mbox{Tr}\Big({Y_u  Y_{u}^{\dagger}  T_u  Y_{u}^{\dagger}}\Big) \Big)\\ 
\beta_{B_{\nu}}^{(1)} & =  
\frac{12}{5} g_{1}^{2} \Big(2 M_1 \nu  - B_{\nu} \Big)  \ \ \ ,   \ \
\beta_{B_{\nu}}^{(2)} =  
-\frac{108}{5} g_{1}^{4} \Big(4 M_1 \nu  - B_{\nu} \Big).
\end{align}} 

\subsection{Soft-Breaking Scalar Masses}
\begin{align} 
\sigma_{1,1} & = \sqrt{\frac{3}{5}} g_1 \Big(-2 \mbox{Tr}\Big({m_u^2}\Big)  - \mbox{Tr}\Big({m_l^2}\Big)  - m_{H_d}^2  - m_{F_-}^2 + m_{H_u}^2 + m_{F_+}^2 + \mbox{Tr}\Big({m_d^2}\Big) \nonumber \\ 
& \ \ \ \ + \mbox{Tr}\Big({m_e^2}\Big) + \mbox{Tr}\Big({m_q^2}\Big)\Big)\\ 
\sigma_{2,11} & = \frac{1}{10} g_{1}^{2} \Big(2 \mbox{Tr}\Big({m_d^2}\Big)  + 3 \mbox{Tr}\Big({m_l^2}\Big)  + 3 m_{H_d}^2  + 3 m_{H_u}^2  + 6 \mbox{Tr}\Big({m_e^2}\Big)  + 6 m_{F^-}^2  + 6 m_{F^+}^2 \nonumber \\ 
&  \ \ \ \ + 8 \mbox{Tr}\Big({m_u^2}\Big)  + \mbox{Tr}\Big({m_q^2}\Big)\Big)\\ 
\sigma_{3,1} & = \frac{1}{20} \frac{1}{\sqrt{15}} g_1 \Big(-9 g_{1}^{2} m_{H_d}^2 -45 g_{2}^{2} m_{H_d}^2 +9 g_{1}^{2} m_{H_u}^2 +45 g_{2}^{2} m_{H_u}^2 -36 g_{1}^{2} m_{F^-}^2 +36 g_{1}^{2} m_{F^+}^2 \nonumber \\
 & \ \ \ \ +4 \Big(20 g_{3}^{2}  + g_{1}^{2}\Big)\mbox{Tr}\Big({m_d^2}\Big) +36 g_{1}^{2} \mbox{Tr}\Big({m_e^2}\Big) -9 g_{1}^{2} \mbox{Tr}\Big({m_l^2}\Big) -45 g_{2}^{2} \mbox{Tr}\Big({m_l^2}\Big) +g_{1}^{2} \mbox{Tr}\Big({m_q^2}\Big) \nonumber \\ 
 & \ \ \ \ +45 g_{2}^{2} \mbox{Tr}\Big({m_q^2}\Big) +80 g_{3}^{2} \mbox{Tr}\Big({m_q^2}\Big) -32 g_{1}^{2} \mbox{Tr}\Big({m_u^2}\Big) -160 g_{3}^{2} \mbox{Tr}\Big({m_u^2}\Big) +90 m_{H_d}^2 \mbox{Tr}\Big({Y_d  Y_{d}^{\dagger}}\Big) \nonumber \\ 
 & \ \ \ \ +30 m_{H_d}^2 \mbox{Tr}\Big({Y_e  Y_{e}^{\dagger}}\Big) -90 m_{H_u}^2 \mbox{Tr}\Big({Y_u  Y_{u}^{\dagger}}\Big) 
-60 \mbox{Tr}\Big({Y_d  Y_{d}^{\dagger}  m_d^{2 *}}\Big) -30 \mbox{Tr}\Big({Y_d  m_q^{2 *}  Y_{d}^{\dagger}}\Big)\nonumber \\ 
& \ \ \ \  -60 \mbox{Tr}\Big({Y_e  Y_{e}^{\dagger}  m_e^{2 *}}\Big) +30 \mbox{Tr}\Big({Y_e  m_l^{2 *}  Y_{e}^{\dagger}}\Big)  +120 \mbox{Tr}\Big({Y_u  Y_{u}^{\dagger}  m_u^{2 *}}\Big) -30 \mbox{Tr}\Big({Y_u  m_q^{2 *}  Y_{u}^{\dagger}}\Big) \Big)\\ 
\sigma_{2,2} & = \frac{1}{2} \Big(3 \mbox{Tr}\Big({m_q^2}\Big)  + m_{H_d}^2 + m_{H_u}^2 + \mbox{Tr}\Big({m_l^2}\Big)\Big)\\ 
\sigma_{2,3} & = \frac{1}{2} \Big(2 \mbox{Tr}\Big({m_q^2}\Big)  + \mbox{Tr}\Big({m_d^2}\Big) + \mbox{Tr}\Big({m_u^2}\Big)\Big)
\end{align}
 
{\allowdisplaybreaks  \begin{align} 
\beta_{m_q^2}^{(1)} & =  
-\frac{2}{15} g_{1}^{2} {\bf 1} |M_1|^2 -\frac{32}{3} g_{3}^{2} {\bf 1} |M_3|^2 -6 g_{2}^{2} {\bf 1} |M_2|^2 +2 m_{H_d}^2 {Y_{d}^{\dagger}  Y_d} +2 m_{H_u}^2 {Y_{u}^{\dagger}  Y_u} +2 {T_{d}^{\dagger}  T_d} \nonumber \\ 
 & \ \ \ \ +2 {T_{u}^{\dagger}  T_u} +{m_q^2  Y_{d}^{\dagger}  Y_d}+{m_q^2  Y_{u}^{\dagger}  Y_u}+2 {Y_{d}^{\dagger}  m_d^2  Y_d} +{Y_{d}^{\dagger}  Y_d  m_q^2}+2 {Y_{u}^{\dagger}  m_u^2  Y_u} \nonumber \\ 
 & \ \ \ \ +{Y_{u}^{\dagger}  Y_u  m_q^2}+\frac{1}{\sqrt{15}} g_1 {\bf 1} \sigma_{1,1} \\ 
\beta_{m_q^2}^{(2)} & =  
+\frac{2}{5} g_{1}^{2} g_{2}^{2} {\bf 1} |M_2|^2 +33 g_{2}^{4} {\bf 1} |M_2|^2 +32 g_{2}^{2} g_{3}^{2} {\bf 1} |M_2|^2 \nonumber \\ 
 & \ \ \ \ +\frac{16}{45} g_{3}^{2} \Big(15 \Big(3 g_{2}^{2} \Big(2 M_3  + M_2\Big) -8 g_{3}^{2} M_3 \Big) + g_{1}^{2} \Big(2 M_3  + M_1\Big)\Big){\bf 1} M_3^*  \nonumber \\ 
 & \ \ \ \ +\frac{1}{5} g_{1}^{2} g_{2}^{2} M_1 {\bf 1} M_2^* +16 g_{2}^{2} g_{3}^{2} M_3 {\bf 1} M_2^* +\frac{4}{5} g_{1}^{2} m_{H_d}^2 {Y_{d}^{\dagger}  Y_d} +\frac{8}{5} g_{1}^{2} m_{H_u}^2 {Y_{u}^{\dagger}  Y_u} \nonumber \\ 
 & \ \ \ \ +\frac{1}{45} g_{1}^{2} M_1^* \Big(\Big(141 g_{1}^{2} M_1  + 16 g_{3}^{2} \Big(2 M_1  + M_3\Big) + 9 g_{2}^{2} \Big(2 M_1  + M_2\Big)\Big){\bf 1} \nonumber \\ 
 & \ \ \ \ +36 \Big(2 M_1 {Y_{d}^{\dagger}  Y_d}  -2 {Y_{u}^{\dagger}  T_u}  + 4 M_1 {Y_{u}^{\dagger}  Y_u}  - {Y_{d}^{\dagger}  T_d} \Big)\Big)\nonumber \\ 
 & \ \ \ \ -\frac{4}{5} g_{1}^{2} M_1 {T_{d}^{\dagger}  Y_d} +\frac{4}{5} g_{1}^{2} {T_{d}^{\dagger}  T_d} -\frac{8}{5} g_{1}^{2} M_1 {T_{u}^{\dagger}  Y_u} +\frac{8}{5} g_{1}^{2} {T_{u}^{\dagger}  T_u} \nonumber \\ 
 & \ \ \ \ +\frac{2}{5} g_{1}^{2} {m_q^2  Y_{d}^{\dagger}  Y_d} +\frac{4}{5} g_{1}^{2} {m_q^2  Y_{u}^{\dagger}  Y_u} +\frac{4}{5} g_{1}^{2} {Y_{d}^{\dagger}  m_d^2  Y_d} +\frac{2}{5} g_{1}^{2} {Y_{d}^{\dagger}  Y_d  m_q^2} \nonumber \\ 
 & \ \ \ \ +\frac{8}{5} g_{1}^{2} {Y_{u}^{\dagger}  m_u^2  Y_u} +\frac{4}{5} g_{1}^{2} {Y_{u}^{\dagger}  Y_u  m_q^2} -8 m_{H_d}^2 {Y_{d}^{\dagger}  Y_d  Y_{d}^{\dagger}  Y_d} -4 {Y_{d}^{\dagger}  Y_d  T_{d}^{\dagger}  T_d} \nonumber \\ 
 & \ \ \ \ -4 {Y_{d}^{\dagger}  T_d  T_{d}^{\dagger}  Y_d} -8 m_{H_u}^2 {Y_{u}^{\dagger}  Y_u  Y_{u}^{\dagger}  Y_u} -4 {Y_{u}^{\dagger}  Y_u  T_{u}^{\dagger}  T_u} -4 {Y_{u}^{\dagger}  T_u  T_{u}^{\dagger}  Y_u} \nonumber \\ 
 & \ \ \ \ -4 {T_{d}^{\dagger}  Y_d  Y_{d}^{\dagger}  T_d} -4 {T_{d}^{\dagger}  T_d  Y_{d}^{\dagger}  Y_d} -4 {T_{u}^{\dagger}  Y_u  Y_{u}^{\dagger}  T_u} -4 {T_{u}^{\dagger}  T_u  Y_{u}^{\dagger}  Y_u} \nonumber \\ 
 & \ \ \ \ -2 {m_q^2  Y_{d}^{\dagger}  Y_d  Y_{d}^{\dagger}  Y_d} -2 {m_q^2  Y_{u}^{\dagger}  Y_u  Y_{u}^{\dagger}  Y_u} -4 {Y_{d}^{\dagger}  m_d^2  Y_d  Y_{d}^{\dagger}  Y_d} -4 {Y_{d}^{\dagger}  Y_d  m_q^2  Y_{d}^{\dagger}  Y_d} \nonumber \\ 
 & \ \ \ \ -4 {Y_{d}^{\dagger}  Y_d  Y_{d}^{\dagger}  m_d^2  Y_d} -2 {Y_{d}^{\dagger}  Y_d  Y_{d}^{\dagger}  Y_d  m_q^2} -4 {Y_{u}^{\dagger}  m_u^2  Y_u  Y_{u}^{\dagger}  Y_u} -4 {Y_{u}^{\dagger}  Y_u  m_q^2  Y_{u}^{\dagger}  Y_u} \nonumber \\ 
 & \ \ \ \ -4 {Y_{u}^{\dagger}  Y_u  Y_{u}^{\dagger}  m_u^2  Y_u} -2 {Y_{u}^{\dagger}  Y_u  Y_{u}^{\dagger}  Y_u  m_q^2} +6 g_{2}^{4} {\bf 1} \sigma_{2,2} +\frac{32}{3} g_{3}^{4} {\bf 1} \sigma_{2,3} +\frac{2}{15} g_{1}^{2} {\bf 1} \sigma_{2,11} +4 \frac{1}{\sqrt{15}} g_1 {\bf 1} \sigma_{3,1} \nonumber \\ 
 & \ \ \ \ -12 m_{H_d}^2 {Y_{d}^{\dagger}  Y_d} \mbox{Tr}\Big({Y_d  Y_{d}^{\dagger}}\Big) -6 {T_{d}^{\dagger}  T_d} \mbox{Tr}\Big({Y_d  Y_{d}^{\dagger}}\Big) -3 {m_q^2  Y_{d}^{\dagger}  Y_d} \mbox{Tr}\Big({Y_d  Y_{d}^{\dagger}}\Big) \nonumber \\ 
 & \ \ \ \ -6 {Y_{d}^{\dagger}  m_d^2  Y_d} \mbox{Tr}\Big({Y_d  Y_{d}^{\dagger}}\Big) -3 {Y_{d}^{\dagger}  Y_d  m_q^2} \mbox{Tr}\Big({Y_d  Y_{d}^{\dagger}}\Big) -4 m_{H_d}^2 {Y_{d}^{\dagger}  Y_d} \mbox{Tr}\Big({Y_e  Y_{e}^{\dagger}}\Big) \nonumber \\ 
 & \ \ \ \ -2 {T_{d}^{\dagger}  T_d} \mbox{Tr}\Big({Y_e  Y_{e}^{\dagger}}\Big) - {m_q^2  Y_{d}^{\dagger}  Y_d} \mbox{Tr}\Big({Y_e  Y_{e}^{\dagger}}\Big) -2 {Y_{d}^{\dagger}  m_d^2  Y_d} \mbox{Tr}\Big({Y_e  Y_{e}^{\dagger}}\Big) \nonumber \\ 
 & \ \ \ \ - {Y_{d}^{\dagger}  Y_d  m_q^2} \mbox{Tr}\Big({Y_e  Y_{e}^{\dagger}}\Big) -12 m_{H_u}^2 {Y_{u}^{\dagger}  Y_u} \mbox{Tr}\Big({Y_u  Y_{u}^{\dagger}}\Big) -6 {T_{u}^{\dagger}  T_u} \mbox{Tr}\Big({Y_u  Y_{u}^{\dagger}}\Big) \nonumber \\ 
 & \ \ \ \ -3 {m_q^2  Y_{u}^{\dagger}  Y_u} \mbox{Tr}\Big({Y_u  Y_{u}^{\dagger}}\Big) -6 {Y_{u}^{\dagger}  m_u^2  Y_u} \mbox{Tr}\Big({Y_u  Y_{u}^{\dagger}}\Big) -3 {Y_{u}^{\dagger}  Y_u  m_q^2} \mbox{Tr}\Big({Y_u  Y_{u}^{\dagger}}\Big) \nonumber \\ 
 & \ \ \ \ -6 {T_{d}^{\dagger}  Y_d} \mbox{Tr}\Big({Y_{d}^{\dagger}  T_d}\Big) -2 {T_{d}^{\dagger}  Y_d} \mbox{Tr}\Big({Y_{e}^{\dagger}  T_e}\Big) -6 {T_{u}^{\dagger}  Y_u} \mbox{Tr}\Big({Y_{u}^{\dagger}  T_u}\Big) \nonumber \\ 
 & \ \ \ \ -6 {Y_{d}^{\dagger}  T_d} \mbox{Tr}\Big({T_d^*  Y_{d}^{T}}\Big) -6 {Y_{d}^{\dagger}  Y_d} \mbox{Tr}\Big({T_d^*  T_{d}^{T}}\Big) -2 {Y_{d}^{\dagger}  T_d} \mbox{Tr}\Big({T_e^*  Y_{e}^{T}}\Big) \nonumber \\ 
 & \ \ \ \ -2 {Y_{d}^{\dagger}  Y_d} \mbox{Tr}\Big({T_e^*  T_{e}^{T}}\Big) -6 {Y_{u}^{\dagger}  T_u} \mbox{Tr}\Big({T_u^*  Y_{u}^{T}}\Big) -6 {Y_{u}^{\dagger}  Y_u} \mbox{Tr}\Big({T_u^*  T_{u}^{T}}\Big) \nonumber \\ 
 & \ \ \ \ -6 {Y_{d}^{\dagger}  Y_d} \mbox{Tr}\Big({m_d^2  Y_d  Y_{d}^{\dagger}}\Big) -2 {Y_{d}^{\dagger}  Y_d} \mbox{Tr}\Big({m_e^2  Y_e  Y_{e}^{\dagger}}\Big) -2 {Y_{d}^{\dagger}  Y_d} \mbox{Tr}\Big({m_l^2  Y_{e}^{\dagger}  Y_e}\Big) \nonumber \\ 
 & \ \ \ \ -6 {Y_{d}^{\dagger}  Y_d} \mbox{Tr}\Big({m_q^2  Y_{d}^{\dagger}  Y_d}\Big) -6 {Y_{u}^{\dagger}  Y_u} \mbox{Tr}\Big({m_q^2  Y_{u}^{\dagger}  Y_u}\Big) -6 {Y_{u}^{\dagger}  Y_u} \mbox{Tr}\Big({m_u^2  Y_u  Y_{u}^{\dagger}}\Big) \\ 
\beta_{m_l^2}^{(1)} & =  
-\frac{6}{5} g_{1}^{2} {\bf 1} |M_1|^2 -6 g_{2}^{2} {\bf 1} |M_2|^2 +2 m_{H_d}^2 {Y_{e}^{\dagger}  Y_e} +2 {T_{e}^{\dagger}  T_e} +{m_l^2  Y_{e}^{\dagger}  Y_e}+2 {Y_{e}^{\dagger}  m_e^2  Y_e} \nonumber \\ 
 & \ \ \ \ +{Y_{e}^{\dagger}  Y_e  m_l^2}- \sqrt{\frac{3}{5}} g_1 {\bf 1} \sigma_{1,1} \\ 
\beta_{m_l^2}^{(2)} & =  
+\frac{3}{5} g_{2}^{2} \Big(3 g_{1}^{2} \Big(2 M_2  + M_1\Big) + 55 g_{2}^{2} M_2 \Big){\bf 1} M_2^* +\frac{12}{5} g_{1}^{2} m_{H_d}^2 {Y_{e}^{\dagger}  Y_e} \nonumber \\ 
 & \ \ \ \ +\frac{3}{25} g_{1}^{2} M_1^* \Big(-20 {Y_{e}^{\dagger}  T_e}  + 3 \Big(5 g_{2}^{2} \Big(2 M_1  + M_2\Big) + 81 g_{1}^{2} M_1 \Big){\bf 1}  + 40 M_1 {Y_{e}^{\dagger}  Y_e} \Big)-\frac{12}{5} g_{1}^{2} M_1 {T_{e}^{\dagger}  Y_e} \nonumber \\ 
 & \ \ \ \ +\frac{12}{5} g_{1}^{2} {T_{e}^{\dagger}  T_e} +\frac{6}{5} g_{1}^{2} {m_l^2  Y_{e}^{\dagger}  Y_e} +\frac{12}{5} g_{1}^{2} {Y_{e}^{\dagger}  m_e^2  Y_e} +\frac{6}{5} g_{1}^{2} {Y_{e}^{\dagger}  Y_e  m_l^2} \nonumber \\ 
 & \ \ \ \ -8 m_{H_d}^2 {Y_{e}^{\dagger}  Y_e  Y_{e}^{\dagger}  Y_e} -4 {Y_{e}^{\dagger}  Y_e  T_{e}^{\dagger}  T_e} -4 {Y_{e}^{\dagger}  T_e  T_{e}^{\dagger}  Y_e} -4 {T_{e}^{\dagger}  Y_e  Y_{e}^{\dagger}  T_e} \nonumber \\ 
 & \ \ \ \ -4 {T_{e}^{\dagger}  T_e  Y_{e}^{\dagger}  Y_e} -2 {m_l^2  Y_{e}^{\dagger}  Y_e  Y_{e}^{\dagger}  Y_e} -4 {Y_{e}^{\dagger}  m_e^2  Y_e  Y_{e}^{\dagger}  Y_e} -4 {Y_{e}^{\dagger}  Y_e  m_l^2  Y_{e}^{\dagger}  Y_e} \nonumber \\ 
 & \ \ \ \ -4 {Y_{e}^{\dagger}  Y_e  Y_{e}^{\dagger}  m_e^2  Y_e} -2 {Y_{e}^{\dagger}  Y_e  Y_{e}^{\dagger}  Y_e  m_l^2} +6 g_{2}^{4} {\bf 1} \sigma_{2,2} +\frac{6}{5} g_{1}^{2} {\bf 1} \sigma_{2,11} -4 \sqrt{\frac{3}{5}} g_1 {\bf 1} \sigma_{3,1} \nonumber \\ 
 & \ \ \ \ -12 m_{H_d}^2 {Y_{e}^{\dagger}  Y_e} \mbox{Tr}\Big({Y_d  Y_{d}^{\dagger}}\Big) -6 {T_{e}^{\dagger}  T_e} \mbox{Tr}\Big({Y_d  Y_{d}^{\dagger}}\Big) -3 {m_l^2  Y_{e}^{\dagger}  Y_e} \mbox{Tr}\Big({Y_d  Y_{d}^{\dagger}}\Big) \nonumber \\ 
 & \ \ \ \ -6 {Y_{e}^{\dagger}  m_e^2  Y_e} \mbox{Tr}\Big({Y_d  Y_{d}^{\dagger}}\Big) -3 {Y_{e}^{\dagger}  Y_e  m_l^2} \mbox{Tr}\Big({Y_d  Y_{d}^{\dagger}}\Big) -4 m_{H_d}^2 {Y_{e}^{\dagger}  Y_e} \mbox{Tr}\Big({Y_e  Y_{e}^{\dagger}}\Big) \nonumber \\ 
 & \ \ \ \ -2 {T_{e}^{\dagger}  T_e} \mbox{Tr}\Big({Y_e  Y_{e}^{\dagger}}\Big) - {m_l^2  Y_{e}^{\dagger}  Y_e} \mbox{Tr}\Big({Y_e  Y_{e}^{\dagger}}\Big) -2 {Y_{e}^{\dagger}  m_e^2  Y_e} \mbox{Tr}\Big({Y_e  Y_{e}^{\dagger}}\Big) \nonumber \\ 
 & \ \ \ \ - {Y_{e}^{\dagger}  Y_e  m_l^2} \mbox{Tr}\Big({Y_e  Y_{e}^{\dagger}}\Big) -6 {T_{e}^{\dagger}  Y_e} \mbox{Tr}\Big({Y_{d}^{\dagger}  T_d}\Big) -2 {T_{e}^{\dagger}  Y_e} \mbox{Tr}\Big({Y_{e}^{\dagger}  T_e}\Big) \nonumber \\ 
 & \ \ \ \ -6 {Y_{e}^{\dagger}  T_e} \mbox{Tr}\Big({T_d^*  Y_{d}^{T}}\Big) -6 {Y_{e}^{\dagger}  Y_e} \mbox{Tr}\Big({T_d^*  T_{d}^{T}}\Big) -2 {Y_{e}^{\dagger}  T_e} \mbox{Tr}\Big({T_e^*  Y_{e}^{T}}\Big) \nonumber \\ 
 & \ \ \ \ -2 {Y_{e}^{\dagger}  Y_e} \mbox{Tr}\Big({T_e^*  T_{e}^{T}}\Big) -6 {Y_{e}^{\dagger}  Y_e} \mbox{Tr}\Big({m_d^2  Y_d  Y_{d}^{\dagger}}\Big) -2 {Y_{e}^{\dagger}  Y_e} \mbox{Tr}\Big({m_e^2  Y_e  Y_{e}^{\dagger}}\Big) \nonumber \\ 
 & \ \ \ \ -2 {Y_{e}^{\dagger}  Y_e} \mbox{Tr}\Big({m_l^2  Y_{e}^{\dagger}  Y_e}\Big) -6 {Y_{e}^{\dagger}  Y_e} \mbox{Tr}\Big({m_q^2  Y_{d}^{\dagger}  Y_d}\Big) \\ 
\beta_{m_{H_d}^2}^{(1)} & =  
-\frac{6}{5} g_{1}^{2} |M_1|^2 -6 g_{2}^{2} |M_2|^2 - \sqrt{\frac{3}{5}} g_1 \sigma_{1,1} +6 m_{H_d}^2 \mbox{Tr}\Big({Y_d  Y_{d}^{\dagger}}\Big) +2 m_{H_d}^2 \mbox{Tr}\Big({Y_e  Y_{e}^{\dagger}}\Big) +6 \mbox{Tr}\Big({T_d^*  T_{d}^{T}}\Big) \nonumber \\ 
 & \ \ \ \ +2 \mbox{Tr}\Big({T_e^*  T_{e}^{T}}\Big) +6 \mbox{Tr}\Big({m_d^2  Y_d  Y_{d}^{\dagger}}\Big) +2 \mbox{Tr}\Big({m_e^2  Y_e  Y_{e}^{\dagger}}\Big) +2 \mbox{Tr}\Big({m_l^2  Y_{e}^{\dagger}  Y_e}\Big) +6 \mbox{Tr}\Big({m_q^2  Y_{d}^{\dagger}  Y_d}\Big) \\ 
\beta_{m_{H_d}^2}^{(2)} & =  
\frac{1}{25} \Big(15 g_{2}^{2} \Big(3 g_{1}^{2} \Big(2 M_2  + M_1\Big) + 55 g_{2}^{2} M_2 \Big)M_2^* \nonumber \\ 
 & \ \ \ \ +g_{1}^{2} M_1^* \Big(729 g_{1}^{2} M_1 +90 g_{2}^{2} M_1 +45 g_{2}^{2} M_2 -40 M_1 \mbox{Tr}\Big({Y_d  Y_{d}^{\dagger}}\Big) +120 M_1 \mbox{Tr}\Big({Y_e  Y_{e}^{\dagger}}\Big) +20 \mbox{Tr}\Big({Y_{d}^{\dagger}  T_d}\Big) \nonumber \\ 
 & \ \ \ \ -60 \mbox{Tr}\Big({Y_{e}^{\dagger}  T_e}\Big) \Big)\nonumber \\ 
 & \ \ \ \ +10 \Big(15 g_{2}^{4} \sigma_{2,2} +3 g_{1}^{2} \sigma_{2,11} -2 \sqrt{15} g_1 \sigma_{3,1} +\Big(160 g_{3}^{2} |M_3|^2  -2 g_{1}^{2} m_{H_d}^2  + 80 g_{3}^{2} m_{H_d}^2 \Big)\mbox{Tr}\Big({Y_d  Y_{d}^{\dagger}}\Big) \nonumber \\ 
 & \ \ \ \ +6 g_{1}^{2} m_{H_d}^2 \mbox{Tr}\Big({Y_e  Y_{e}^{\dagger}}\Big) -80 g_{3}^{2} M_3^* \mbox{Tr}\Big({Y_{d}^{\dagger}  T_d}\Big) +2 g_{1}^{2} M_1 \mbox{Tr}\Big({T_d^*  Y_{d}^{T}}\Big) -80 g_{3}^{2} M_3 \mbox{Tr}\Big({T_d^*  Y_{d}^{T}}\Big) \nonumber \\ 
 & \ \ \ \ -2 g_{1}^{2} \mbox{Tr}\Big({T_d^*  T_{d}^{T}}\Big) +80 g_{3}^{2} \mbox{Tr}\Big({T_d^*  T_{d}^{T}}\Big) -6 g_{1}^{2} M_1 \mbox{Tr}\Big({T_e^*  Y_{e}^{T}}\Big) +6 g_{1}^{2} \mbox{Tr}\Big({T_e^*  T_{e}^{T}}\Big) \nonumber \\ 
 & \ \ \ \ -2 g_{1}^{2} \mbox{Tr}\Big({m_d^2  Y_d  Y_{d}^{\dagger}}\Big) +80 g_{3}^{2} \mbox{Tr}\Big({m_d^2  Y_d  Y_{d}^{\dagger}}\Big) +6 g_{1}^{2} \mbox{Tr}\Big({m_e^2  Y_e  Y_{e}^{\dagger}}\Big) +6 g_{1}^{2} \mbox{Tr}\Big({m_l^2  Y_{e}^{\dagger}  Y_e}\Big) \nonumber \\ 
 & \ \ \ \ -2 g_{1}^{2} \mbox{Tr}\Big({m_q^2  Y_{d}^{\dagger}  Y_d}\Big) +80 g_{3}^{2} \mbox{Tr}\Big({m_q^2  Y_{d}^{\dagger}  Y_d}\Big) -90 m_{H_d}^2 \mbox{Tr}\Big({Y_d  Y_{d}^{\dagger}  Y_d  Y_{d}^{\dagger}}\Big) -90 \mbox{Tr}\Big({Y_d  Y_{d}^{\dagger}  T_d  T_{d}^{\dagger}}\Big) \nonumber \\ 
 & \ \ \ \ -15 m_{H_d}^2 \mbox{Tr}\Big({Y_d  Y_{u}^{\dagger}  Y_u  Y_{d}^{\dagger}}\Big) -15 m_{H_u}^2 \mbox{Tr}\Big({Y_d  Y_{u}^{\dagger}  Y_u  Y_{d}^{\dagger}}\Big) -15 \mbox{Tr}\Big({Y_d  Y_{u}^{\dagger}  T_u  T_{d}^{\dagger}}\Big) \nonumber \\ 
 & \ \ \ \ -90 \mbox{Tr}\Big({Y_d  T_{d}^{\dagger}  T_d  Y_{d}^{\dagger}}\Big) -15 \mbox{Tr}\Big({Y_d  T_{u}^{\dagger}  T_u  Y_{d}^{\dagger}}\Big) -30 m_{H_d}^2 \mbox{Tr}\Big({Y_e  Y_{e}^{\dagger}  Y_e  Y_{e}^{\dagger}}\Big) -30 \mbox{Tr}\Big({Y_e  Y_{e}^{\dagger}  T_e  T_{e}^{\dagger}}\Big) \nonumber \\ 
 & \ \ \ \ -30 \mbox{Tr}\Big({Y_e  T_{e}^{\dagger}  T_e  Y_{e}^{\dagger}}\Big) -15 \mbox{Tr}\Big({Y_u  Y_{d}^{\dagger}  T_d  T_{u}^{\dagger}}\Big) -15 \mbox{Tr}\Big({Y_u  T_{d}^{\dagger}  T_d  Y_{u}^{\dagger}}\Big) -90 \mbox{Tr}\Big({m_d^2  Y_d  Y_{d}^{\dagger}  Y_d  Y_{d}^{\dagger}}\Big) \nonumber \\ 
 & \ \ \ \ -15 \mbox{Tr}\Big({m_d^2  Y_d  Y_{u}^{\dagger}  Y_u  Y_{d}^{\dagger}}\Big) -30 \mbox{Tr}\Big({m_e^2  Y_e  Y_{e}^{\dagger}  Y_e  Y_{e}^{\dagger}}\Big) -30 \mbox{Tr}\Big({m_l^2  Y_{e}^{\dagger}  Y_e  Y_{e}^{\dagger}  Y_e}\Big) -90 \mbox{Tr}\Big({m_q^2  Y_{d}^{\dagger}  Y_d  Y_{d}^{\dagger}  Y_d}\Big) \nonumber \\ 
 & \ \ \ \ -15 \mbox{Tr}\Big({m_q^2  Y_{d}^{\dagger}  Y_d  Y_{u}^{\dagger}  Y_u}\Big) -15 \mbox{Tr}\Big({m_q^2  Y_{u}^{\dagger}  Y_u  Y_{d}^{\dagger}  Y_d}\Big) -15 \mbox{Tr}\Big({m_u^2  Y_u  Y_{d}^{\dagger}  Y_d  Y_{u}^{\dagger}}\Big) \Big)\Big)\\ 
\beta_{m_{H_u}^2}^{(1)} & =  
-\frac{6}{5} g_{1}^{2} |M_1|^2 -6 g_{2}^{2} |M_2|^2
 +\sqrt{\frac{3}{5}} g_1 \sigma_{1,1} 
+6 m_{H_u}^2 \mbox{Tr}\Big({Y_u  Y_{u}^{\dagger}}\Big) 
+6 \mbox{Tr}\Big({T_u^*  T_{u}^{T}}\Big) +6 \mbox{Tr}\Big({m_q^2  Y_{u}^{\dagger}  Y_u}\Big) \nonumber \\ 
 & \ \ \ \ +6 \mbox{Tr}\Big({m_u^2  Y_u  Y_{u}^{\dagger}}\Big) \\ 
\beta_{m_{H_u}^2}^{(2)} & =  
+\frac{3}{5} g_{2}^{2} \Big(3 g_{1}^{2} \Big(2 M_2  + M_1\Big) + 55 g_{2}^{2} M_2 \Big)M_2^* +6 g_{2}^{4} \sigma_{2,2} +\frac{6}{5} g_{1}^{2} \sigma_{2,11} +4 \sqrt{\frac{3}{5}} g_1 \sigma_{3,1} \nonumber \\ 
 & \ \ \ \ +\frac{8}{5} g_{1}^{2} m_{H_u}^2 \mbox{Tr}\Big({Y_u  Y_{u}^{\dagger}}\Big) +32 g_{3}^{2} m_{H_u}^2 \mbox{Tr}\Big({Y_u  Y_{u}^{\dagger}}\Big) +64 g_{3}^{2} |M_3|^2 \mbox{Tr}\Big({Y_u  Y_{u}^{\dagger}}\Big) \nonumber \\ 
 & \ \ \ \ +\frac{1}{25} g_{1}^{2} M_1^* \Big(-40 \mbox{Tr}\Big({Y_{u}^{\dagger}  T_u}\Big)  + 45 g_{2}^{2} M_2  + 729 g_{1}^{2} M_1  + 80 M_1 \mbox{Tr}\Big({Y_u  Y_{u}^{\dagger}}\Big)  + 90 g_{2}^{2} M_1 \Big)\nonumber \\ 
 & \ \ \ \ -32 g_{3}^{2} M_3^* \mbox{Tr}\Big({Y_{u}^{\dagger}  T_u}\Big) -\frac{8}{5} g_{1}^{2} M_1 \mbox{Tr}\Big({T_u^*  Y_{u}^{T}}\Big) -32 g_{3}^{2} M_3 \mbox{Tr}\Big({T_u^*  Y_{u}^{T}}\Big) +\frac{8}{5} g_{1}^{2} \mbox{Tr}\Big({T_u^*  T_{u}^{T}}\Big) \nonumber \\ 
 & \ \ \ \ +32 g_{3}^{2} \mbox{Tr}\Big({T_u^*  T_{u}^{T}}\Big) +\frac{8}{5} g_{1}^{2} \mbox{Tr}\Big({m_q^2  Y_{u}^{\dagger}  Y_u}\Big) +32 g_{3}^{2} \mbox{Tr}\Big({m_q^2  Y_{u}^{\dagger}  Y_u}\Big) +\frac{8}{5} g_{1}^{2} \mbox{Tr}\Big({m_u^2  Y_u  Y_{u}^{\dagger}}\Big) \nonumber \\ 
 & \ \ \ \ +32 g_{3}^{2} \mbox{Tr}\Big({m_u^2  Y_u  Y_{u}^{\dagger}}\Big) -6 m_{H_d}^2 \mbox{Tr}\Big({Y_d  Y_{u}^{\dagger}  Y_u  Y_{d}^{\dagger}}\Big) -6 m_{H_u}^2 \mbox{Tr}\Big({Y_d  Y_{u}^{\dagger}  Y_u  Y_{d}^{\dagger}}\Big) \nonumber \\ 
 & \ \ \ \ -6 \mbox{Tr}\Big({Y_d  Y_{u}^{\dagger}  T_u  T_{d}^{\dagger}}\Big) -6 \mbox{Tr}\Big({Y_d  T_{u}^{\dagger}  T_u  Y_{d}^{\dagger}}\Big) -6 \mbox{Tr}\Big({Y_u  Y_{d}^{\dagger}  T_d  T_{u}^{\dagger}}\Big) -36 m_{H_u}^2 \mbox{Tr}\Big({Y_u  Y_{u}^{\dagger}  Y_u  Y_{u}^{\dagger}}\Big) \nonumber \\ 
 & \ \ \ \ -36 \mbox{Tr}\Big({Y_u  Y_{u}^{\dagger}  T_u  T_{u}^{\dagger}}\Big) -6 \mbox{Tr}\Big({Y_u  T_{d}^{\dagger}  T_d  Y_{u}^{\dagger}}\Big) -36 \mbox{Tr}\Big({Y_u  T_{u}^{\dagger}  T_u  Y_{u}^{\dagger}}\Big) \nonumber \\ 
 & \ \ \ \ -6 \mbox{Tr}\Big({m_d^2  Y_d  Y_{u}^{\dagger}  Y_u  Y_{d}^{\dagger}}\Big) -6 \mbox{Tr}\Big({m_q^2  Y_{d}^{\dagger}  Y_d  Y_{u}^{\dagger}  Y_u}\Big) -6 \mbox{Tr}\Big({m_q^2  Y_{u}^{\dagger}  Y_u  Y_{d}^{\dagger}  Y_d}\Big) \nonumber \\ 
 & \ \ \ \ -36 \mbox{Tr}\Big({m_q^2  Y_{u}^{\dagger}  Y_u  Y_{u}^{\dagger}  Y_u}\Big) -6 \mbox{Tr}\Big({m_u^2  Y_u  Y_{d}^{\dagger}  Y_d  Y_{u}^{\dagger}}\Big) -36 \mbox{Tr}\Big({m_u^2  Y_u  Y_{u}^{\dagger}  Y_u  Y_{u}^{\dagger}}\Big) \\ 
\beta_{m_d^2}^{(1)} & =  
-\frac{8}{15} g_{1}^{2} {\bf 1} |M_1|^2 -\frac{32}{3} g_{3}^{2} {\bf 1} |M_3|^2 +4 m_{H_d}^2 {Y_d  Y_{d}^{\dagger}} +4 {T_d  T_{d}^{\dagger}} +2 {m_d^2  Y_d  Y_{d}^{\dagger}} +4 {Y_d  m_q^2  Y_{d}^{\dagger}} \nonumber \\ 
 & \ \ \ \ +2 {Y_d  Y_{d}^{\dagger}  m_d^2} +2 \frac{1}{\sqrt{15}} g_1 {\bf 1} \sigma_{1,1} \\ 
\beta_{m_d^2}^{(2)} & =  
+\frac{64}{45} g_{3}^{2} \Big(-30 g_{3}^{2} M_3  + g_{1}^{2} \Big(2 M_3  + M_1\Big)\Big){\bf 1} M_3^* +\frac{4}{5} g_{1}^{2} m_{H_d}^2 {Y_d  Y_{d}^{\dagger}} +12 g_{2}^{2} m_{H_d}^2 {Y_d  Y_{d}^{\dagger}} \nonumber \\ 
 & \ \ \ \ +24 g_{2}^{2} |M_2|^2 {Y_d  Y_{d}^{\dagger}} -\frac{4}{5} g_{1}^{2} M_1 {Y_d  T_{d}^{\dagger}} -12 g_{2}^{2} M_2 {Y_d  T_{d}^{\dagger}} \nonumber \\ 
 & \ \ \ \ +\frac{4}{225} g_{1}^{2} M_1^* \Big(2 \Big(357 g_{1}^{2} M_1  + 40 g_{3}^{2} \Big(2 M_1  + M_3\Big)\Big){\bf 1}  -45 {T_d  Y_{d}^{\dagger}}  + 90 M_1 {Y_d  Y_{d}^{\dagger}} \Big)-12 g_{2}^{2} M_2^* {T_d  Y_{d}^{\dagger}} \nonumber \\ 
 & \ \ \ \ +\frac{4}{5} g_{1}^{2} {T_d  T_{d}^{\dagger}} +12 g_{2}^{2} {T_d  T_{d}^{\dagger}} +\frac{2}{5} g_{1}^{2} {m_d^2  Y_d  Y_{d}^{\dagger}} +6 g_{2}^{2} {m_d^2  Y_d  Y_{d}^{\dagger}} \nonumber \\ 
 & \ \ \ \ +\frac{4}{5} g_{1}^{2} {Y_d  m_q^2  Y_{d}^{\dagger}} +12 g_{2}^{2} {Y_d  m_q^2  Y_{d}^{\dagger}} +\frac{2}{5} g_{1}^{2} {Y_d  Y_{d}^{\dagger}  m_d^2} +6 g_{2}^{2} {Y_d  Y_{d}^{\dagger}  m_d^2} \nonumber \\ 
 & \ \ \ \ -8 m_{H_d}^2 {Y_d  Y_{d}^{\dagger}  Y_d  Y_{d}^{\dagger}} -4 {Y_d  Y_{d}^{\dagger}  T_d  T_{d}^{\dagger}} -4 m_{H_d}^2 {Y_d  Y_{u}^{\dagger}  Y_u  Y_{d}^{\dagger}} \nonumber \\ 
 & \ \ \ \ -4 m_{H_u}^2 {Y_d  Y_{u}^{\dagger}  Y_u  Y_{d}^{\dagger}} -4 {Y_d  Y_{u}^{\dagger}  T_u  T_{d}^{\dagger}} -4 {Y_d  T_{d}^{\dagger}  T_d  Y_{d}^{\dagger}} -4 {Y_d  T_{u}^{\dagger}  T_u  Y_{d}^{\dagger}} \nonumber \\ 
 & \ \ \ \ -4 {T_d  Y_{d}^{\dagger}  Y_d  T_{d}^{\dagger}} -4 {T_d  Y_{u}^{\dagger}  Y_u  T_{d}^{\dagger}} -4 {T_d  T_{d}^{\dagger}  Y_d  Y_{d}^{\dagger}} -4 {T_d  T_{u}^{\dagger}  Y_u  Y_{d}^{\dagger}} \nonumber \\ 
 & \ \ \ \ -2 {m_d^2  Y_d  Y_{d}^{\dagger}  Y_d  Y_{d}^{\dagger}} -2 {m_d^2  Y_d  Y_{u}^{\dagger}  Y_u  Y_{d}^{\dagger}} -4 {Y_d  m_q^2  Y_{d}^{\dagger}  Y_d  Y_{d}^{\dagger}} -4 {Y_d  m_q^2  Y_{u}^{\dagger}  Y_u  Y_{d}^{\dagger}} \nonumber \\ 
 & \ \ \ \ -4 {Y_d  Y_{d}^{\dagger}  m_d^2  Y_d  Y_{d}^{\dagger}} -4 {Y_d  Y_{d}^{\dagger}  Y_d  m_q^2  Y_{d}^{\dagger}} -2 {Y_d  Y_{d}^{\dagger}  Y_d  Y_{d}^{\dagger}  m_d^2} -4 {Y_d  Y_{u}^{\dagger}  m_u^2  Y_u  Y_{d}^{\dagger}} \nonumber \\ 
 & \ \ \ \ -4 {Y_d  Y_{u}^{\dagger}  Y_u  m_q^2  Y_{d}^{\dagger}} -2 {Y_d  Y_{u}^{\dagger}  Y_u  Y_{d}^{\dagger}  m_d^2} +\frac{32}{3} g_{3}^{4} {\bf 1} \sigma_{2,3} +\frac{8}{15} g_{1}^{2} {\bf 1} \sigma_{2,11} +8 \frac{1}{\sqrt{15}} g_1 {\bf 1} \sigma_{3,1} \nonumber \\ 
 & \ \ \ \ -24 m_{H_d}^2 {Y_d  Y_{d}^{\dagger}} \mbox{Tr}\Big({Y_d  Y_{d}^{\dagger}}\Big) -12 {T_d  T_{d}^{\dagger}} \mbox{Tr}\Big({Y_d  Y_{d}^{\dagger}}\Big) -6 {m_d^2  Y_d  Y_{d}^{\dagger}} \mbox{Tr}\Big({Y_d  Y_{d}^{\dagger}}\Big) \nonumber \\ 
 & \ \ \ \ -12 {Y_d  m_q^2  Y_{d}^{\dagger}} \mbox{Tr}\Big({Y_d  Y_{d}^{\dagger}}\Big) -6 {Y_d  Y_{d}^{\dagger}  m_d^2} \mbox{Tr}\Big({Y_d  Y_{d}^{\dagger}}\Big) -8 m_{H_d}^2 {Y_d  Y_{d}^{\dagger}} \mbox{Tr}\Big({Y_e  Y_{e}^{\dagger}}\Big) \nonumber \\ 
 & \ \ \ \ -4 {T_d  T_{d}^{\dagger}} \mbox{Tr}\Big({Y_e  Y_{e}^{\dagger}}\Big) -2 {m_d^2  Y_d  Y_{d}^{\dagger}} \mbox{Tr}\Big({Y_e  Y_{e}^{\dagger}}\Big) -4 {Y_d  m_q^2  Y_{d}^{\dagger}} \mbox{Tr}\Big({Y_e  Y_{e}^{\dagger}}\Big) \nonumber \\ 
 & \ \ \ \ -2 {Y_d  Y_{d}^{\dagger}  m_d^2} \mbox{Tr}\Big({Y_e  Y_{e}^{\dagger}}\Big) -12 {Y_d  T_{d}^{\dagger}} \mbox{Tr}\Big({Y_{d}^{\dagger}  T_d}\Big) -4 {Y_d  T_{d}^{\dagger}} \mbox{Tr}\Big({Y_{e}^{\dagger}  T_e}\Big) \nonumber \\ 
 & \ \ \ \ -12 {T_d  Y_{d}^{\dagger}} \mbox{Tr}\Big({T_d^*  Y_{d}^{T}}\Big) -12 {Y_d  Y_{d}^{\dagger}} \mbox{Tr}\Big({T_d^*  T_{d}^{T}}\Big) -4 {T_d  Y_{d}^{\dagger}} \mbox{Tr}\Big({T_e^*  Y_{e}^{T}}\Big) \nonumber \\ 
 & \ \ \ \ -4 {Y_d  Y_{d}^{\dagger}} \mbox{Tr}\Big({T_e^*  T_{e}^{T}}\Big) -12 {Y_d  Y_{d}^{\dagger}} \mbox{Tr}\Big({m_d^2  Y_d  Y_{d}^{\dagger}}\Big) -4 {Y_d  Y_{d}^{\dagger}} \mbox{Tr}\Big({m_e^2  Y_e  Y_{e}^{\dagger}}\Big) \nonumber \\ 
 & \ \ \ \ -4 {Y_d  Y_{d}^{\dagger}} \mbox{Tr}\Big({m_l^2  Y_{e}^{\dagger}  Y_e}\Big) -12 {Y_d  Y_{d}^{\dagger}} \mbox{Tr}\Big({m_q^2  Y_{d}^{\dagger}  Y_d}\Big) \\ 
\beta_{m_u^2}^{(1)} & =  
-\frac{32}{15} g_{1}^{2} {\bf 1} |M_1|^2 -\frac{32}{3} g_{3}^{2} {\bf 1} |M_3|^2 +4 m_{H_u}^2 {Y_u  Y_{u}^{\dagger}} +4 {T_u  T_{u}^{\dagger}} +2 {m_u^2  Y_u  Y_{u}^{\dagger}} +4 {Y_u  m_q^2  Y_{u}^{\dagger}} \nonumber \\ 
 & \ \ \ \ +2 {Y_u  Y_{u}^{\dagger}  m_u^2} -4 \frac{1}{\sqrt{15}} g_1 {\bf 1} \sigma_{1,1} \\ 
\beta_{m_u^2}^{(2)} & =  
-\frac{128}{45} g_{3}^{2} \Big(15 g_{3}^{2} M_3  -2 g_{1}^{2} \Big(2 M_3  + M_1\Big)\Big){\bf 1} M_3^* -\frac{4}{5} g_{1}^{2} m_{H_u}^2 {Y_u  Y_{u}^{\dagger}} +12 g_{2}^{2} m_{H_u}^2 {Y_u  Y_{u}^{\dagger}} \nonumber \\ 
 & \ \ \ \ +24 g_{2}^{2} |M_2|^2 {Y_u  Y_{u}^{\dagger}} +\frac{4}{5} g_{1}^{2} M_1 {Y_u  T_{u}^{\dagger}} -12 g_{2}^{2} M_2 {Y_u  T_{u}^{\dagger}} -12 g_{2}^{2} M_2^* {T_u  Y_{u}^{\dagger}} \nonumber \\ 
 & \ \ \ \ +\frac{4}{45} g_{1}^{2} M_1^* \Big(8 \Big(75 g_{1}^{2} M_1  + 8 g_{3}^{2} \Big(2 M_1  + M_3\Big)\Big){\bf 1}  + 9 \Big(-2 M_1 {Y_u  Y_{u}^{\dagger}}  + {T_u  Y_{u}^{\dagger}}\Big)\Big)-\frac{4}{5} g_{1}^{2} {T_u  T_{u}^{\dagger}} \nonumber \\ 
 & \ \ \ \ +12 g_{2}^{2} {T_u  T_{u}^{\dagger}} -\frac{2}{5} g_{1}^{2} {m_u^2  Y_u  Y_{u}^{\dagger}} +6 g_{2}^{2} {m_u^2  Y_u  Y_{u}^{\dagger}} -\frac{4}{5} g_{1}^{2} {Y_u  m_q^2  Y_{u}^{\dagger}} \nonumber \\ 
 & \ \ \ \ +12 g_{2}^{2} {Y_u  m_q^2  Y_{u}^{\dagger}} -\frac{2}{5} g_{1}^{2} {Y_u  Y_{u}^{\dagger}  m_u^2} +6 g_{2}^{2} {Y_u  Y_{u}^{\dagger}  m_u^2} -4 m_{H_d}^2 {Y_u  Y_{d}^{\dagger}  Y_d  Y_{u}^{\dagger}} \nonumber \\ 
 & \ \ \ \ -4 m_{H_u}^2 {Y_u  Y_{d}^{\dagger}  Y_d  Y_{u}^{\dagger}} -4 {Y_u  Y_{d}^{\dagger}  T_d  T_{u}^{\dagger}} -8 m_{H_u}^2 {Y_u  Y_{u}^{\dagger}  Y_u  Y_{u}^{\dagger}} -4 {Y_u  Y_{u}^{\dagger}  T_u  T_{u}^{\dagger}} \nonumber \\ 
 & \ \ \ \ -4 {Y_u  T_{d}^{\dagger}  T_d  Y_{u}^{\dagger}} -4 {Y_u  T_{u}^{\dagger}  T_u  Y_{u}^{\dagger}} -4 {T_u  Y_{d}^{\dagger}  Y_d  T_{u}^{\dagger}} -4 {T_u  Y_{u}^{\dagger}  Y_u  T_{u}^{\dagger}} \nonumber \\ 
 & \ \ \ \ -4 {T_u  T_{d}^{\dagger}  Y_d  Y_{u}^{\dagger}} -4 {T_u  T_{u}^{\dagger}  Y_u  Y_{u}^{\dagger}} -2 {m_u^2  Y_u  Y_{d}^{\dagger}  Y_d  Y_{u}^{\dagger}} -2 {m_u^2  Y_u  Y_{u}^{\dagger}  Y_u  Y_{u}^{\dagger}} \nonumber \\ 
 & \ \ \ \ -4 {Y_u  m_q^2  Y_{d}^{\dagger}  Y_d  Y_{u}^{\dagger}} -4 {Y_u  m_q^2  Y_{u}^{\dagger}  Y_u  Y_{u}^{\dagger}} -4 {Y_u  Y_{d}^{\dagger}  m_d^2  Y_d  Y_{u}^{\dagger}} \nonumber \\ 
 & \ \ \ \ -4 {Y_u  Y_{d}^{\dagger}  Y_d  m_q^2  Y_{u}^{\dagger}} -2 {Y_u  Y_{d}^{\dagger}  Y_d  Y_{u}^{\dagger}  m_u^2} -4 {Y_u  Y_{u}^{\dagger}  m_u^2  Y_u  Y_{u}^{\dagger}} -4 {Y_u  Y_{u}^{\dagger}  Y_u  m_q^2  Y_{u}^{\dagger}} \nonumber \\ 
 & \ \ \ \ -2 {Y_u  Y_{u}^{\dagger}  Y_u  Y_{u}^{\dagger}  m_u^2} +\frac{32}{3} g_{3}^{4} {\bf 1} \sigma_{2,3} +\frac{32}{15} g_{1}^{2} {\bf 1} \sigma_{2,11} -16 \frac{1}{\sqrt{15}} g_1 {\bf 1} \sigma_{3,1} -24 m_{H_u}^2 {Y_u  Y_{u}^{\dagger}} \mbox{Tr}\Big({Y_u  Y_{u}^{\dagger}}\Big) \nonumber \\ 
 & \ \ \ \ -12 {T_u  T_{u}^{\dagger}} \mbox{Tr}\Big({Y_u  Y_{u}^{\dagger}}\Big) -6 {m_u^2  Y_u  Y_{u}^{\dagger}} \mbox{Tr}\Big({Y_u  Y_{u}^{\dagger}}\Big) -12 {Y_u  m_q^2  Y_{u}^{\dagger}} \mbox{Tr}\Big({Y_u  Y_{u}^{\dagger}}\Big) \nonumber \\ 
 & \ \ \ \ -6 {Y_u  Y_{u}^{\dagger}  m_u^2} \mbox{Tr}\Big({Y_u  Y_{u}^{\dagger}}\Big) -12 {Y_u  T_{u}^{\dagger}} \mbox{Tr}\Big({Y_{u}^{\dagger}  T_u}\Big) -12 {T_u  Y_{u}^{\dagger}} \mbox{Tr}\Big({T_u^*  Y_{u}^{T}}\Big) \nonumber \\ 
 & \ \ \ \ -12 {Y_u  Y_{u}^{\dagger}} \mbox{Tr}\Big({T_u^*  T_{u}^{T}}\Big) -12 {Y_u  Y_{u}^{\dagger}} \mbox{Tr}\Big({m_q^2  Y_{u}^{\dagger}  Y_u}\Big) -12 {Y_u  Y_{u}^{\dagger}} \mbox{Tr}\Big({m_u^2  Y_u  Y_{u}^{\dagger}}\Big) \\ 
\beta_{m_e^2}^{(1)} & =  
-\frac{24}{5} g_{1}^{2} {\bf 1} |M_1|^2 +2 \Big(2 m_{H_d}^2 {Y_e  Y_{e}^{\dagger}}  + 2 {T_e  T_{e}^{\dagger}}  + 2 {Y_e  m_l^2  Y_{e}^{\dagger}}  + {m_e^2  Y_e  Y_{e}^{\dagger}} + {Y_e  Y_{e}^{\dagger}  m_e^2}\Big)\nonumber \\ 
 & \ \ \ \ +2 \sqrt{\frac{3}{5}} g_1 {\bf 1} \sigma_{1,1} \\ 
\beta_{m_e^2}^{(2)} & =  
\frac{2}{5} \Big(-6 g_{1}^{2} m_{H_d}^2 {Y_e  Y_{e}^{\dagger}} +30 g_{2}^{2} m_{H_d}^2 {Y_e  Y_{e}^{\dagger}} +60 g_{2}^{2} |M_2|^2 {Y_e  Y_{e}^{\dagger}} +6 g_{1}^{2} M_1 {Y_e  T_{e}^{\dagger}} \nonumber \\ 
 & \ \ \ \ -30 g_{2}^{2} M_2 {Y_e  T_{e}^{\dagger}} -30 g_{2}^{2} M_2^* {T_e  Y_{e}^{\dagger}} +6 g_{1}^{2} M_1^* \Big(-2 M_1 {Y_e  Y_{e}^{\dagger}}  + 54 g_{1}^{2} M_1 {\bf 1}  + {T_e  Y_{e}^{\dagger}}\Big)\nonumber \\ 
 & \ \ \ \ -6 g_{1}^{2} {T_e  T_{e}^{\dagger}} +30 g_{2}^{2} {T_e  T_{e}^{\dagger}} -3 g_{1}^{2} {m_e^2  Y_e  Y_{e}^{\dagger}} +15 g_{2}^{2} {m_e^2  Y_e  Y_{e}^{\dagger}} \nonumber \\ 
 & \ \ \ \ -6 g_{1}^{2} {Y_e  m_l^2  Y_{e}^{\dagger}} +30 g_{2}^{2} {Y_e  m_l^2  Y_{e}^{\dagger}} -3 g_{1}^{2} {Y_e  Y_{e}^{\dagger}  m_e^2} +15 g_{2}^{2} {Y_e  Y_{e}^{\dagger}  m_e^2} \nonumber \\ 
 & \ \ \ \ -20 m_{H_d}^2 {Y_e  Y_{e}^{\dagger}  Y_e  Y_{e}^{\dagger}} -10 {Y_e  Y_{e}^{\dagger}  T_e  T_{e}^{\dagger}} -10 {Y_e  T_{e}^{\dagger}  T_e  Y_{e}^{\dagger}} -10 {T_e  Y_{e}^{\dagger}  Y_e  T_{e}^{\dagger}} \nonumber \\ 
 & \ \ \ \ -10 {T_e  T_{e}^{\dagger}  Y_e  Y_{e}^{\dagger}} -5 {m_e^2  Y_e  Y_{e}^{\dagger}  Y_e  Y_{e}^{\dagger}} -10 {Y_e  m_l^2  Y_{e}^{\dagger}  Y_e  Y_{e}^{\dagger}} -10 {Y_e  Y_{e}^{\dagger}  m_e^2  Y_e  Y_{e}^{\dagger}} \nonumber \\ 
 & \ \ \ \ -10 {Y_e  Y_{e}^{\dagger}  Y_e  m_l^2  Y_{e}^{\dagger}} -5 {Y_e  Y_{e}^{\dagger}  Y_e  Y_{e}^{\dagger}  m_e^2} +4 g_1 {\bf 1} \Big(3 g_1 \sigma_{2,11}  + \sqrt{15} \sigma_{3,1} \Big)\nonumber \\
 &  \ \ \ \ -60 m_{H_d}^2 {Y_e  Y_{e}^{\dagger}} \mbox{Tr}\Big({Y_d  Y_{d}^{\dagger}}\Big) -30 {T_e  T_{e}^{\dagger}} \mbox{Tr}\Big({Y_d  Y_{d}^{\dagger}}\Big) -15 {m_e^2  Y_e  Y_{e}^{\dagger}} \mbox{Tr}\Big({Y_d  Y_{d}^{\dagger}}\Big) \nonumber \\ 
 & \ \ \ \ -30 {Y_e  m_l^2  Y_{e}^{\dagger}} \mbox{Tr}\Big({Y_d  Y_{d}^{\dagger}}\Big) -15 {Y_e  Y_{e}^{\dagger}  m_e^2} \mbox{Tr}\Big({Y_d  Y_{d}^{\dagger}}\Big) -20 m_{H_d}^2 {Y_e  Y_{e}^{\dagger}} \mbox{Tr}\Big({Y_e  Y_{e}^{\dagger}}\Big) \nonumber \\ 
 & \ \ \ \ -10 {T_e  T_{e}^{\dagger}} \mbox{Tr}\Big({Y_e  Y_{e}^{\dagger}}\Big) -5 {m_e^2  Y_e  Y_{e}^{\dagger}} \mbox{Tr}\Big({Y_e  Y_{e}^{\dagger}}\Big) -10 {Y_e  m_l^2  Y_{e}^{\dagger}} \mbox{Tr}\Big({Y_e  Y_{e}^{\dagger}}\Big) \nonumber \\ 
 & \ \ \ \ -5 {Y_e  Y_{e}^{\dagger}  m_e^2} \mbox{Tr}\Big({Y_e  Y_{e}^{\dagger}}\Big) -30 {Y_e  T_{e}^{\dagger}} \mbox{Tr}\Big({Y_{d}^{\dagger}  T_d}\Big) -10 {Y_e  T_{e}^{\dagger}} \mbox{Tr}\Big({Y_{e}^{\dagger}  T_e}\Big) \nonumber \\ 
 & \ \ \ \ -30 {T_e  Y_{e}^{\dagger}} \mbox{Tr}\Big({T_d^*  Y_{d}^{T}}\Big) -30 {Y_e  Y_{e}^{\dagger}} \mbox{Tr}\Big({T_d^*  T_{d}^{T}}\Big) -10 {T_e  Y_{e}^{\dagger}} \mbox{Tr}\Big({T_e^*  Y_{e}^{T}}\Big) \nonumber \\ 
 & \ \ \ \ -10 {Y_e  Y_{e}^{\dagger}} \mbox{Tr}\Big({T_e^*  T_{e}^{T}}\Big) -30 {Y_e  Y_{e}^{\dagger}} \mbox{Tr}\Big({m_d^2  Y_d  Y_{d}^{\dagger}}\Big) -10 {Y_e  Y_{e}^{\dagger}} \mbox{Tr}\Big({m_e^2  Y_e  Y_{e}^{\dagger}}\Big) \nonumber \\ 
 & \ \ \ \ -10 {Y_e  Y_{e}^{\dagger}} \mbox{Tr}\Big({m_l^2  Y_{e}^{\dagger}  Y_e}\Big) -30 {Y_e  Y_{e}^{\dagger}} \mbox{Tr}\Big({m_q^2  Y_{d}^{\dagger}  Y_d}\Big) \Big)\\ 
\beta_{m_{F_+}^2}^{(1)} & =  
\frac{2}{5} g_1 \Big(-12 g_1 |M_1|^2  + \sqrt{15} \sigma_{1,1} \Big)\\ 
\beta_{m_{F_+}^2}^{(2)} & =  
\frac{8}{5} g_1 \Big(3 g_1 \sigma_{2,11}  + 81 g_{1}^{3} |M_1|^2  + \sqrt{15} \sigma_{3,1} \Big)\\ 
\beta_{m_{F_-}^2}^{(1)} & =  
-\frac{2}{5} g_1 \Big(12 g_1 |M_1|^2  + \sqrt{15} \sigma_{1,1} \Big)\\ 
\beta_{m_{F_-}^2}^{(2)} & =  
\frac{8}{5} g_1 \Big(3 g_1 \sigma_{2,11}  + 81 g_{1}^{3} |M_1|^2  - \sqrt{15} \sigma_{3,1} \Big)
\end{align}} 


\section{Renormalisation group equations for the 5D-SSM+$F^{\pm}$}\label{RGES5D}

\par In this appendix we supply the one-loop beta functions used in the main paper for the five dimensional model 1 and model 2, including the five dimensional Kaluza-Klein states and extra fields. We define $t=Log_{10}Q$ and $\beta_A=16\pi^2 d A/dt$.  It is useful to also define the power law contribution, which may be written equivalently as 
\be
(QR)^d=  10^t R\; .
\ee

\subsection{Gauge couplings}

\par The one-loop beta function for the gauge couplings if $t>\ln(1/R)/\ln(10)$ are given by
\be
16\pi^2  \frac{d g_i(t) }{dt}= b_{MSSM}^i g^3_i(t) + b^i_{5D}g^3_i(t)(S(t)-1),
\ee
where $i=1,2,3$ and $S( t )=R 10^t$, the power law contribution. For the $4DSSM+F^{\pm}$, $b^i=(39/5,1,-3)$ and for five dimensions $b^i_{5D}=(18/5,-2,-6)+4\eta$, where $\eta$ is the number of fermion generation in the bulk. The fine structure constants may be defined from $\alpha_i= g^2_i/4\pi$.

\subsection{Yukawa couplings}

\par The beta functions for the Yukawa couplings may be related to the matrices of anomalous dimensions
\be
\beta_Y^{ijk}=\gamma^i_n Y^{njk}+\gamma^i_n Y^{ink}+\gamma^k_n Y^{ijn}.
\ee

\subsubsection{Anomalous dimensions for model 1}

\begin{eqnarray}
\gamma_{\tilde{H}_u} &=& 3\mbox{Tr}\Big({Y_u  Y_{u}^{\dagger}}\Big)- \Big(\frac{3}{10} g_{1}^{2}+\frac{3}{2} g_{2}^{2} \Big) S(t)\\
\gamma_{\tilde{H}_d} &=& 3\mbox{Tr}\Big({Y_d  Y_{d}^{\dagger}}\Big)+\mbox{Tr}\Big({Y_e  Y_{e}^{\dagger}}\Big)- \Big(\frac{3}{10} g_{1}^{2}+\frac{3}{2} g_{2}^{2} \Big) S(t)\\
\gamma_{\tilde{F}^{\pm}} &=& {Y_F  Y_{F}^{\dagger}}-\frac{12}{10} g_{1}^{2} S(t)\\
\gamma_{\tilde{q}} &=& \Big(2\Big({Y_u  Y_{u}^{\dagger}}+{Y_d  Y_{d}^{\dagger}}\Big)- \Big(\frac{1}{15} g_{1}^{2}+3 g_{2}^{2}+\frac{16}{3} g_{3}^{2} \Big)\Big)S(t)\\
\gamma_{\tilde{u}} &=& \Big(4{Y_u  Y_{u}^{\dagger}}- \Big(\frac{16}{15} g_{1}^{2}+\frac{16}{3} g_{3}^{2} \Big)\Big)S(t)\\
\gamma_{\tilde{d}} &=& \Big(4{Y_d  Y_{d}^{\dagger}}- \Big(\frac{4}{15} g_{1}^{2}+\frac{16}{3} g_{3}^{2} \Big)\Big)S(t)\\
\gamma_{\tilde{l}} &=& \Big(2{Y_e  Y_{e}^{\dagger}}- \Big(\frac{3}{5} g_{1}^{2}+3 g_{2}^{2} \Big)\Big)S(t)\\
\gamma_{\tilde{e}} &=& \Big(4{Y_e  Y_{e}^{\dagger}}- \frac{12}{5} g_{1}^{2}\Big)S(t).
\end{eqnarray}

\subsubsection{Anomalous dimensions for model 2}

\begin{eqnarray}
\gamma_{\tilde{H}_u} &=& 3\mbox{Tr}\Big({Y_u  Y_{u}^{\dagger}}\Big)\pi S(t)^2- \Big(\frac{3}{10} g_{1}^{2}+\frac{3}{2} g_{2}^{2} \Big) S(t)\\
\gamma_{\tilde{H}_d} &=& \Big(3\mbox{Tr}\Big({Y_d  Y_{d}^{\dagger}}\Big)+\mbox{Tr}\Big({Y_e  Y_{e}^{\dagger}}\Big)\Big)\pi S(t)^2- \Big(\frac{3}{10} g_{1}^{2}+\frac{3}{2} g_{2}^{2} \Big) S(t)\\
\gamma_{\tilde{F}^{\pm}} &=& {Y_F  Y_{F}^{\dagger}}\pi S(t)^2-\frac{12}{10} g_{1}^{2} S(t)\\
\gamma_{\tilde{q}^3} &=& \Big({Y_t  Y_{t}^{\dagger}}+{Y_b  Y_{b}^{\dagger}}\Big)\pi S(t)^2- \Big(\frac{1}{30} g_{1}^{2}+\frac{3}{2} g_{2}^{2}+\frac{8}{3} g_{3}^{2} \Big)S(t)\\
\gamma_{\tilde{u}^3} &=& 2{Y_t  Y_{t}^{\dagger}}\pi S(t)^2- \Big(\frac{8}{15} g_{1}^{2}+\frac{8}{3} g_{3}^{2} \Big)S(t)\\
\gamma_{\tilde{d}^3} &=& 2{Y_b  Y_{b}^{\dagger}}\pi S(t)^2- \Big(\frac{2}{15} g_{1}^{2}+\frac{8}{3} g_{3}^{2} \Big)S(t)\\
\gamma_{\tilde{l}^3} &=& {Y_{\tau}  Y_{\tau}^{\dagger}}\pi S(t)^2- \Big(\frac{3}{10} g_{1}^{2}+\frac{3}{2} g_{2}^{2} \Big)S(t)\\
\gamma_{\tilde{e}^3} &=& 2{Y_{\tau}  Y_{\tau}^{\dagger}}\pi S(t)^2- \frac{6}{5} g_{1}^{2}S(t).
\end{eqnarray}

\subsubsection{Yukawa coupling RGEs for model 1}

\par The five dimensional contributions for model 1 are given by
\bea\label{beta_model1}
\beta^{(1)}_{(5D)Y_u} &=& Y_u \left( \left(6Y^{\dagger}_u Y_u + 2Y^{\dagger}_d Y_d +2{Y_{F}^{\dagger}  Y_F}\right)-  \left(\frac{79}{30}g_1^2+\frac{9}{2}g_2^2+\frac{32}{3}g_3^2\right) \right)S(t) \\
\beta^{(1)}_{(5D)Y_d} &=&  Y_d \left( \Big(6Y^{\dagger}_d Y_d + 2Y^{\dagger}_u Y_u+2{Y_{F}^{\dagger}  Y_F}\Big) -  \left(\frac{55}{30}g_1^2+\frac{9}{2}g_2^2+\frac{32}{3}g_3^2\right) \right)S(t) \\
\beta^{(1)}_{(5D)Y_e} &=&  Y_e \left( 6Y^{\dagger}_e Y_e+2{Y_{F}^{\dagger}  Y_F} -\left(\frac{9}{2}g_1^2+\frac{9}{2}g_2^2\right) \right)S(t)\\
\beta^{(1)}_{(5D)Y_F} &=&  Y_F \left( 4{Y_{F}^{\dagger}  Y_F}-\left(3g_1^2+3g_2^2\right) \right)S(t).
\eea

\subsubsection{Yukawa coupling RGEs for model 2}

\par The five dimensional contributions for model 2 are given by
\bea
\beta^{(1)}_{(5D)Y_t} &=& Y_t \left(3 \mbox{Tr}\Big(Y^{\dagger}_t Y_t\Big) + 3Y^{\dagger}_t Y_t+Y^{\dagger}_b Y_b +{Y_{F}^{\dagger}  Y_F}\right)\pi S(t)^2\nonumber\\
&&- Y_t\left(\frac{31}{15}g_1^2+3g_2^2+\frac{16}{3}g_3^2\right)S(t) \\
\beta^{(1)}_{(5D)Y_b} &=&  Y_b \left(3 \mbox{Tr}\Big(Y^{\dagger}_b Y_b\Big)+\mbox{Tr}\Big(Y^{\dagger}_{\tau} Y_{\tau}\Big)+ 3Y^{\dagger}_b Y_b+Y^{\dagger}_t Y_t +{Y_{F}^{\dagger}  Y_F}\right)\pi S(t)^2\nonumber\\
&&- Y_b\left(\frac{25}{15}g_1^2+3g_2^2+\frac{16}{3}g_3^2\right) S(t)\\
\beta^{(1)}_{(5D)Y_{\tau}} &=&  Y_{\tau} \left( 3 \mbox{Tr}\Big(Y^{\dagger}_b Y_b\Big)+\mbox{Tr}\Big(Y^{\dagger}_{\tau} Y_{\tau}\Big)+3Y^{\dagger}_{\tau} Y_{\tau}+{Y_{F}^{\dagger}  Y_F}\right)\pi S(t)^2\nonumber\\
&& -Y_{\tau}\left(\frac{15}{5}g_1^2+3g_2^2\right)S(t)\\
\beta^{(1)}_{(5D)Y_F} &=&  Y_F \left( 2{Y_{F}^{\dagger}  Y_F}+ 3 \mbox{Tr}\Big(Y^{\dagger}_b Y_b\Big)+3 \mbox{Tr}\Big(Y^{\dagger}_t Y_t\Big)+\mbox{Tr}\Big(Y^{\dagger}_{\tau}Y_{\tau}\Big)\right)\pi S(t)^2\nonumber\\
&&-\left(3g_1^2+3g_2^2\right)S(t).
\eea
Note that the evolution equations for $Y_{u,c}$, $Y_{d,s}$ and $Y_{e,\mu}$  can be read from Eq. \eqref{beta_model1}, since the first and second generation live on the brane.

\subsection{Trilinear soft breaking parameters}

\subsubsection{Trilinear soft breaking parameters for model 1}

\begin{eqnarray}
\beta^{(1)}_{(5D) T_u} &=& T_u \left(\left(18Y^{\dagger}_u Y_u + 2Y^{\dagger}_d Y_d +2Y^{\dagger}_FY_F\right)- \left(\frac{79}{30}g_1^2+\frac{9}{2}g_2^2+\frac{32}{3}g_3^2\right) \right)S(t)\nonumber\\
&&+ Y_u \left(4T_d Y^{\dagger}_d +4Y^{\dagger}_F T_F+ \frac{79}{15}g_1^2 M_1 + 9g_2^2 M_2 +\frac{64}{3}g_3^2 M_3 \right)S(t)\\
\beta^{(1)}_{(5D) T_d} &=& T_d \left(\left(18Y^{\dagger}_d Y_d + 2Y^{\dagger}_u Y_u +2Y^{\dagger}_F Y_F\right)-\left(\frac{55}{30}g_1^2+\frac{9}{2}g_2^2+\frac{32}{3}g_3^2\right)\right)S(t)\\
&& + Y_d \left(4Y^{\dagger}_F T_F+ 4T_u Y^{\dagger}_u  + 2T_e Y^{\dagger}_e+\frac{55}{15}g_1^2 M_1 + 9g_2^2 M_2 +\frac{64}{3}g_3^2 M_3 \right)S(t)\nonumber\\
\beta^{(1)}_{(5D) T_e} &=& T_e \left( 18Y^{\dagger}_e Y_e +2Y^{\dagger}_F Y_F- \left(\frac{9}{2}g_1^2+\frac{9}{2}g_2^2\right) \right)S(t)\nonumber\\
 &&+Y_e \left(6T_d Y^{\dagger}_d +4Y^{\dagger}_F T_F+\frac{18}{2}g_1^2 M_1 + 9g_2^2 M_2 \right)S(t)\\
\beta^{(1)}_{(5D) T_F} &=& T_F \left(12Y^{\dagger}_F Y_F- \left(3g_1^2+3g_2^2\right) \right)S(t)+ Y_F \left(6g_1^2 M_1 + 6g_2^2 M_2 \right)S(t).
\end{eqnarray}

\subsubsection{Trilinear soft breaking parameters for model 2}

\begin{eqnarray}
\beta^{(1)}_{(5D) T_t} &=& -T_t\left(\frac{31}{15}g_1^2+3g_2^2+\frac{16}{3}g_3^2\right)S(t)+ Y_t \left(\frac{62}{15}g_1^2 M_1 + 6g_2^2 M_2 +\frac{32}{3}g_3^2 M_3 \right)S(t)\nonumber\\
&&+Y_t\left(6 \mbox{Tr}\Big(Y^{\dagger}_t T_t\Big) + 6Y^{\dagger}_t T_t+2Y^{\dagger}_b T_b +2{Y_{F}^{\dagger}  T_F}\right)\pi S(t)^2\nonumber\\
&&+T_t \left(3 \mbox{Tr}\Big(Y^{\dagger}_t Y_t\Big) + 3Y^{\dagger}_t Y_t+Y^{\dagger}_b Y_b +{Y_{F}^{\dagger}  Y_F}\right)\pi S(t)^2 \\
\beta^{(1)}_{(5D) T_b} &=&-T_b\left(\frac{25}{15}g_1^2+3g_2^2+\frac{16}{3}g_3^2\right)S(t)+ Y_b \left(\frac{50}{15}g_1^2 M_1 + 6g_2^2 M_2 +\frac{32}{3}g_3^2 M_3 \right)S(t)\nonumber\\
&&+Y_b\left(6 \mbox{Tr}\Big(Y^{\dagger}_b T_b\Big)+2\mbox{Tr}\Big(Y^{\dagger}_{\tau} T_{\tau}\Big) + 6Y^{\dagger}_b T_b+2Y^{\dagger}_t T_t +2{Y_{F}^{\dagger}  T_F}\right)\pi S(t)^2\nonumber\\
&&+T_b \left(3 \mbox{Tr}\Big(Y^{\dagger}_b Y_b\Big)+\mbox{Tr}\Big(Y^{\dagger}_{\tau} Y_{\tau}\Big) + 3Y^{\dagger}_b Y_b+Y^{\dagger}_t Y_t +{Y_{F}^{\dagger}  Y_F}\right)\pi S(t)^2 \\
\beta^{(1)}_{(5D) T_{\tau}} &=&-T_{\tau}\left(3g_1^2+3g_2^2\right)S(t)+ Y_{\tau} \left(6g_1^2 M_1 + 6g_2^2 M_2 \right)S(t)\nonumber\\
&&+Y_{\tau}\left(6 \mbox{Tr}\Big(Y^{\dagger}_b T_b\Big)+2\mbox{Tr}\Big(Y^{\dagger}_{\tau} T_{\tau}\Big) + 6Y^{\dagger}_{\tau} T_{\tau}+2{Y_{F}^{\dagger}  T_F}\right)\pi S(t)^2\nonumber\\
&&+T_{\tau} \left(3 \mbox{Tr}\Big(Y^{\dagger}_b Y_b\Big)+\mbox{Tr}\Big(Y^{\dagger}_{\tau} Y_{\tau}\Big) + 3Y^{\dagger}_{\tau} Y_{\tau} +{Y_{F}^{\dagger}  Y_F}\right)\pi S(t)^2 \\
\beta^{(1)}_{(5D) T_F} &=&-T_F\left(3g_1^2+3g_2^2\right)S(t)+ Y_F \left(6g_1^2 M_1 + 6g_2^2 M_2 \right)S(t)\nonumber\\
&&+Y_F\left(6 \mbox{Tr}\Big(Y^{\dagger}_b T_b\Big)+6 \mbox{Tr}\Big(Y^{\dagger}_t T_t\Big)+2\mbox{Tr}\Big(Y^{\dagger}_{\tau} T_{\tau}\Big)+4{Y_{F}^{\dagger}  T_F}\right)\pi S(t)^2\nonumber\\
&&+T_F \left(3 \mbox{Tr}\Big(Y^{\dagger}_b Y_b\Big)+3 \mbox{Tr}\Big(Y^{\dagger}_t Y_t\Big)+\mbox{Tr}\Big(Y^{\dagger}_{\tau} Y_{\tau}\Big)+2{Y_{F}^{\dagger}  Y_F}\right)\pi S(t)^2.
\end{eqnarray}

\subsection{Soft mass parameters}

\subsubsection{Gaugino soft mass parameters}

\par The gaugino soft masses in 5D run following
\be
 \beta^{(1)}_{M_i}= 2b^{i} g^2_i M_i+ 2(S(t)-1)b^i_{5D}g^2_i M_i.
\ee

\subsubsection{Scalar soft mass parameters for model 1}

\bea
\beta_{m_q^2}^{(1)} & = & 
\Big(-\frac{8}{15} g_{1}^{2} {\bf 1} |M_1|^2 -\frac{64}{3} g_{3}^{2} {\bf 1} |M_3|^2 -12 g_{2}^{2} {\bf 1} |M_2|^2 +4 m_{H_d}^2 {Y_{d}^{\dagger} Y_d} +4 m_{H_u}^2 {Y_{u}^{\dagger} Y_u}\Big)S(t) \nonumber\\
&&+\Big(4 {T_{d}^{\dagger} T_d}+4 {T_{u}^{\dagger} T_u} +2{m_q^2 Y_{d}^{\dagger} Y_d}+2{m_q^2 Y_{u}^{\dagger} Y_u}+4 {Y_{d}^{\dagger} m_d^2 Y_d} +2{Y_{d}^{\dagger}  Y_d  m_q^2}\Big)S(t)\nonumber\\
&&+\Big(4 {Y_{u}^{\dagger} m_u^2 Y_u} +2{Y_{u}^{\dagger} Y_u  m_q^2}+\frac{2}{\sqrt{15}} g_1 {\bf 1} \sigma_{1,1}\Big)S(t) \\ 
\beta_{m_u^2}^{(1)} & = & 
\Big(-\frac{64}{15} g_{1}^{2} {\bf 1} |M_1|^2 -\frac{64}{3} g_{3}^{2} {\bf 1} |M_3|^2 +8 m_{H_u}^2 {Y_u Y_{u}^{\dagger}} +8 {T_u T_{u}^{\dagger}} +4 {m_u^2 Y_u Y_{u}^{\dagger}}\Big)S(t)\nonumber\\
 &&+\Big(8 Y_u m_q^2 Y_{u}^{\dagger}+ 4{Y_u Y_{u}^{\dagger} m_u^2} -4 \sqrt{\frac{2}{15}} g_1 {\bf 1} \sigma_{1,1}\Big)S(t) \\ 
\beta_{m_d^2}^{(1)} & =  &
\Big(-\frac{16}{15} g_{1}^{2} {\bf 1} |M_1|^2 -\frac{64}{3} g_{3}^{2} {\bf 1} |M_3|^2 +8 m_{H_d}^2 {Y_d Y_{d}^{\dagger}} +8 {T_d T_{d}^{\dagger}} +4 {m_d^2 Y_d Y_{d}^{\dagger}}\Big)S(t)\nonumber\\
&& +\Big(8 {Y_d m_q^2 Y_{d}^{\dagger}} +4 {Y_d Y_{d}^{\dagger} m_d^2} +2 \sqrt{\frac{2}{15}} g_1 {\bf 1} \sigma_{1,1}\Big)S(t) \\ 
\beta_{m_l^2}^{(1)} & = & 
\Big(-\frac{12}{5} g_{1}^{2} {\bf 1} |M_1|^2 -12 g_{2}^{2} {\bf 1} |M_2|^2 +4 m_{H_d}^2 {Y_{e}^{\dagger} Y_e} +4 {T_{e}^{\dagger} T_e} +2{m_l^2 Y_{e}^{\dagger} Y_e}\Big)S(t)\nonumber\\
&&+\Big(4 {Y_{e}^{\dagger} m_e^2 Y_e} +2{Y_{e}^{\dagger} Y_e m_l^2}- \sqrt{\frac{6}{5}} g_1 {\bf 1} \sigma_{1,1}\Big)S(t) \\ 
\beta_{m_e^2}^{(1)} & =  &
2 \Big( 4 m_{H_d}^2 {Y_e Y_{e}^{\dagger}} + 4 {T_e  T_{e}^{\dagger}} + 4 {Y_e m_l^2 Y_{e}^{\dagger}}  + 2{m_e^2  Y_e  Y_{e}^{\dagger}} + 2{Y_e  Y_{e}^{\dagger}  m_e^2}\Big)S(t)\nonumber \\ 
& &+\Big(2 \sqrt{\frac{6}{5}} g_1 {\bf 1} \sigma_{1,1} -\frac{48}{5} g_{1}^{2} {\bf 1} |M_1|^2\Big)S(t)
\eea 
In model 1 the two Higgs doublet soft masses obey the RGE's 
\bea
\beta_{m_{H_d}^2}^{(1)} & =  &
\Big(-\frac{6}{5} g_{1}^{2} |M_1|^2 -6 g_{2}^{2} |M_2|^2 - \sqrt{\frac{3}{5}} g_1 \sigma_{1,1} +6 m_{H_d}^2 \mbox{Tr}\Big({Y_d Y_{d}^{\dagger}}\Big)\Big)S(t)\nonumber\\
&&+\Big(2 m_{H_d}^2 \mbox{Tr}\Big({Y_e Y_{e}^{\dagger}}\Big)+6 \mbox{Tr}\Big({T_d^* T_{d}^{T}}\Big)+2 \mbox{Tr}\Big({T_e^* T_{e}^{T}}\Big)+6 \mbox{Tr}\Big({m_d^2 Y_d Y_{d}^{\dagger}}\Big)\Big)S(t)\nonumber\\
&&+\Big(2 \mbox{Tr}\Big({m_e^2 Y_e Y_{e}^{\dagger}}\Big)+2 \mbox{Tr}\Big({m_l^2 Y_{e}^{\dagger} Y_e}\Big) +6 \mbox{Tr}\Big({m_q^2 Y_{d}^{\dagger} Y_d}\Big)\Big)S(t)\\
\beta_{m_{H_u}^2}^{(1)} & =  &
\Big(-\frac{6}{5} g_{1}^{2} |M_1|^2 -6 g_{2}^{2} |M_2|^2 +\sqrt{\frac{3}{5}} g_1 \sigma_{1,1} +6 m_{H_u}^2 \mbox{Tr}\Big({Y_u Y_{u}^{\dagger}}\Big) \Big)S(t)\nonumber\\
&&+\Big(6 \mbox{Tr}\Big({T_u^* T_{u}^{T}}\Big) +6 \mbox{Tr}\Big({m_q^2 Y_{u}^{\dagger} Y_u}\Big) +6 \mbox{Tr}\Big({m_u^2 Y_u Y_{u}^{\dagger}}\Big)\Big)S(t)\\
\beta_{m_{F^{\pm}}^2}^{(1)} & = &
\Big(-\frac{24}{5} g_{1}^{2} |M_1|^2 + 2m_{H_{u,d}}^2 {Y_{F}^{\dagger}Y_F }+2{T_{F}^{\dagger} T_{F}}+m_{F^{\pm}}^2 {Y_{F}^{\dagger}Y_F }\Big)S(t)\nonumber\\
&&+\Big( {Y_{F}^{\dagger}Y_F }m_{F^{\pm}}^2\Big) S(t)
\eea

\subsubsection{Scalar soft mass parameters for model 2}

\bea
\beta_{m_{q_3}^2}^{(1)} & = & 
\Big(-\frac{2}{15} g_{1}^{2} {\bf 1} |M_1|^2 -\frac{32}{3} g_{3}^{2} {\bf 1} |M_3|^2 -6 g_{2}^{2} {\bf 1} |M_2|^2 \Big)S(t)+\Big(2 m_{H_d}^2 {Y_{b}^{\dagger} Y_b} +2 m_{H_u}^2 {Y_{t}^{\dagger} Y_t}\Big)\pi S(t)^2 \nonumber\\
&&+\Big(2 {T_{b}^{\dagger} T_b}+2 {T_{t}^{\dagger} T_t} +{m_{q_3}^2 Y_{b}^{\dagger} Y_b}+{m_{q_3}^2 Y_{t}^{\dagger} Y_t}+2 {Y_{b}^{\dagger} m_{d_3}^2 Y_b} +{Y_{b}^{\dagger}  Y_b  m_{q_3}^2}\Big)\pi S(t)^2\nonumber\\
&&+\Big(2 {Y_{t}^{\dagger} m_{u_3}^2 Y_t} +{Y_{t}^{\dagger} Y_t  m_{q_3}^2}\Big)\pi S(t)^2+\Big(\frac{1}{\sqrt{15}} g_1 {\bf 1} \sigma_{1,1}\Big)S(t) \\ 
\beta_{m_{u_3}^2}^{(1)} & = & 
\Big(-\frac{32}{15} g_{1}^{2} {\bf 1} |M_1|^2 -\frac{32}{3} g_{3}^{2} {\bf 1} |M_3|^2\Big)S(t) +\Big(4 m_{H_u}^2 {Y_t Y_{t}^{\dagger}} +4 {T_t T_{t}^{\dagger}} +2 {m_{u_3}^2 Y_t Y_{t}^{\dagger}}\Big)\pi S(t)^2\nonumber\\
 &&+\Big(4 Y_t m_{q_3}^2 Y_{t}^{\dagger}+ 2{Y_t Y_{t}^{\dagger} m_{u_3}^2}\Big)\pi S(t)^2 -\Big(4\frac{1}{\sqrt{15}} g_1 {\bf 1} \sigma_{1,1}\Big)S(t) \\ 
\beta_{m_{d_3}^2}^{(1)} & =  &
\Big(-\frac{8}{15} g_{1}^{2} {\bf 1} |M_1|^2 -\frac{32}{3} g_{3}^{2} {\bf 1} |M_3|^2\Big)S(t) +\Big(4 m_{H_d}^2 {Y_b Y_{b}^{\dagger}} +4 {T_b T_{b}^{\dagger}} +2 {m_{d_3}^2 Y_b Y_{b}^{\dagger}}\Big)\pi S(t)^2\nonumber\\
&& +\Big(4 {Y_b m_{q_3}^2 Y_{b}^{\dagger}} +2 {Y_b Y_{b}^{\dagger} m_{d_3}^2}\Big)\pi S(t)^2 + \Big(2\sqrt{\frac{1}{15}} g_1 {\bf 1} \sigma_{1,1}\Big)S(t) \\ 
\beta_{m_{l_3}^2}^{(1)} & = & 
\Big(-\frac{6}{5} g_{1}^{2} {\bf 1} |M_1|^2 -6 g_{2}^{2} {\bf 1} |M_2|^2\Big)S(t) +\Big(4 m_{H_d}^2 {Y_{\tau}^{\dagger} Y_{\tau}} +4 {T_{\tau}^{\dagger} T_{\tau}} +2{m_{l_3}^2 Y_{\tau}^{\dagger} Y_{\tau}}\Big)\pi S(t)^2\nonumber\\
&&+\Big(2 {Y_{\tau}^{\dagger} m_{e_3}^2 Y_{\tau}} +{Y_{\tau}^{\dagger} Y_{\tau} m_{l_3}^2}\Big)\pi S(t)^2- \Big(\sqrt{\frac{3}{5}} g_1 {\bf 1} \sigma_{1,1}\Big)S(t) \\ 
\beta_{m_{e_3}^2}^{(1)} & =  &
2 \Big( 2 m_{H_d}^2 {Y_{\tau} Y_{\tau}^{\dagger}} + 2 {T_{\tau}  T_{\tau}^{\dagger}} + 2 {Y_{\tau} m_{l_3}^2 Y_{\tau}^{\dagger}}  + {m_{e_3}^2  Y_{\tau}  Y_{\tau}^{\dagger}} + {Y_{\tau}  Y_{\tau}^{\dagger}  m_{e_3}^2}\Big)\pi S(t)^2\nonumber \\ 
& &+\Big(2 \sqrt{\frac{3}{5}} g_1 {\bf 1} \sigma_{1,1} -\frac{24}{5} g_{1}^{2} {\bf 1} |M_1|^2\Big)S(t)
\eea 
In model 2 the two Higgs doublet soft masses obey the RGE's 
\bea
\beta_{m_{H_d}^2}^{(1)} & =  &
\Big(-\frac{6}{5} g_{1}^{2} |M_1|^2 -6 g_{2}^{2} |M_2|^2 - \sqrt{\frac{3}{5}} g_1 \sigma_{1,1}\Big)S(t)  +\Big(6 m_{H_d}^2 \mbox{Tr}\Big({Y_b Y_{b}^{\dagger}}\Big)\Big)\pi S(t)^2\nonumber\\
&&+\Big(2 m_{H_d}^2 \mbox{Tr}\Big({Y_{\tau} Y_{\tau}^{\dagger}}\Big)+6 \mbox{Tr}\Big({T_b^* T_{b}^{T}}\Big)+2 \mbox{Tr}\Big({T_{\tau}^* T_{\tau}^{T}}\Big)+6 \mbox{Tr}\Big({m_{d_3}^2 Y_b Y_{b}^{\dagger}}\Big)\Big)\pi S(t)^2\nonumber\\
&&+\Big(2 \mbox{Tr}\Big({m_{e_3}^2 Y_{\tau} Y_{\tau}^{\dagger}}\Big)+2 \mbox{Tr}\Big({m_{l_3}^2 Y_{\tau}^{\dagger} Y_{\tau}}\Big) +6 \mbox{Tr}\Big({m_{q_3}^2 Y_{b}^{\dagger} Y_b}\Big)\Big)\pi S(t)^2\\
\beta_{m_{H_u}^2}^{(1)} & =  &
\Big(-\frac{6}{5} g_{1}^{2} |M_1|^2 -6 g_{2}^{2} |M_2|^2 +\sqrt{\frac{3}{5}} g_1 \sigma_{1,1}\Big)S(t) +\Big(6 m_{H_u}^2 \mbox{Tr}\Big({Y_t Y_{t}^{\dagger}}\Big) \Big)\pi S(t)^2\nonumber\\
&&+\Big(6 \mbox{Tr}\Big({T_t^* T_{t}^{T}}\Big) +6 \mbox{Tr}\Big({m_{q_3}^2 Y_{t}^{\dagger} Y_t}\Big) +6 \mbox{Tr}\Big({m_{u_3}^2 Y_t Y_{t}^{\dagger}}\Big)\Big)\pi S(t)^2\\
\beta_{m_{F^{\pm}}^2}^{(1)} & = &
-\frac{24}{5} g_{1}^{2} |M_1|^2S(t) + \Big(2m_{H_{u,d}}^2 {Y_{F}^{\dagger}Y_F }+2{T_{F}^{\dagger} T_{F}}+m_{F^{\pm}}^2 {Y_{F}^{\dagger}Y_F }\Big)\pi S(t)^2\nonumber\\
&&+\Big( {Y_{F}^{\dagger}Y_F }m_{F^{\pm}}^2\Big)\pi S(t)^2
\eea 

\subsection{Bilinear parameters $\mu$ and $B_{\mu}$}

\par In 5D  these are given by:
\bea 
\beta_{\mu}^{(1)} & = & \mu \Big(3 \mbox{Tr}\Big(Y^{\dagger}_u Y_u\Big)+3 \mbox{Tr}\Big(Y^{\dagger}_d Y_d\Big)+ \mbox{Tr}\Big(Y^{\dagger}_e Y_e\Big)-\frac{3}{5} g_{1}^{2} -3 g_{2}^{2} \Big) S(t)\\
\beta_{\acute{\mu}}^{(1)} & =  &
\Big(2 \acute{\mu}\Big({Y_F Y_{F}^{\dagger}}\Big) -\frac{12}{5} \acute{\mu} g_{1}^{2}\Big) S(t)\\
\beta^{(1)}_{B_{\mu}} & = &
  B_{\mu} \Big(-3 g_{2}^{2} - \frac{3}{5} g_{1}^{2}+ 3 \mbox{Tr}\Big(Y^{\dagger}_u Y_u\Big)+3 \mbox{Tr}\Big(Y^{\dagger}_d Y_d\Big)+ \mbox{Tr}\Big(Y^{\dagger}_e Y_e\Big)\Big)S(t)\nonumber\\
	&&+ \mu \Big(6 g_{2}^{2} M_2+\frac{6}{5}g_{1}^{2} M_1 +6 \mbox{Tr}\Big(Y^{\dagger}_u T_u\Big)+6 \mbox{Tr}\Big(Y^{\dagger}_d T_d\Big)+ 2\mbox{Tr}\Big(Y^{\dagger}_e T_e\Big)\Big)S(t)\\
\beta^{(1)}_{B_{\acute{\mu}}} & =  &
\Big( -\frac{12}{5} B_{\acute{\mu}} g_{1}^{2} +\frac{24}{5} \acute{\mu} g_{1}^{2} M_1+2B_{\acute{\mu}}Y^{\dagger}_FY_F +4\acute{\mu}Y^{\dagger}_FY_F\Big) S(t).
\eea 


\bibliographystyle{JHEP}
\bibliography{twoloop}

\end{document}